\pdfoutput=1 
\documentclass[aps,groupedaddress]{revtex4-2} % with prl after aps, sections are not numbered
\usepackage{array}
\usepackage{graphicx}
\usepackage[input-open-uncertainty =, input-close-uncertainty = , table-align-text-pre = false, tight-spacing = true]{siunitx} % for aligning table entries with decimal point, the extra options were for dealing with parenthesis ()
\usepackage{mathrsfs}
\usepackage{float}
\usepackage{color}
\usepackage{amssymb}
\usepackage{bm}
\usepackage{natbib}

\definecolor{burnt_orange}{RGB}{255,140,0}
\definecolor{royal_blue}{RGB}{0, 0, 204}
\definecolor{teal}{RGB}{7, 122, 147}
\definecolor{crimson}{RGB}{165, 62, 82}
\definecolor{purple}{RGB}{118, 9, 165}
\definecolor{magenta}{RGB}{255, 0, 255}
\definecolor{light_blue}{RGB}{91, 207, 244}
\definecolor{mygray}{gray}{0.5}
\definecolor{indigo}{RGB}{75,0,130}

\def\drawline#1#2{\raise 2.5pt\vbox{\hrule width #1pt height #2pt}}
\def\spacce#1{\hskip #1pt}

\def\bdash{\hbox{\spacce{2}\drawline{6}{0.25}}}

\def\dashed{\bdash\bdash\bdash}

\newcommand\Rey{\mbox{\textit{Re}}}  % Reynolds number

\begin{document}

% Use the \preprint command to place your local institutional report
% number in the upper righthand corner of the title page in preprint mode.
% Multiple \preprint commands are allowed.
% Use the 'preprintnumbers' class option to override journal defaults
% to display numbers if necessary
%\preprint{}

%Title of paper
\title{Aerodynamically-driven rupture of a liquid film by turbulent shear flow}

% repeat the \author .. \affiliation  etc. as needed
% \email, \thanks, \homepage, \altaffiliation all apply to the current
% author. Explanatory text should go in the []'s, actual e-mail
% address or url should go in the {}'s for \email and \homepage.
% Please use the appropriate macro foreach each type of information

% \affiliation command applies to all authors since the last
% \affiliation command. The \affiliation command should follow the
% other information
% \affiliation can be followed by \email, \homepage, \thanks as well.
\author{Melissa Kozul}
\email[]{melissa.kozul@ntnu.no}
%\homepage[]{Your web page}
%\thanks{}
%\altaffiliation{}
\affiliation{Department of Energy and Process Eng., NTNU, N-7491 Trondheim, Norway}

\author{Pedro S. Costa}
%\homepage[]{Your web page}
%\thanks{}
%\altaffiliation{}
\altaffiliation{Linn\'e FLOW Centre and SeRC (Swedish e-Science Research Centre), KTH Mechanics, SE-100 44 Stockholm, Sweden}
\affiliation{Faculty of Industrial Eng., Mechanical Eng. and Computer Science, 107 University of Iceland, Reykjav\'{\i}k, Iceland}

\author{James R. Dawson}
%\homepage[]{Your web page}
%\thanks{}
%\altaffiliation{}
\affiliation{Department of Energy and Process Eng., NTNU, N-7491 Trondheim, Norway}

\author{Luca Brandt}
%\homepage[]{Your web page}
%\thanks{}
\altaffiliation{Department of Energy and Process Eng., NTNU, N-7491 Trondheim, Norway}
\affiliation{Linn\'e FLOW Centre and SeRC (Swedish e-Science Research Centre), KTH Mechanics, SE-100 44 Stockholm, Sweden}

%Collaboration name if desired (requires use of superscriptaddress
%option in \documentclass). \noaffiliation is required (may also be
%used with the \author command).
%\collaboration can be followed by \email, \homepage, \thanks as well.
%\collaboration{}
%\noaffiliation

\date{\today}

\begin{abstract}
The rupture of a liquid film due to co-flowing turbulent shear flows in the gas phase is studied using a volume-of-fluid method. To simulate this multiphase problem, we use a simplified numerical setup where the liquid film is `sandwiched' between two fully developed boundary layers from a turbulent channel simulation. The film deforms and eventually ruptures within the shear zone created by the co-flows. This efficient setup allows systematic variation of physical parameters to gauge their role in the aerodynamically-driven deformation and rupture of a liquid film under fully developed sheared turbulence. The present work presents a detailed study of the developing pressure field over the deforming film and related aerodynamic effects, as previously suggested by other authors, in particular the role of the inviscid lift and drag forces. A cumulative lift force is introduced to capture the effect of the alternating pressure minima and maxima forming over the film which amplify and eventually rupture the film. A velocity scale derived from the lift-induced drag force reflects the state of the turbulent boundary layer over the film and collapses the temporal development of this cumulative lift force as well as the amplitude of film deformation with some success for the different film thicknesses and Reynolds numbers.

\end{abstract}

% insert suggested keywords - APS authors don't need to do this
%\keywords{}

%\maketitle must follow title, authors, abstract, and keywords
\maketitle

\section{Introduction} \label{sec:introduction}

Liquid fuel injection in the gas turbines of aircraft commonly utilize airblast atomization. This process exploits the kinetic energy of a flowing airstream to shatter the liquid fuel jet first into ligaments and then onward to drops \cite{lefebvre2010gas}. Thus the breakup of a liquid film (or jet) is commonly demarcated into two phases: primary atomization, where a coherent liquid film disintegrates into ligaments and then drops, and secondary atomization, the breakup of liquid drops into yet smaller liquid drops \cite{lefebvre2010gas, desjardins2010detailed}. Most modern airblast atomizers are of the prefilming type where, supplied from holes or slits upstream, the liquid fuel forms a thin film over the prefilming surface before being driven to the atomizing edge by turbulent flow. The use of a second air stream on the other side of the prefilmer to prevent fuel accumulation means the breakup and eventual atomization of the liquid film occurs in the shear zone formed by the co-flowing air streams \cite{aigner1988swirl}, akin to a classical mixing layer. To maximise combustion efficiency and minimise pollutant production, the liquid fuel should be well atomised as quickly as possible as it enters the combustion chamber. This will promote evaporation and mixing in the gas phase resulting in stable, low-emission combustion \cite{lefebvre2010gas}. The current work seeks to pare down this complex physical problem to the core process of liquid sheet breakup under the effect of a turbulent sheared flow. 

Many theoretical studies have investigated the shear-driven instability arising at the interface between a liquid film or sheet of finite thickness and a parallel-flowing gas. Given the symmetry of the problem about the central plane of the liquid, it is well-established only two types of waves are possible at the surface of a flat liquid sheet at any given wavenumber within the linear stability framework that is mostly used. The first type considered is the antisymmetric (sinusoidal) mode where the two sheet surfaces are in phase, the second type being the symmetric (dilational) mode where the two sheet surfaces are out-of-phase \cite{squire1953investigation}. Ref. \cite{rangel1991linear} expanded the range of density ratios considered by previous authors to show that for low gas-to-liquid density ratios $\rho_g / \rho_\ell \ll 1$, the growth rate of the sinusoidal waves is larger than that of dilational waves, agreeing with previous results in the literature, yet finding additionally that, as the density ratio approaches unity, dilational waves become more unstable than sinusoidal ones. Whilst earlier stability studies (e.g. \cite{squire1953investigation,rangel1991linear}) focused on inviscid approaches to predict liquid sheet oscillation frequencies, wavelengths and growth rates, more recent efforts have accounted for viscous effects (e.g. \cite{lozano2001longitudinal, otto2013viscous, agarwal2018closer}), the inclusion of which generally improves predictions when compared to experimental work. However, the study \cite{lozano2001longitudinal} found that even when viscosity is considered, quantitative agreement is not satisfactory, indicating that the true phenomenon is more complex and probably beyond the capabilities of linear analysis. The same study also demonstrated the inherently three-dimensional nature of the atomization process, with active participation of transverse waves and longitudinal filaments. The numerical study \cite{deshpande2015computational} concluded that the length scale of the most unstable mode in the linear regime is too small to fragment the entire liquid sheet and that it therefore cannot be directly responsible for its atomization. By considering the spectrum of surface wavelengths preceding primary atomization, they found standard linear theories tend to underestimate the dominant length scales associated with atomization by approximately 2 to 3 orders of magnitude, constituting a serious discrepancy from their volume-of-fluid (VoF) simulations. It was emphasised in Ref. \cite{deshpande2015computational} that their numerical results displayed good agreement with linear theories during the initial development of interfacial instabilities within the linear regime. Nevertheless, they showed these dominant modes were clearly not the ones responsible for atomization. The authors point instead to the emergence of a much larger mode following the linear window, with a wavelength approximately an order of magnitude larger than the most unstable wavelength predicted in the linear regime, as being that which eventually causes breakup. Initial (linear) instabilities do lead to the breakup of a small portion of the liquid sheet (i.e. the bottom and top surfaces), however, due to their small magnitude, these instabilities leave the liquid core intact and therefore do not directly lead to atomization of the sheet. In a subsequent study of a liquid jet \cite{agarwal2018closer}, it was confirmed that, while the most unstable modes are captured in the simulations and agree with theoretical predictions, these modes were once again found to not be directly responsible for fragmenting the liquid core or causing primary atomization. Their action is limited to stripping off the surface of the liquid jet, while its liquid core remained intact for another 20 nozzle diameters downstream.
% PhysFluids_1991_3_2392_Rangel_Sirignano: excellent reference showing clear images of dilatational/sinusoidal sheet instabilities.

% note here modelling work for laminar gas flows: check Villermaux, "Fragmentation" Ann. Rev. Fluid Mech.

For turbulent shear flows over liquid films leading to breakup, experiments are largely confined to observations of the disintegrating sheet to determine breakup length (e.g. \cite{carvalho1998liquid, mansour1991dynamic}), documentation of liquid sheet topology as well as determining broad breakup processes or regimes. There has additionally been a heavy focus on measurements of the liquid particle diameter far downstream once the sheet has atomised \cite{gepperth2010pre}. The phenomenological study in \cite{mansour1991dynamic} measured the frequency of an oscillating sheet as well as breakup length deduced from experimental photographs. The chief experimental difficulty in investigating the primary breakup phase lies in obtaining an accurate description of the flow field across the gas-liquid interface. More recently, novel interfacial velocimetry techniques have been developed \cite{duke2012experimental, chaussonnet2020influence} allowing observation of primary breakup, where laboratory measurements have typically been restricted to secondary breakup, which shows more generality. Ref. \cite{chaussonnet2020influence} found the aerodynamic stress $\tau_g = \rho_g U_g^2$, where $\rho_g$ is the gas density and $U_g$ the mean gas velocity, to be an important quantity governing primary breakup. The study emphasised the phenomenon of accumulation breakup, where the liquid sheet always disintegrates at the edge of the atomizer, and closely investigated different breakup sequences. The experimental study of Ref. \cite{dejean2016part2} created a regime map delineating liquid sheet breakup into three different regimes (`smooth', `waves', `accumulation') depending on operating conditions, and developed a correlation 
%(their equation 8) 
between breakup length and film thickness. 
%and furthermore showed that the liquid sheet dynamics have similar behaviour to flapping flags. 
The experiment of Ref. \cite{shavit2001gas} injected a water jet coaxially into a high-velocity gas jet, and increased the air turbulence while maintaining the mean air velocity, finding increased lateral motion to be the most obvious visual effect. The authors found jet dynamics to be less symmetric, with a shortened breakup length and more complex liquid ligament shapes. Moreover, Ref. \cite{shavit2001gas} understood the breakup mechanism under high turbulence intensity to be driven by `overturning' of the liquid jet, which leads to liquid perpendicularly aligned with the gas stream, resulting in strong normal forces that `blow up' the liquid to generate thin membranes that then proceed to secondary breakup. The experiments of Ref. \cite{kourmatzis2015air} investigated the role of shear-driven gas-phase turbulence on the primary atomization and break-up morphology of non-turbulent liquid jets. They found the ratio of the turbulent Weber number to the mean Weber number to be a relevant parameter, as was the turbulence intensity, in dictating the primary break-up length. An alternate breakup regime was described by the experimental study of Ref. \cite{park2004experimental} where the emergence of cellular structures was considered as being the dominant rupture mechanism of the sheet. The emerging `cells' in the deforming liquid sheet were described as comprising thin films of a higher amplitude surrounded by ligaments of a lower amplitude.

Given the limitations of laboratory measurements and theoretical modelling, numerical simulations are well-poised to make a substantial contribution, particularly in the primary breakup stage. The rapid development of multiphase simulation methods combined with increasing computing power means interface capturing simulations of multiphase flows within the fully turbulent regime are now feasible, to capture the three-dimensionality and unsteadiness of realistic flows. Recently, simulations have been conducted of a semi-realistic film breakup system \cite{ling2019two} including a splitter plate, with a turbulent gas only on one side of the film and a prescribed error function profile for the gas-phase velocity over the liquid film to model the boundary layer coming off the splitter plate. This study investigated the grid resolution necessary to faithfully capture downstream development of terms in the kinetic energy budget and the smallest size of droplets formed in atomization, finding converged results of the turbulent dissipation only with the finest mesh they used. They found agreement with linear stability theory both in the dominant frequency for the wave formation, deduced from spectra of the interfacial height near the edge of the splitter plate where wave amplitudes remain small, as well as in the exponential growth in the mixing layer thickness for an intermediate distance downstream of the splitter plate. They reported that simulation results deviated from linear theory as the amplitude of the interfacial wave grows and propagates downstream and non-linear effects become important. Despite making approximations, this study required a costly numerical setup where it is possibly unfeasible to perform parametric studies to investigate the influence of flow parameters and material properties. Another recent numerical study used a similar setup including a splitter plate \cite{jiang2020destabilization}, once again imposing a turbulent gas only on one side of the film, and investigated the role of turbulence by superimposing different intensities of disturbances on the error function profiles for the mean streamwise inlet gas velocity. They compared their simulations results to predictions of linear stability analysis incorporating a simple turbulent eddy viscosity model, and found that while linear theory was able to capture the trend of increasing dominant frequency and spatial growth rate with increasing gas inlet turbulence intensity, it significantly underestimated the values themselves. Recently, \cite{warncke2019new} used direct numerical simulation (DNS) of the flow in the vicinity of the splitter plate, which, embedded within a larger large-eddy simulation (LES), was used to simulate a realistic turbulent nozzle flow. This was compared with a constant boundary layer flow as used in previous numerical studies. The study found the Sauter mean diameter (SMD) of the spray was reduced by almost 20\% if the turbulence was taken into account demonstrating that the turbulent scales in the gaseous flow increase the amount of fine droplets. 

%\cite{krolick2018primary} applied dynamic mode decomposition (DMD) to the outer core radius of data resulting from simulation of an atomizing liquid jet in order to deduce the dominate spatial modes of the system and their temporal characteristics. 

Quantifying the forces acting on a liquid film requires access to the pressure field which is difficult, if not impossible to deduce experimentally, yet the enhancement of liquid structure breakup through the aerodynamic lift effect has been previously suggested \cite{wu1993aerodynamic, desjardins2010detailed} underlining the potentially key role of aerodynamic forces in atomization. It is well understood that, for $\rho_\ell/\rho_g < 500$, aerodynamic effects greatly influence liquid breakup \cite{wu1993aerodynamic}, this ratio being on the order of 100 for aircraft engines \cite{desjardins2010detailed}. Unlike in laboratory experiments of this physical problem, the present simulations have full access to the velocity and pressure fields. We will therefore directly investigate principally inviscid aerodynamic effects previously only suggested by other authors, our focus being on large film deformations beyond the linear regime which directly result in film rupture. 
%, seeking to quantify the inviscid lift and drag forces resulting from the developing pressure field over the deforming film. 
To do this, we consider a simplified numerical setup using a recently developed VoF method \cite{ii2012interface}. A stationary liquid film is sandwiched between fully developed sheared turbulent gas flows from a precursor simulation of a turbulent channel flow. This setup allows us to systematically vary parameters such as film thickness and turbulent gas flow Reynolds number to gauge the effect upon momentum transfer into the initially stationary liquid film. Aerodynamic quantities such as lift and lift-induced drag are considered to describe the film while it is still intact. The recent simulations of Ref. \cite{zandian2018understanding} utilized as similar `sandwich' setup but prescribed initial interfacial perturbations, for example a sinusoidal surface on a liquid film, as did Ref. \cite{lozano1998instability}. The numerical study of Ref. \cite{trontin2010direct} place a liquid sheet between homogeneous isotropic turbulence with no liquid or density jumps. They demonstrated that, even in the absence of mean shear, turbulence will provoke primary breakup in a liquid sheet if the Weber number is high enough, therefore suggesting the importance of realistically representing the gas-phase turbulence. There has been a keen focus on primary instabilities and minute topological changes \cite{zandian2018understanding,desjardins2010detailed}, whereas the present work seeks to investigate the influence of aerodynamic effects and material properties on global phenomena such as deformation amplitude and time to breakup. 
% The experiment of Ref. \cite{chaussonnet2020influence} utilized the friction velocity $u_\tau$ whose value was close to that of the upstream boundary layer, pointing to the physical relevance of the present approach. 
This novel numerical setup using mirrored turbulent channel velocity fields is both simple and effective. The present technique considers a more realistic turbulent gas phase acting directly on the liquid surface without pre-determined surface perturbations or introduced scales, aiming to understand how the large deformations of the liquid film leading to rupture arise spontaneously from the turbulence-laden gas phase, as they do in practical engineering applications. 

\section{Mathematical formulation} \label{sec:mathFormulation}

The present simulations use a recently-implemented VoF method for multiphase flow simulations \cite{ii2012interface, rosti2019numerical}. This interface-capturing approach has the major strength of ensuring mass conservation by construction, a property particularly desirable for the present physical problem since statistics will be collected over a considerable simulation time. Hereafter, we refer to fluctuating velocities $u$, $v$ and $w$ in the $x$-streamwise, $y$-wall- or film-normal and $z$-spanwise directions. In that which follows, the subscript $g$ will refer to the turbulent gas phase, and subscript $\ell$ to the liquid phase, representing the initially quiescent film. The following is a summary of the formulation presented in Ref. \cite{rosti2019numerical}. The gas-liquid (two-fluid) motion is governed by the conservation of momentum and the incompressibility constraint. Additionally, the kinematic and dynamic interactions between the two fluid phases are determined by enforcing continuity of the velocity at the interface between the two phases:
 \begin{equation}
  \mathbf{u}^\ell = \mathbf{u}^g,
 %u_i^\ell = u_i^g,
 \end{equation}
 and imposing the following surface stress boundary condition at the interface \cite{brackbill1992continuum}: 
 \begin{equation}
  \mathbf{\sigma}^\ell \cdot \mathbf{n} = \mathbf{\sigma}^g \cdot \mathbf{n} + \gamma \kappa \mathbf{n},
 %\sigma_{ij}^\ell n_j = \sigma_{ij}^g n_j + \gamma \kappa n_i,
 \end{equation}
 where $\mathbf{\sigma} = \sigma_{ij}$ is the Cauchy stress tensor, $\mathbf{n}$ a vector normal to the interface, $\kappa$ the interface curvature and $\gamma$ the surface tension coefficient, assumed to be constant in the present formulation. The VoF method following Ref. \cite{ii2012interface} is used to numerically solve the two-phase flow problem. An indicator (or colour) function $H$ is introduced to identify each phase, in the present methodology $H = 1$ within the liquid phase, and $H = 0$ within the gas phase. $H$ is updated according to the following advection equation:
 \begin{equation}
  \frac{\partial \phi}{\partial t} + \nabla \cdot (\mathbf{u} H^{ht}) = \phi \nabla \cdot \mathbf{u}, 
% \frac{\partial \phi}{\partial t} + \frac{\partial u_i H^{ht}}{\partial x_i} = \phi \frac{\partial u_i}{\partial x_i}, % not sure if I fully understand this equation: is it because the grad 
 \end{equation}
 where $\mathbf{u}$ is the local fluid velocity and $\phi$ is the cell-averaged value of the indicator function $H$. $H^{ht}$ is the hyperbolic tangent function approximating the indicator function $H$, which is determined along with the curvature $\kappa$ and the surface normal vector $\mathbf{n}$ via the multi-dimensional tangent of hyperbola for interface capturing (MTHINC) method \cite{ii2012interface}. Since we have set $\phi$ to be the cell-averaged volume fraction of the liquid, $\phi = 1$ if a cell is fully filled with the liquid phase, $\phi = 0$ when only gas fills the cell, and $0 < \phi < 1$ at the diffuse interface. % check this, Rosti et al. (2018) had it written the other way around, strange since they wrote H = 1 within f1, and \phi is the cell-average of H, although this here is how it's written in Scapin 2020. 
 Once $\phi$ is known, the two-fluid equations may be written in the so-called one-continuum formulation, meaning only one set of equations is solved over the entire computational domain. Thus the single velocity vector $\mathbf{u}$ field is defined everywhere, governed by the following equations: 
 \begin{equation}
 \rho \left( \frac{\partial \mathbf{u}}{\partial t} + \mathbf{u} \cdot \nabla \mathbf{u} \right)  = -\nabla p + \nabla \cdot\mu(\nabla \mathbf{u} + \nabla \mathbf{u}^T) + \gamma \kappa \mathbf{n} \delta, \hspace{10mm} \nabla \cdot \mathbf{u} = 0,
% \frac{\partial u_i}{\partial t} + \frac{\partial u_i u_j}{\partial x_j} + \frac{1}{\rho} \left( \frac{\partial \gamma_{ij}}{\partial x_j}  + f_i \right), \hspace{5mm} \frac{\partial u_i}{\partial x_i} = 0,
 \end{equation}
and deduced by applying a volume averaging procedure, where $p$ is the pressure, $\rho$ is the density, $\mu$ the dynamic viscosity and $\delta$ is the Dirac delta function at the interface of the two fluids. 
% The stress $\gamma_{ij}$ is written in the mixed form:
%\begin{equation} \label{eqn:mixedStress}
%\gamma_{ij} = \phi\, \gamma_{ij}^\ell + (1-\phi)\,\gamma_{ij}^g.
%\end{equation} 
%Both the liquid and gas are assumed to be Newtonian such that their stress tensors may be written $\gamma_{ij} = -p \delta_{ij} + 2\mu D_{ij}$, for $p$ pressure, $\delta_{ij}$ Kronecker delta, $\mu$ dynamic viscosity and $D_{ij} = (\partial u_i/\partial x_j + \partial u_j/\partial x_i)/2$ is the strain rate tensor. 
The density $\rho$ and dynamic viscosity $\mu$ are defined as mixed forms:
\begin{equation}
\rho = \phi \rho^\ell + (1-\phi)\rho^g,  \hspace{5mm} \mu = \phi \mu^\ell + (1-\phi)\mu^g.
\end{equation}

\section{Numerical methodology} \label{sec:method}

\begin{figure}
	\begin{center}
		
		\includegraphics[scale=0.5,clip]{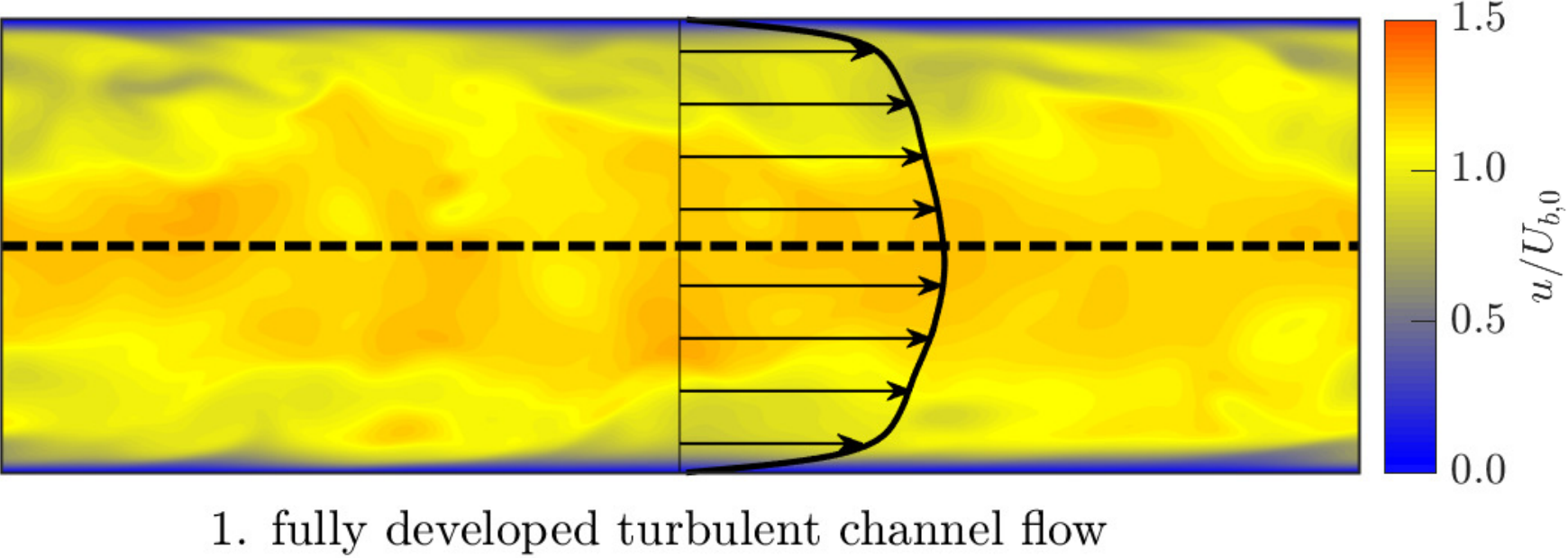}		
		
		\vspace{4mm}
		
		\includegraphics[scale=0.5,clip]{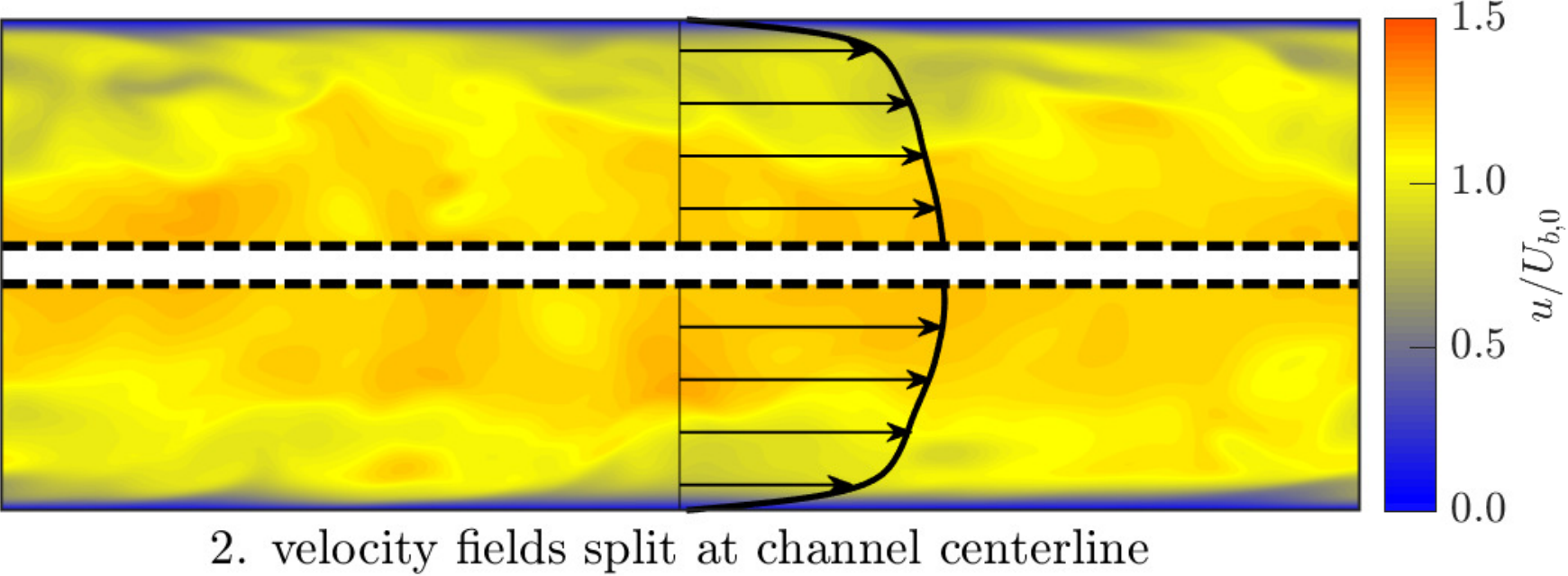}
		
		\vspace{4mm}
		
		\includegraphics[scale=0.5,clip]{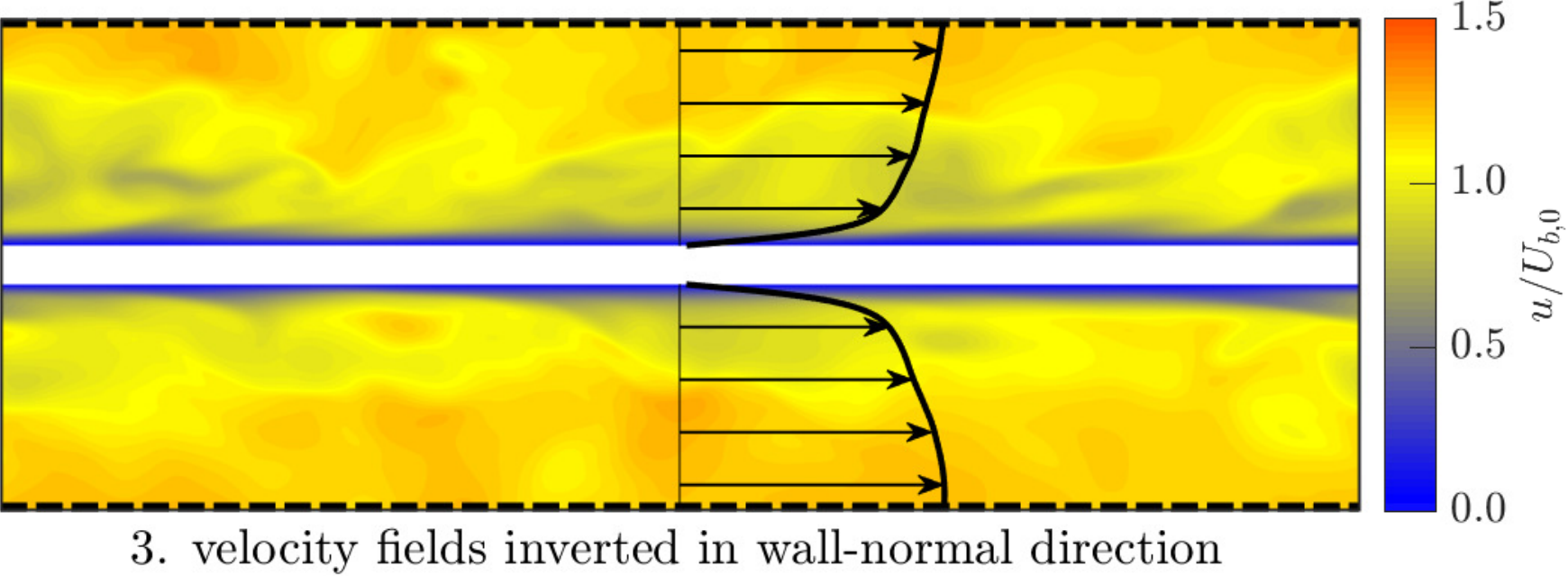}		
		
		\vspace{4mm}
		
		\includegraphics[scale=0.5,clip]{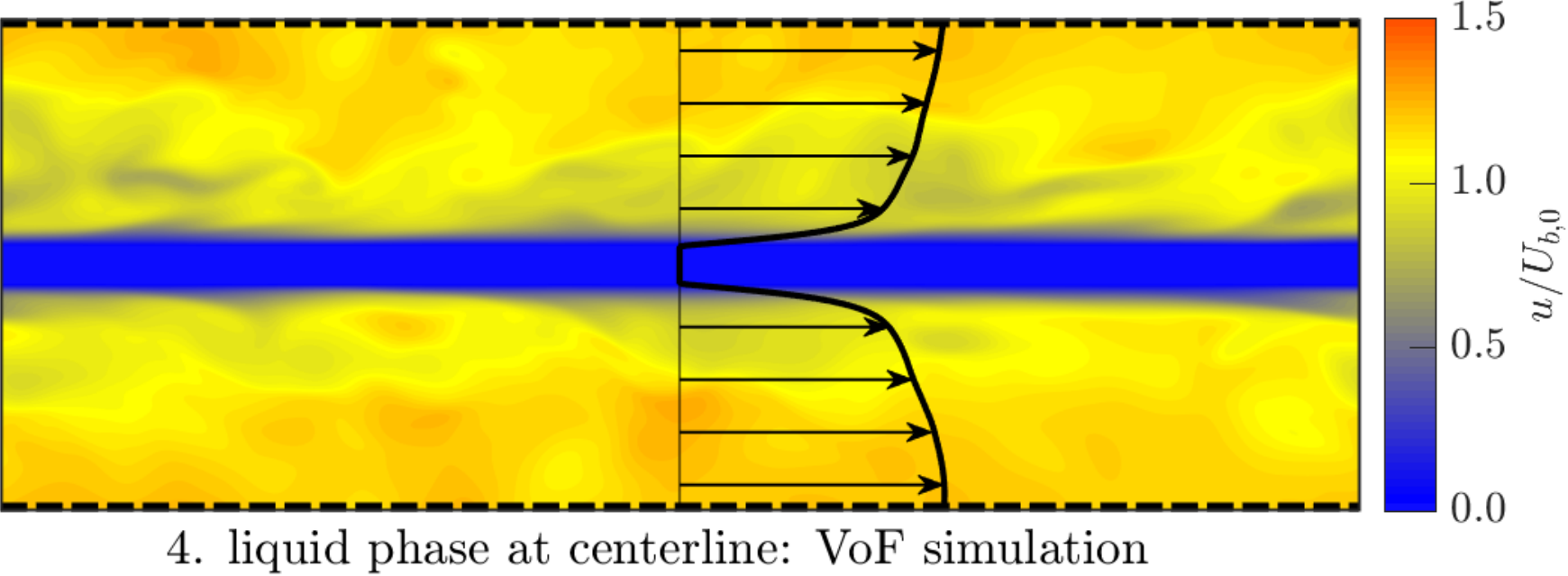}
		
		\caption{Four-step method to set up the VoF simulations: use of velocity fields from fully developed turbulent channel flow simulations. Shown here are contours of the streamwise velocity however the procedure is applied to all three velocity components to form the starting fields of the VoF simulations. Thick black contour indicates the initial mean streamwise velocity profile. Dashed lines indicate how a single plane at the centerline is recast as a periodic boundary condition, meaning the flow field is still numerically contiguous in the `sandwich' VoF setup.}
		\label{fig:simulationSetup}
	\end{center}
\end{figure}

Figure \ref{fig:simulationSetup} shows how the initial velocity fields for the VoF simulations are prepared. The velocity fields are from a fully developed turbulent channel flow with walls at the top and bottom of the domain, simulated using a single-phase channel solver \cite{costa2018fft}. Table \ref{tab:ChannelSimulations} summarizes the numerical parameters of the channel simulations. Split at the channel centerline, the fully developed fields are inverted in the wall-normal (or film-normal) direction $y$ and the liquid film placed between the turbulent shear layers in the $xz$-plane to form the novel `sandwich' setup. 

Table \ref{tab:FilmSimulations} summarises the parameters of liquid the film simulations, where subscript $0$ refers to parameters at the instant the film simulations are started. The VoF is set to 1 in the region between the inverted shear layers, and zero elsewhere. The initial velocity is zero within the film `sandwiched' between the inverted shear layers to model the action of fully developed wall-bounded turbulence on the gas-liquid interface on both sides of the film. Ref. \cite{fuster2013instability} found the liquid boundary layer does not play an important role in the gas-liquid interface interaction. The resulting periodic boundary condition from the inverted channel fields is exploited at the top and bottom of the domain. That is, flow that was contiguous at the centerline is now contiguous via the periodic boundary condition. Periodic boundary conditions are also used in the streamwise and spanwise directions, and no flow forcing is applied once the film simulations are started. The present cases are therefore temporal simulations. In analogy to the spatial system of the physical prefilmer, the flow when the VoF simulation is launched is a model for that at the edge of the atomizer as the fuel film leaves the atomizing edge. The flow's development in time is then analogous to that of the flow as it moves away from the atomizing edge of the prefilmer in the spatial system, where the mean shear of the gas phase decays in the streamwise direction. The initial condition thus formed inherently respects the boundary conditions of the physical problem, ensuring the initial condition rapidly adapts to the full multiphase flow problem. Numerical artefacts due to adjustments of the flow to the initial condition are thus rapidly mitigated.

This numerical setup permits systematic variation of parameters known to play an important role in the liquid film's primary breakup, such as film thickness and Reynolds number of the turbulent gas phase (effectively, the initial shear stresses felt by the surface of the film, which had been at the walls in the channel flow simulations). Liquid-to-gas density and viscosity ratios are $\mu_l/\mu_g = \rho_l/\rho_g = 40$. The initial Weber number is set as $We_0 = \rho_l U_{b,0}^2 w_{l,0} / \gamma = 500$ (note since our liquid is initially at rest, the present Weber number is defined solely in terms of the gas-phase velocity), where $U_{b,0}$ is the bulk velocity of the channel flow, $w_{l,0}$ is the initial film thickness and $\gamma$ is the surface tension coefficient, as a direct comparison to previous numerical work \cite{desjardins2010detailed}.  An alternate, temporally developing Weber number based on the streamwise wavelength of the film's deformation will be considered in Sec. \ref{sec:filmScales} (figure \ref{fig:LambdaA}d). Three film thicknesses are considered (suffixes to case names `thin', `med' and `thick' refer to the physical thickness $w_{\ell,0}$ scaled by the channel simulation half-height $h$). The role of Reynolds number is presently investigated in a limited sense with one of the cases using a turbulent gas phase at a higher starting Reynolds number of $Re_{\tau,0} = 393$ (HiRe-thin); all other cases begin with a turbulent gas phase at $Re_{\tau,0} = 180$.

\begin{table}
	\setlength{\tabcolsep}{8pt} % change this to widen/narrow columns
	\def\arraystretch{1.1}
	\begin{center}	
		\begin{tabular}{cccccS[table-format=1.2]S[table-format=1.2]S[table-format=1.2]}  
			\hline \hline 
			$\Rey_{b}$  & $\Rey_{\tau}$ & $N_x$ & $N_y$ & $N_z$ & $\Delta x^{+}$ & $\Delta y^{+}$  & $\Delta z^{+}$ \\  
			\hline
			5600           & 180             & 1152     & 384      & 576     &  0.94    & 0.94     & 0.94   \\ 
			13750          & 393            & 2496    & 832      &  1248   & 0.94    & 0.94     & 0.94   \\ 
			\hline \hline 			
		\end{tabular}
		\caption{Parameters for the channel simulations used to form the turbulent gas phase in the multiphase simulations. Both channel simulations have a domain size $L_x/h = 6$, $L_z/h = 3$, where $h$ is the channel half-height. A constant grid spacing was used in all three dimensions. Superscript `+' indicates viscous scaling (e.g. $\Delta x^{+} = \Delta_x u_{\tau}/\nu_g$).}
		\label{tab:ChannelSimulations} 
	\end{center}	
\end{table}

A uniform grid is used in all three dimensions, with grid spacing $\Delta x_0^+ = \Delta x\, u_{\tau,0}/\nu_g = \Delta y_0^+ = \Delta z_0^+ < 1$, where $u_{\tau,0}$ is the steady state friction velocity of the gas phase from the channel simulations, and $\nu_g = \nu_l$ is the gas phase kinematic viscosity. The adequacy of the grid resolution, a contentious topic in multiphase simulations \cite{desjardins2010detailed}, was investigated with a companion simulation to LoRe-thin named `LoRe-thin-loRes' using a coarser grid to that of LoRe-thin (see Appendix), equivalent to the grid resolution used herein for the other cases of table \ref{tab:FilmSimulations}. A high level of similarity was found between the two thin film simulations with different resolutions, although some differences emerge once the film ruptures, since topology changes in VoF methods occur when liquid structures approach the grid size \cite{tryggvason2011direct}. Therefore the lower resolution is found to be satisfactory for capturing the film's deformation whilst still intact, being the route to rupture, which is the concern of the present work. We note that the `thin' film case is the most demanding since we can expect smaller liquid phase features to arise for thinner liquid films. This suggests that if the resolution is shown to be adequate for the thin film, this same resolution will be satisfactory for thicker films (i.e. the LoRe-med and  LoRe-thick cases). Unless otherwise stated, simulation times cited throughout this work are normalized by $U_{b,0}/(2h)$. A constant time step $\Delta t = 10^{-4}$ is used for all cases herein, being smaller than that suggested by equation (99) in Ref. \cite{kang2000boundary}.

% add vorticity thickness to table: \delta_w = $\frac{\mu_g(u_g - u_l)}{\tau}$
\begin{table}
	\setlength{\tabcolsep}{7pt} % change this to widen/narrow columns
	\def\arraystretch{1.1}
	\begin{center}	
		\begin{tabular}{ccccS[table-format=1.1]ccccS[table-format=1.2]S[table-format=1.2]S[table-format=1.2]}  
			\hline \hline 
			case name & symbol & $\Rey_{\tau,0}$   & $w_{\ell,0}/h$  & $w_{\ell,0}^{+}$ & $N_x$ & $(N_{\ell, y})_0$  & $N_{y}$ & $N_z$ &  $(\Delta x^{+})_{0}$ & $(\Delta y^{+})_{0}$  & $(\Delta z^{+})_{0}$ \\  
			\hline
			% Film thickness study
%			Thin, lower-res. & $\bullet$ & 180               &   $1/12$         & 15.0      & 1152    & 16             & 400    	      &  576     & 0.94         & 0.94    & 0.94 \\ 
			% high-res case
			LoRe-thin &\textcolor{black}{$\bullet$} & 180               &   $1/12$          & 15.0     & 2304    & 32     	    & 800    		  &  1152    & 0.47         & 0.47   & 0.47\\ 
			LoRe-med & $\blacktriangle$ &180               &    $1/6$          & 30.0  	 & 1152    & 32             & 416    		  &  576     & 0.94         & 0.94    & 0.94 \\ 
			LoRe-thick & \tiny{$\blacksquare$} & 180               &     $1/3$         & 60.0     & 1152    & 64     	    & 448    		  &  576     & 0.94         & 0.94    & 0.94\\ 				
			% high-Reynolds number chase			
			HiRe-thin  & \textcolor{red}{$\bullet$} & 393     &  $1/13$         & 30.4     & 2496   	& 32        & 864   		 & 1248    	& 0.94         & 0.94  & 0.94   \\ 
			\hline \hline 			
		\end{tabular}
		\caption{Parameters for the liquid film simulations. Subscript `$0$' refers to parameters when the film simulations are started. Fully developed turbulent velocity fields from the channel simulations (table \ref{tab:ChannelSimulations}) form the gas phase; $\Rey_{\tau,0}$ refers to their steady state Reynolds number. The viscous scaled initial film thickness is $w_{\ell,0}^+ = \Rey_{\tau,0} (w_{\ell,0}/h)$. $(N_{\ell, y})_0$ is the number of grid points over the liquid phase at the start of the VoF simulations. All simulations have domain size $L_x/h = 6$, $L_y/h = 2 + w_{\ell,0}$, $L_z/h = 3$.}
	    \label{tab:FilmSimulations} 
	\end{center}	
\end{table}

\begin{figure}
	\setlength{\tabcolsep}{0pt} % change this to widen/narrow columns
	\def\arraystretch{1.1}
	\begin{center}	
		\begin{tabular}{ccc} 	
			
			%  trim={<left> <lower> <right> <upper>}
			\includegraphics[trim=50mm 0mm 140mm 0mm,scale=0.12,clip]{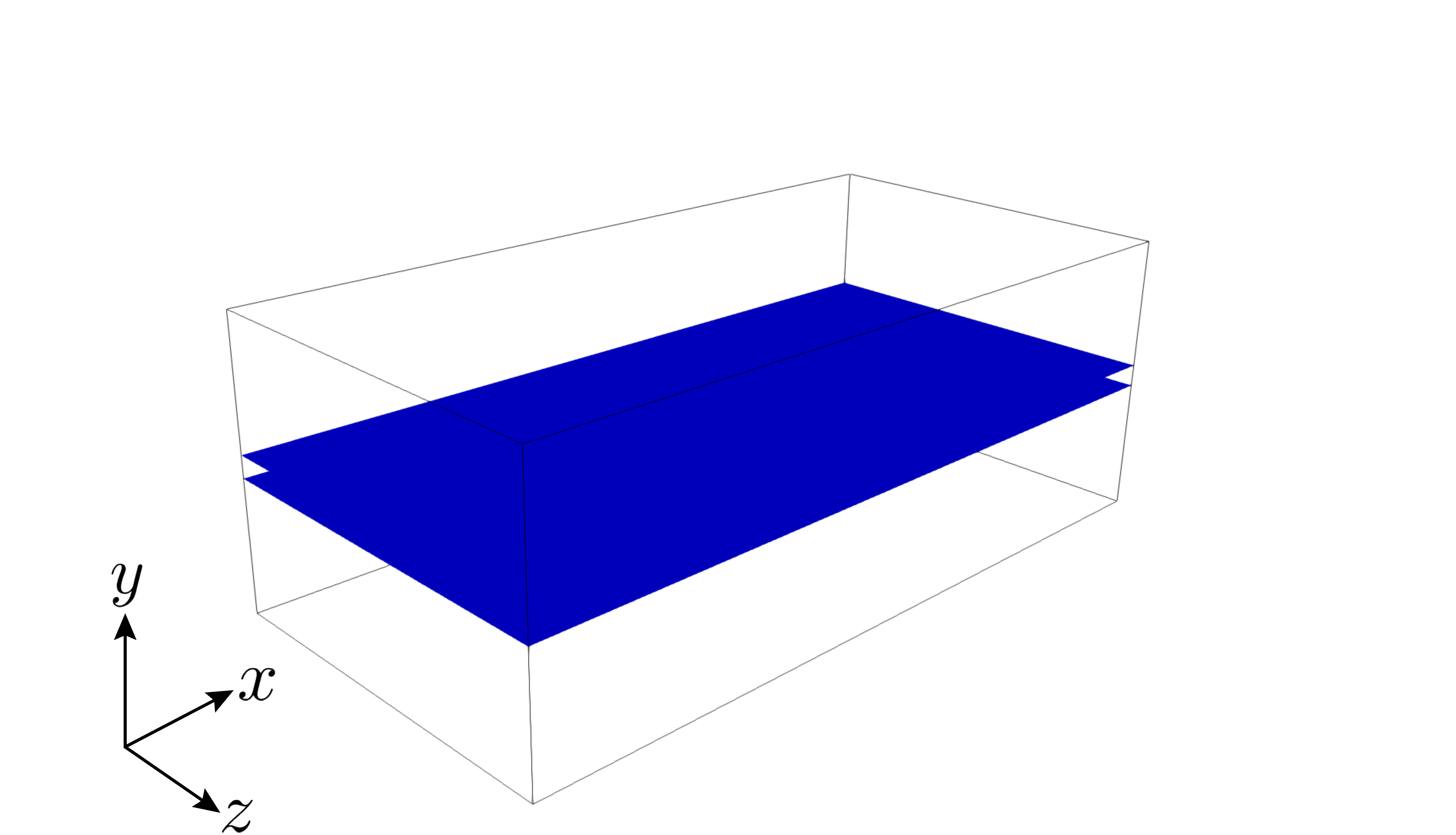}	& 
			\includegraphics[trim=50mm 0mm 140mm 0mm,scale=0.12,clip]{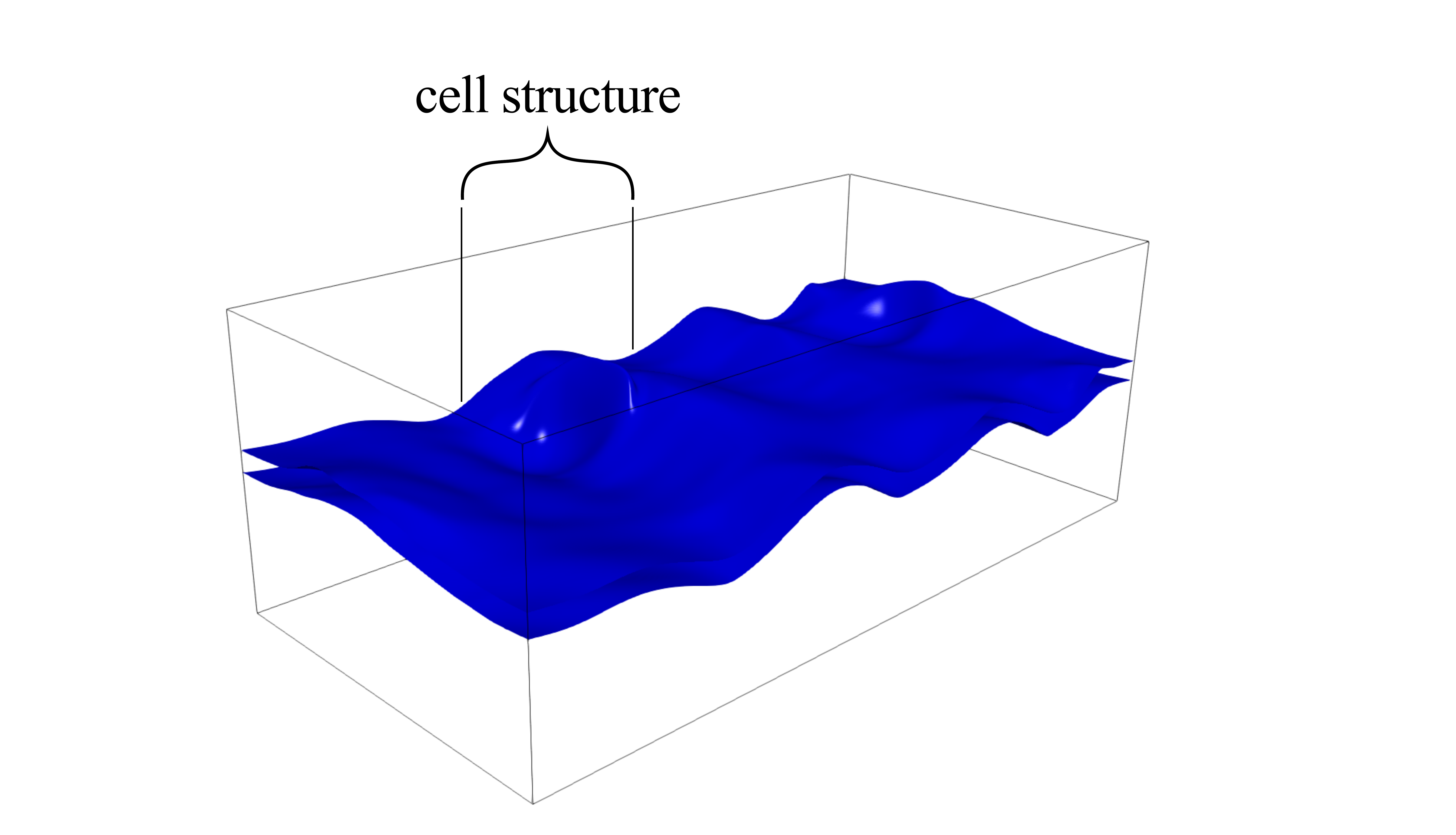} & 
			\includegraphics[trim=50mm 0mm 140mm 0mm,scale=0.12,clip]{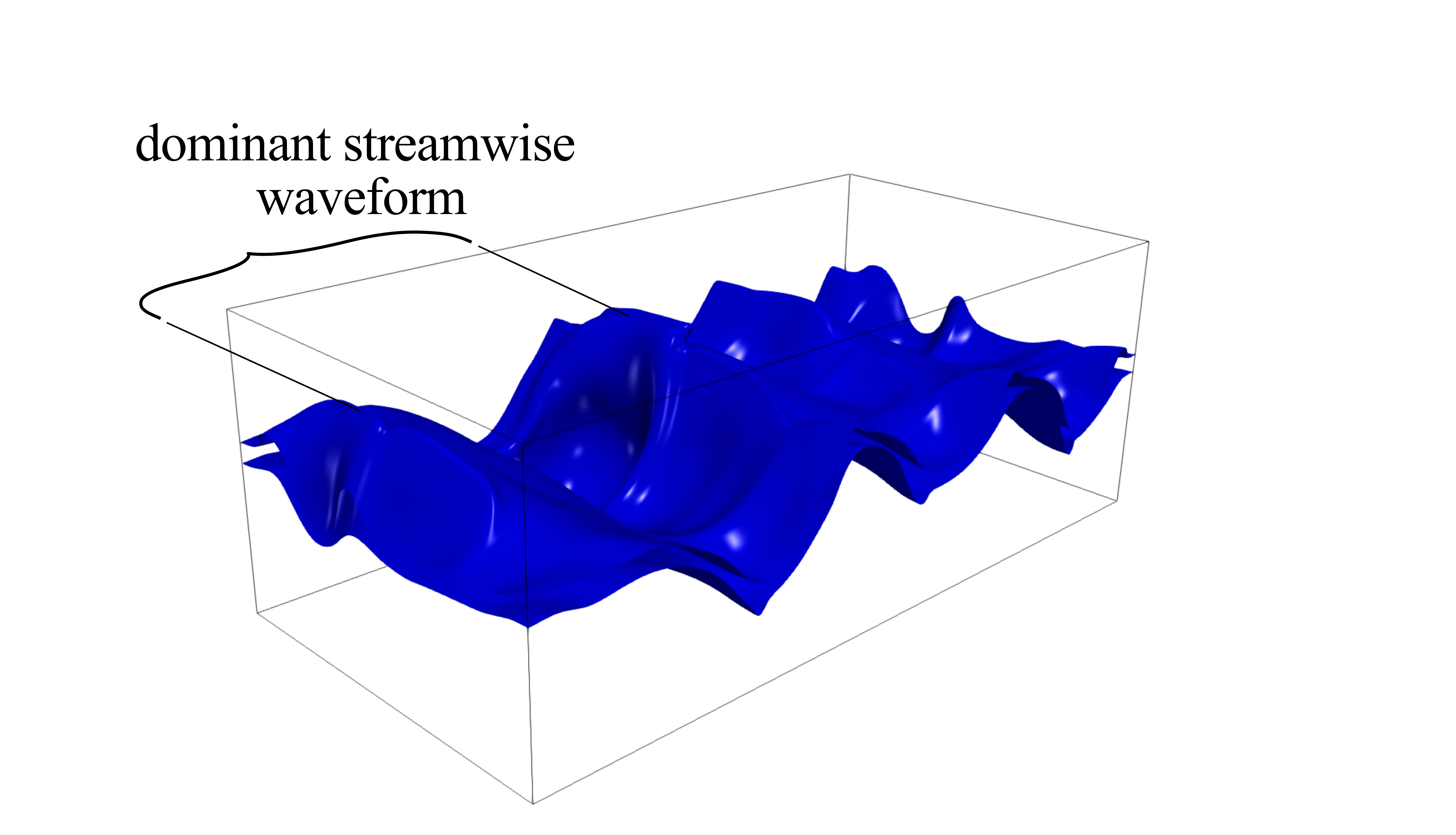} \\			
				
		$t=0$ & $t=6$ & $t=7$ \\	
					
		\vspace{3mm}
					
		\end{tabular}
		\caption{Visualisation of the surface of the liquid film (contours drawn at $\phi = 0.5$) for the LoRe-med case in time. The initially flat film deforms under the action of the fully developed turbulent gas-phase boundary layers.}
		\label{fig:render}
	\end{center}	
\end{figure}

\section{Results}  \label{sec:results}

\subsection{Visualisations}

\begin{figure}
	
	\includegraphics[trim=0mm 0mm 0mm 0mm, scale=0.47,clip]{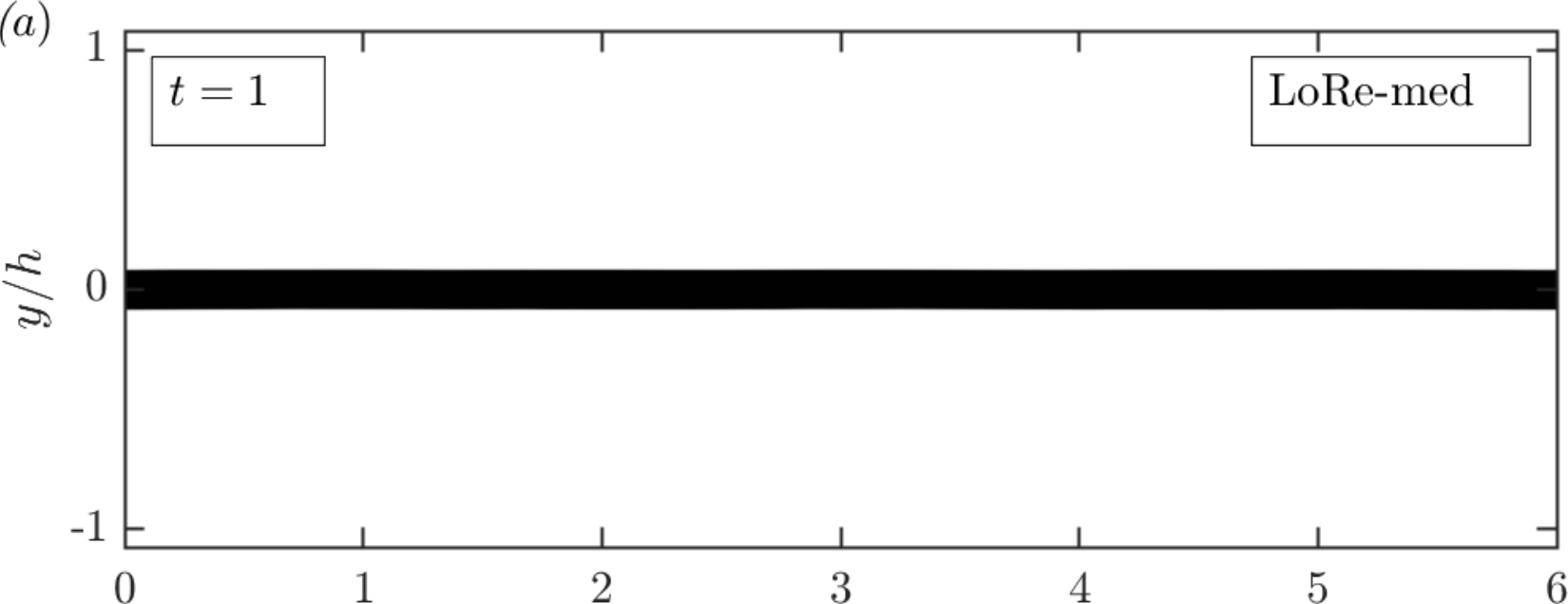} 
	\includegraphics[trim=0mm 0mm 0mm 0mm, scale=0.511,clip]{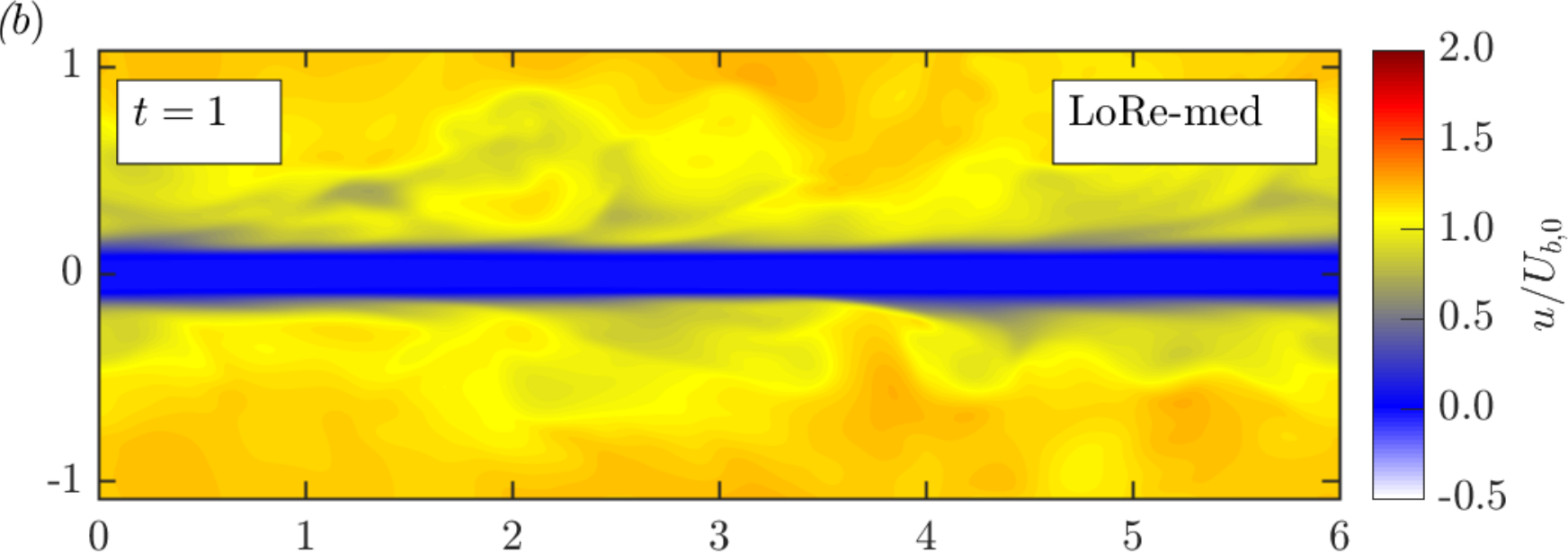} 
	
	\includegraphics[trim=0mm 0mm 0mm 0mm, scale=0.47,clip]{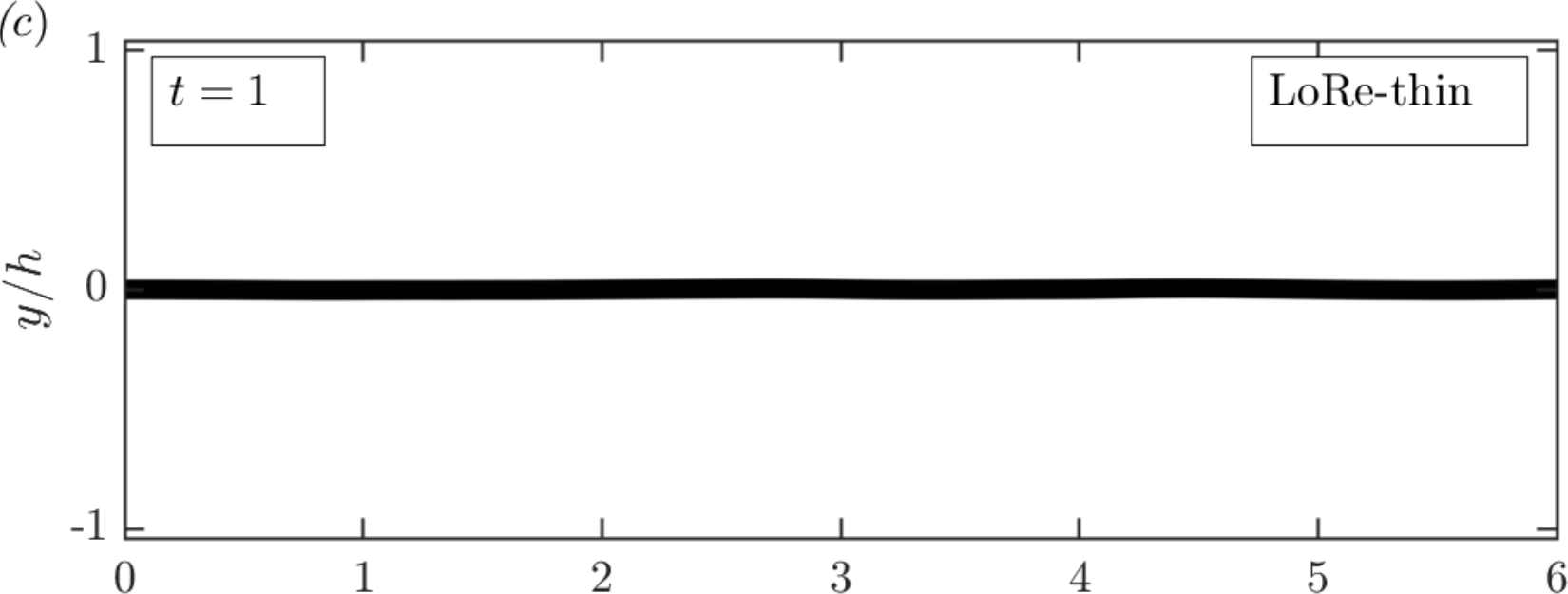} 
	\includegraphics[trim=0mm 0mm 0mm 0mm, scale=0.511,clip]{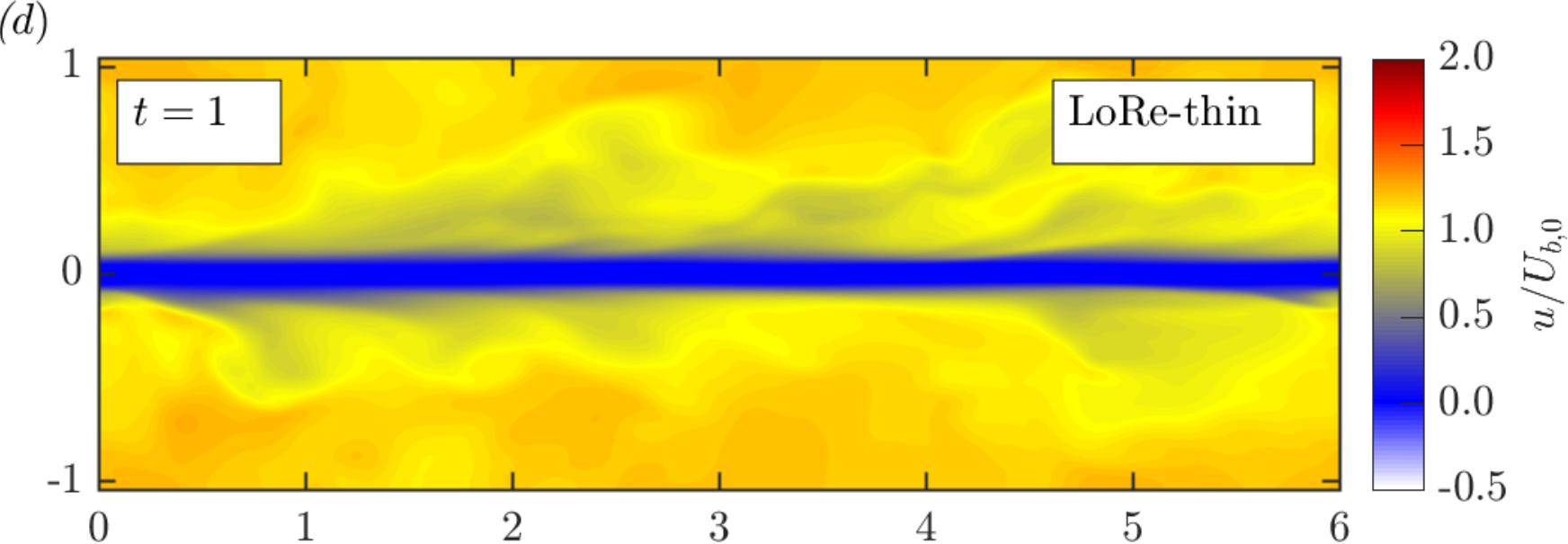} 
	
	\includegraphics[trim=0mm 0mm 0mm 0mm, scale=0.47,clip]{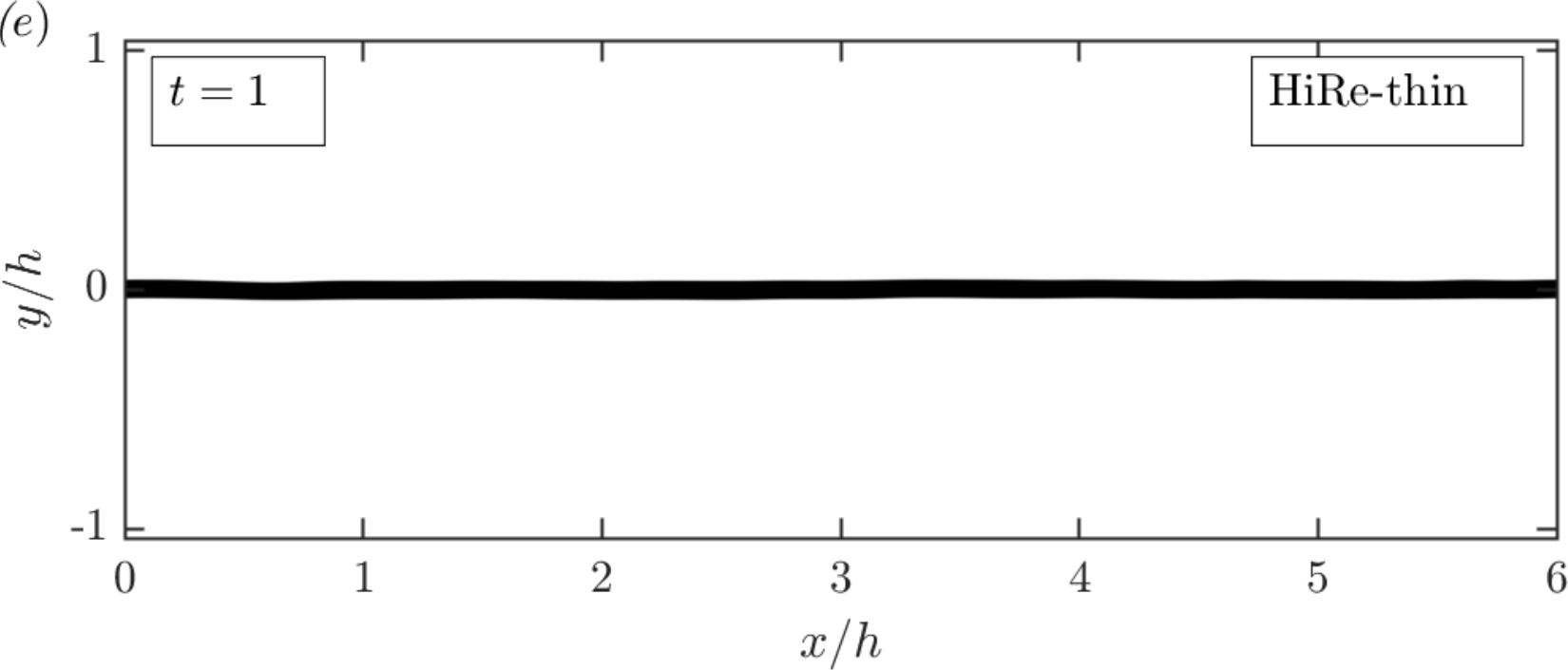} 
	\includegraphics[trim=0mm 0mm 0mm 0mm, scale=0.511,clip]{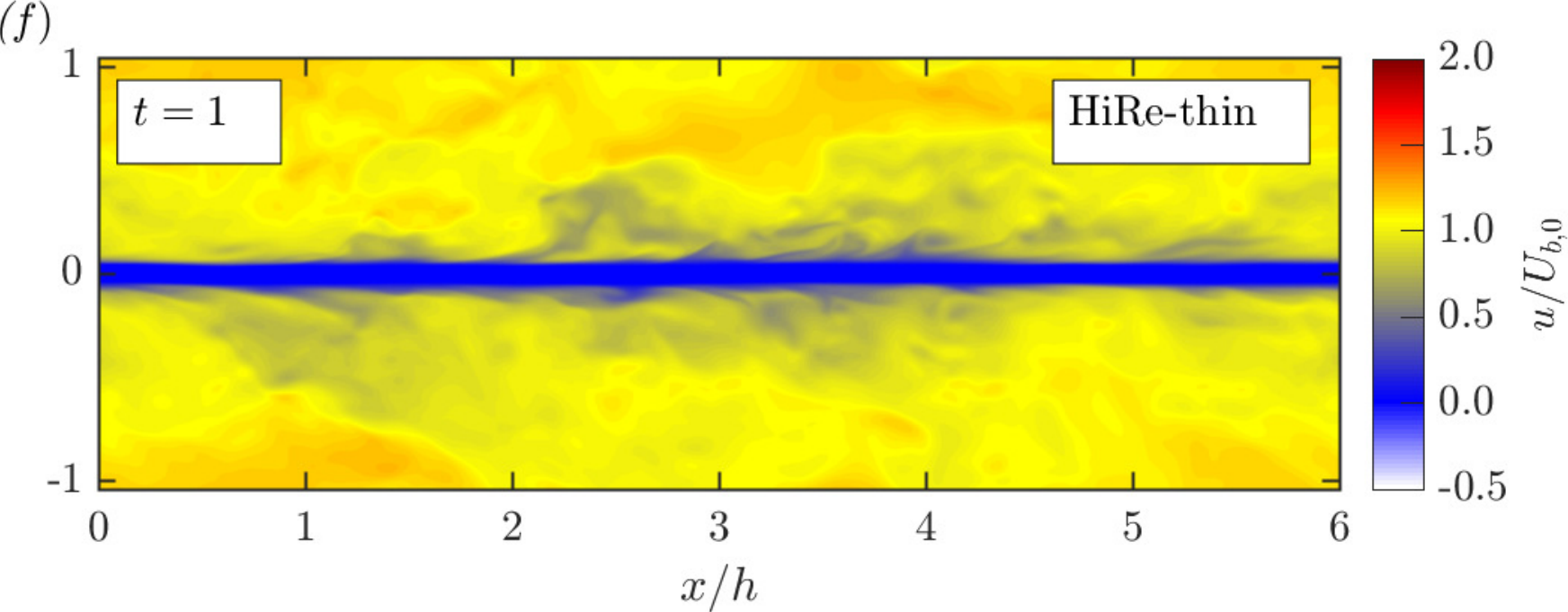} 
		
	\caption{Visualisations of three cases of the VoF simulations at $t = 1$, fields of the `sandwich' film setup at an early time. Top, LoRe-med case; middle, LoRe-thin case; bottom, HiRe-thin case. Left, the cell-averaged volume fraction of the liquid, $\phi$; right, streamwise velocity fields. Bulk velocity is to the right.}
	\label{fig:filmVisA}
\end{figure}

\begin{figure}
	
\includegraphics[trim=0mm 0mm 0mm 0mm, scale=0.47,clip]{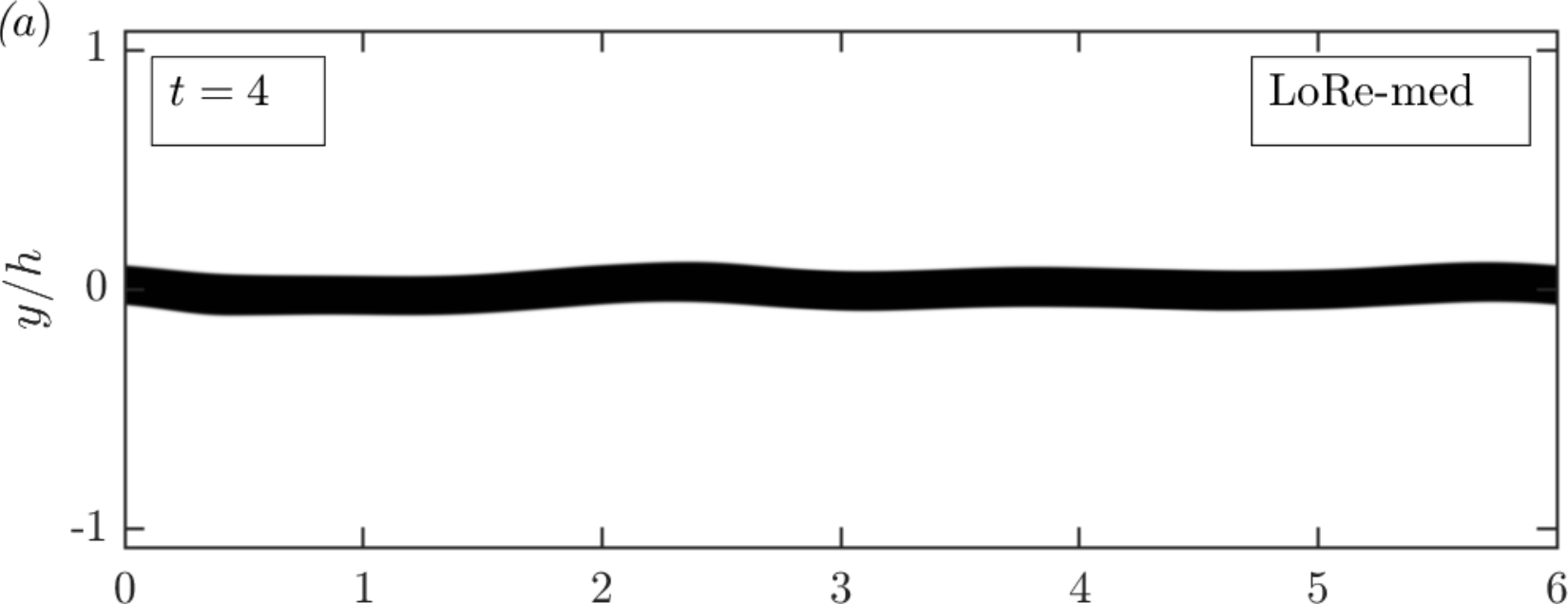} 
\includegraphics[trim=0mm 0mm 0mm 0mm, scale=0.511,clip]{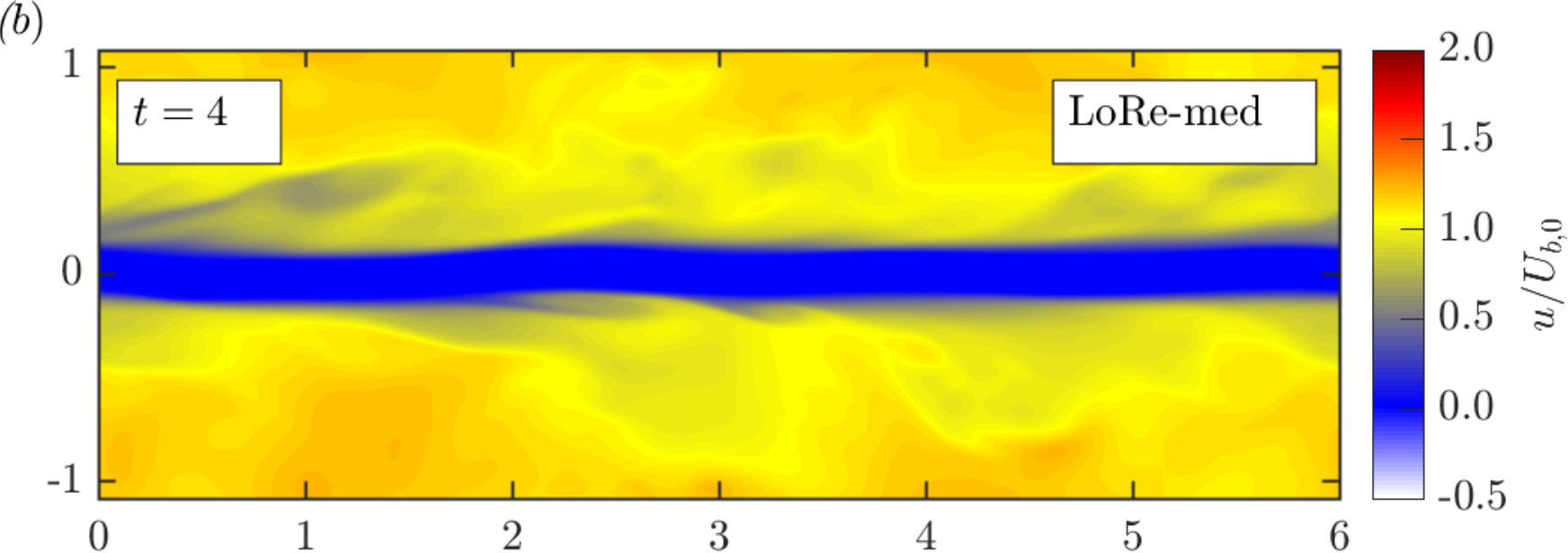} 	
	
\includegraphics[trim=0mm 0mm 0mm 0mm, scale=0.47,clip]{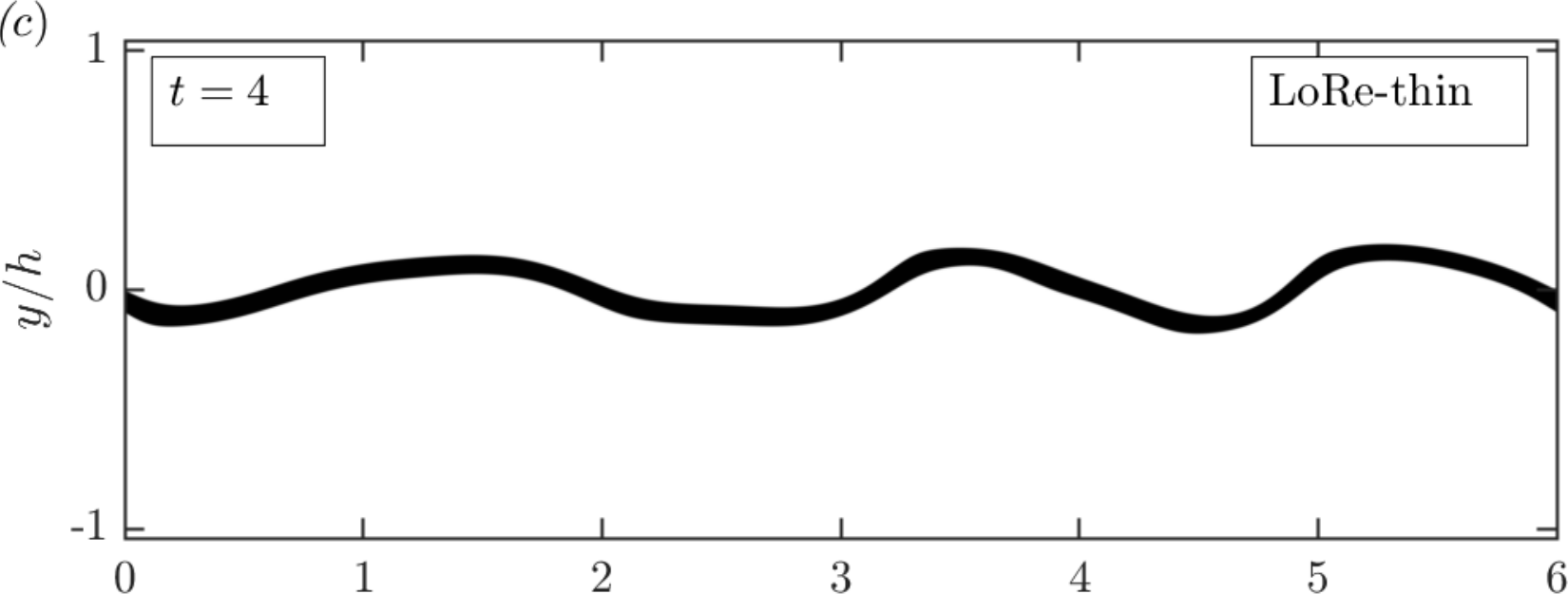} 
\includegraphics[trim=0mm 0mm 0mm 0mm, scale=0.511,clip]{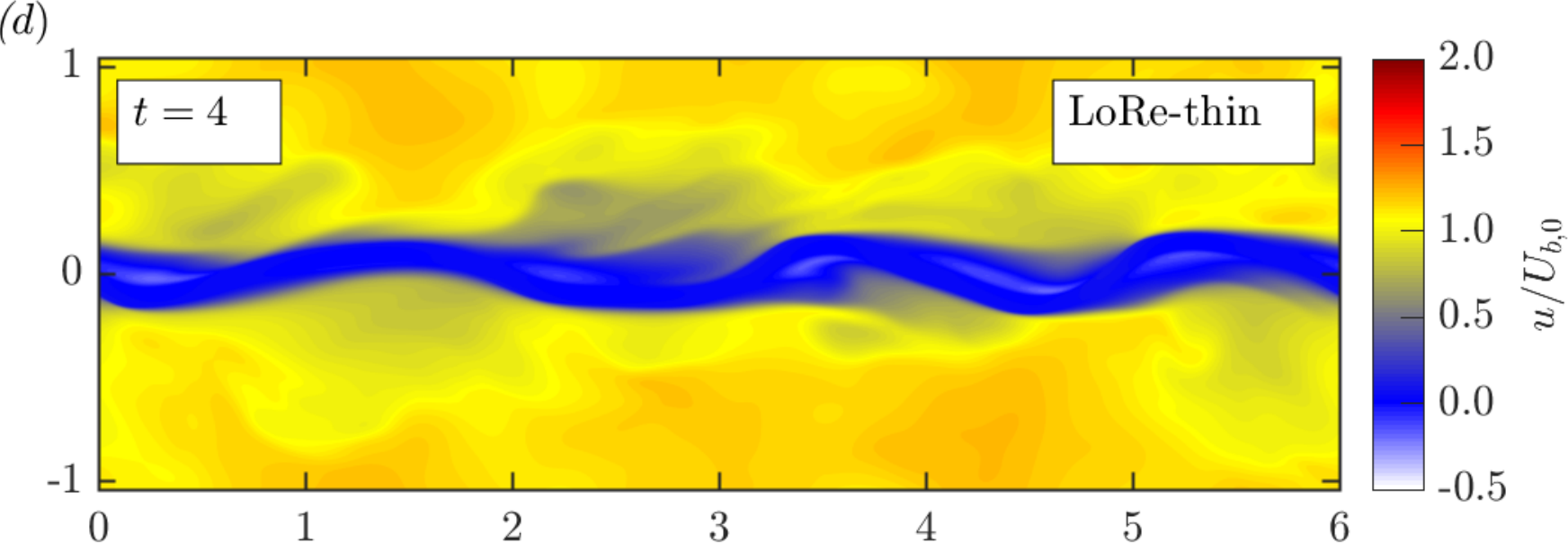} 

\includegraphics[trim=0mm 0mm 0mm 0mm, scale=0.47,clip]{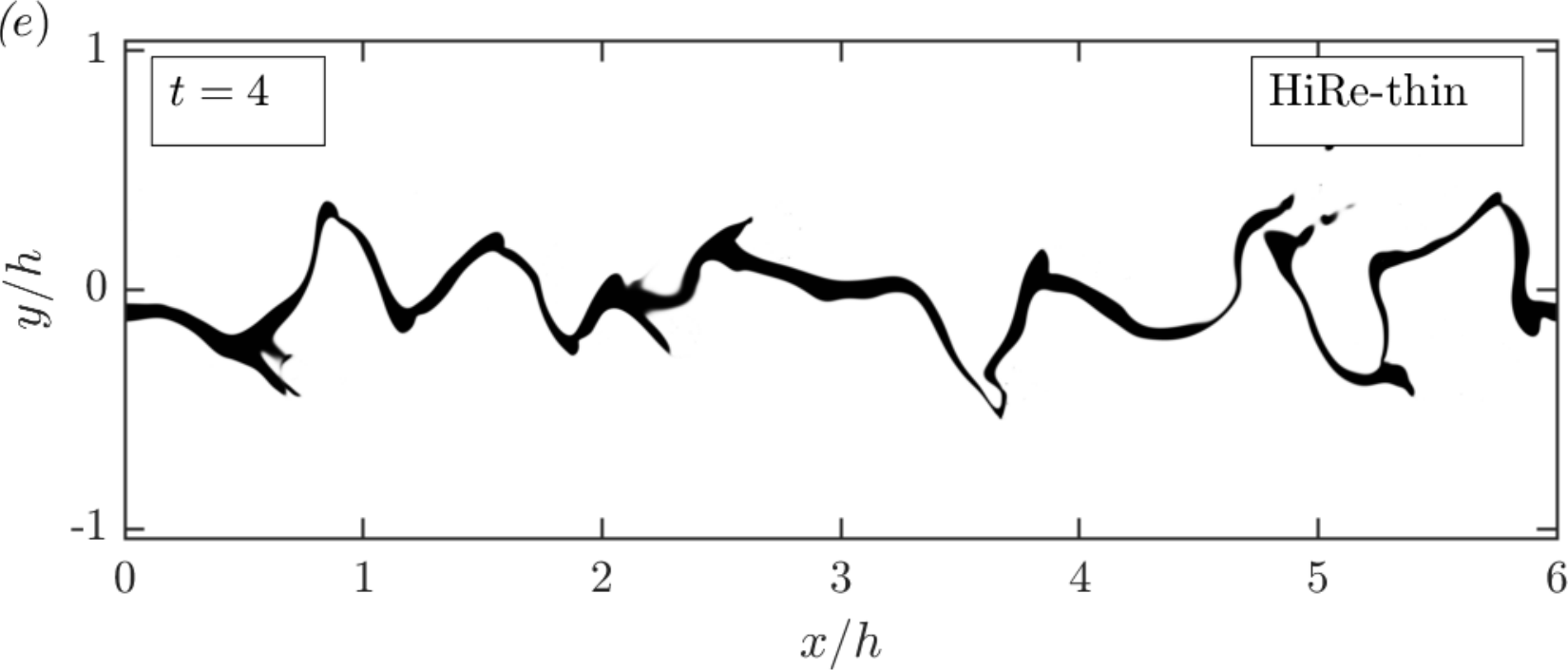} 
\includegraphics[trim=0mm 0mm 0mm 0mm, scale=0.511,clip]{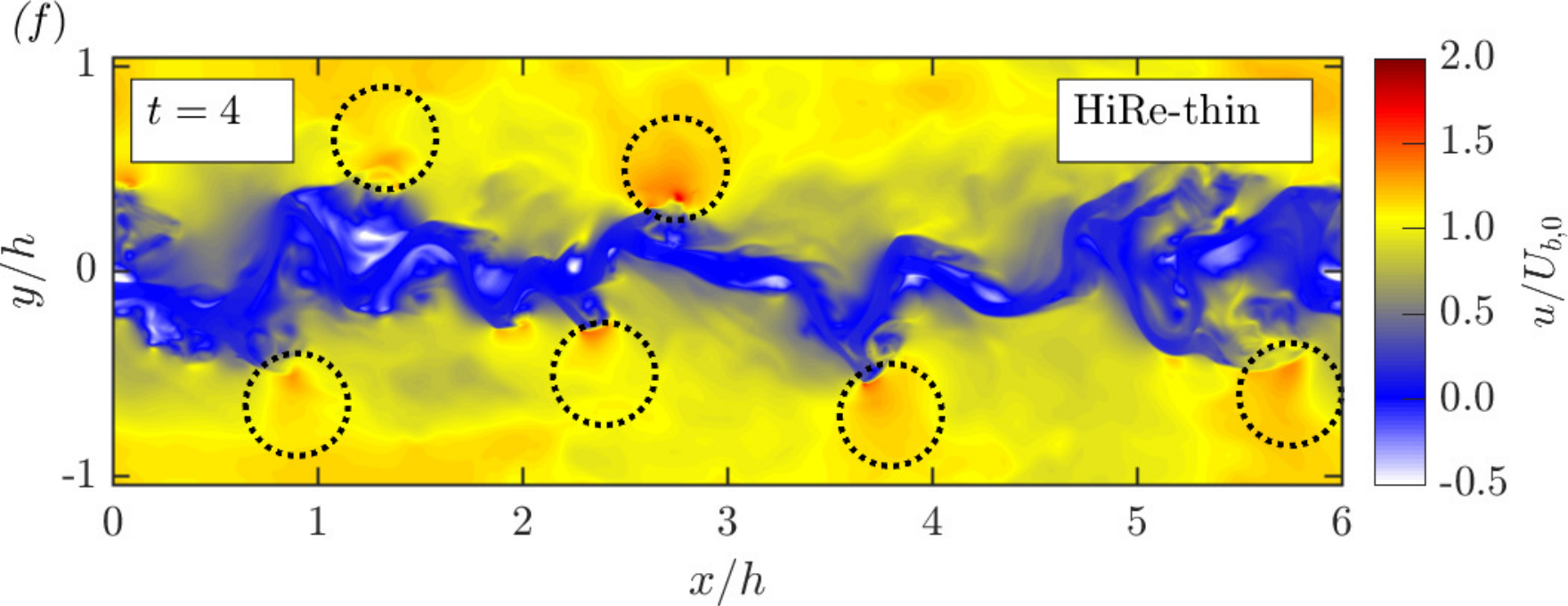} 

\caption{Visualisations of three cases of the VoF simulations at $t = 4$. Top, LoRe-med case; middle, LoRe-thin case; bottom, HiRe-thin case. Left, the cell-averaged volume fraction of the liquid, $\phi$; right, streamwise velocity fields. Dotted circles in (f) mark regions of high streamwise velocity. Bulk velocity is to the right.}
\label{fig:filmVisB}
\end{figure}

In this section we consider important physical features of the liquid film's deformation and rupture, and establish some commonalities with experimental work in the literature. A first view of the physical problem is given in figure \ref{fig:render} where the surface of the liquid film is shown as it deforms during the simulation of the LoRe-med case. We see there is a considerable time delay until significant deformations of the film emerge at $t=6$, which then rapidly grow until the film is at the point of rupture by $t=7$. As well as a dominant waveform in the streamwise direction, there are secondary `cells' forming across the span reminiscent of that seen in the experiments of Ref. \cite{park2004experimental}. A cellular structure with thickening edges clearly emerges at $t=6$ in figure \ref{fig:render}. Figure \ref{fig:filmVisA} (at time $t = 1$) and figure \ref{fig:filmVisB} (at time $t = 4$) plot the cell-average volume fraction of the liquid $\phi$ and the streamwise velocity fields in the $xy$-plane for three of the present cases. This permits a first visual comparison between the  LoRe-thin and LoRe-med cases, both with starting gas phase friction Reynolds number $\Rey_{\tau,0} = 180$, and the HiRe-thin case, with $\Rey_{\tau,0} = 393$. The initially flat film (figure \ref{fig:filmVisA}) is deformed by the fully developed turbulent boundary layers on either side of it after some time at $t= 4$ (figure \ref{fig:filmVisB}).  As expected, a thicker film is more resistant to deformation as only small perturbations are present in the LoRe-med  case at $t= 4$. On the other hand, the LoRe-thin case is highly sinusoidal by this time, whereas the HiRe-thin case is much more contorted by $t= 4$, suggesting non-linear development, and has in fact already ruptured. The influence of the Reynolds number is clearly significant, increasing the deformation and contortion which results in rupture. At higher Reynolds number, a wider spectrum of scales interacts with the film, clear from figure \ref{fig:filmVisA}(f), since $w_{\ell,0}/\delta_\nu$ is larger, where $\delta_\nu = \nu/u_\tau$ is the viscous wall unit. Additionally, liquid ligaments can be seen to be `peeling' from the bulk of the liquid by $t=4$ in figure \ref{fig:filmVisB}(e). These events directly precede film rupture and were found to cause the ensuing rupturing mechanism first identified by Ref. \cite{duke2012experimental}. By $t=4$ in figures \ref{fig:filmVisB}(d) and \ref{fig:filmVisB}(f), some reversed flow (in white) appears for both the LoRe-thin and HiRe-thin cases. For the HiRe-thin case, these high-speed regions forming over alternating extrema (highlighted with dotted circles in figure \ref{fig:filmVisB}f) are especially pronounced given the film's high contortion. Figure \ref{fig:filmVisB} demonstrates the sensitivity of the rate and nature of film deformation to the turbulence in the gas phase and therefore the necessity of realistically representing gas-phase turbulence in numerical studies of this physical problem. 

 Qualitatively, similar behaviour has been observed in experimental investigations. Ref. \cite{lozano2001experimental} found that for velocity and momentum ratios for which the sinusoidal mode is dominant, liquid sheet oscillation can cause air boundary layer separation at the interface for the wave troughs. They postulated that this separation causes a pressure difference on both sides of the liquid sheet, which can be partly responsible for enhancement of the flapping motion. They also found that before breaking, the sheet oscillations grow non-linearly and the sinusoidal wave deforms into a `zigzag' shape, as observed for the HiRe-thin case in figure \ref{fig:filmVisB}(f). The peaks then bend, and point upstream, forcing the air to recirculate. They hypothesised that these recirculation zones may act as a feedback mechanism contributing to increasing wave growth. The semi-realistic simulations of Ref. \cite{ling2019two} also reported that the increasing interfacial wave amplitude in their mixing layer between parallel gas and liquid streams acts as an obstacle to the gas flow, causing the flow to separate at the downstream face of the wave, giving rise to a turbulent wake. The cellular rupture phenomenon in air-blasted liquid sheets was investigated experimentally by Ref. \cite{park2004experimental} where they systematically studied the effect of both gas and liquid velocity. They found that turbulent transition of the liquid jet did not have any noticeable effect on cellular structures and the ensuing sheet breakup, with similar results for similar relative gas-liquid velocities. They deduced the size of `cell' structures in the disintegrating sheet to vary approximately as $U_r^{-2}$, where $U_r$ is the relative velocity between the gas and liquid phases. In the present cases the bulk velocity is the same in all cases, yet smaller `cell' structures form for the HiRe-thin case in figure \ref{fig:filmVisB} due to the higher Reynolds number. We have shown here that our temporal VoF simulations are qualitatively reflective of experimental work on this same physical problem. Our onward analysis is inspired by phenomena such as the reversed and high-speed flow regions in particular, and will borrow ideas from classical aerodynamics. 

\subsection{Temporal development}

\begin{figure}		
\setlength{\tabcolsep}{10pt} % change this to widen/narrow columns
\def\arraystretch{3}
\begin{center}	
	\begin{tabular}{ccc} 		
							
\hspace{7.5mm}	LoRe-med & \hspace{7.5mm} LoRe-thin & \hspace{7.5mm} HiRe-thin \\	
		
\includegraphics[scale=0.68,clip]{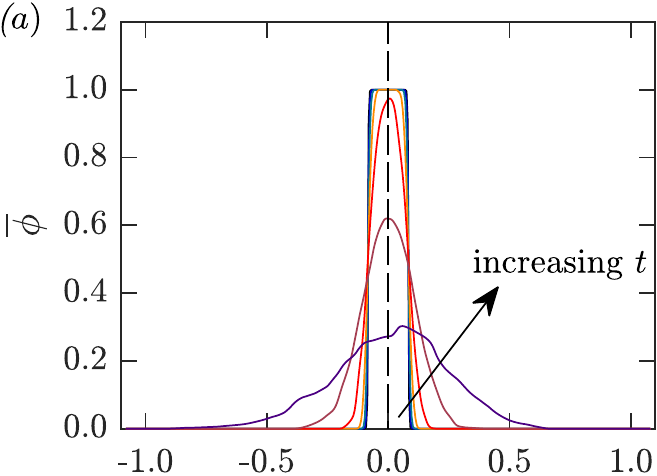} &			
\includegraphics[scale=0.68,clip]{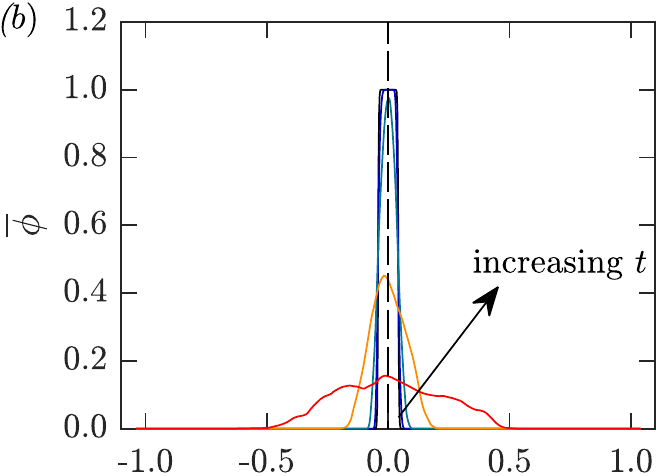}	&
\includegraphics[scale=0.68,clip]{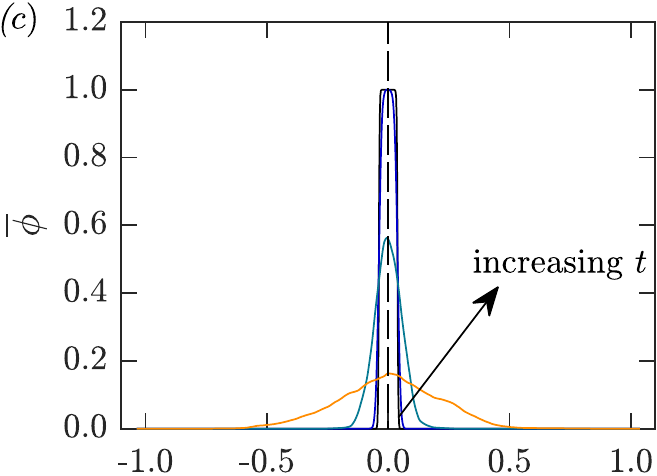}\\		
\includegraphics[scale=0.68,clip]{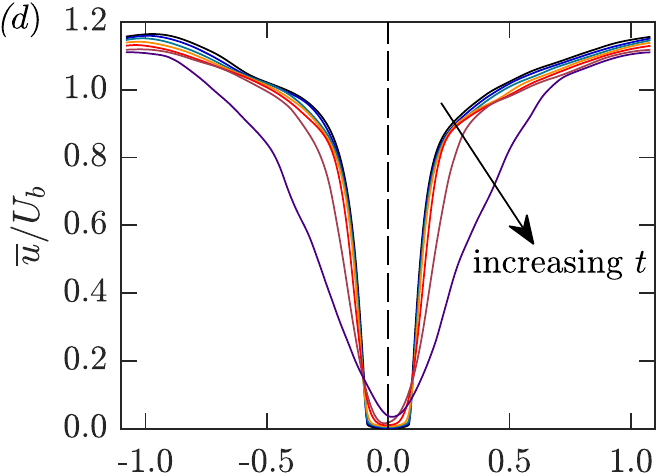} & 
\includegraphics[scale=0.68,clip]{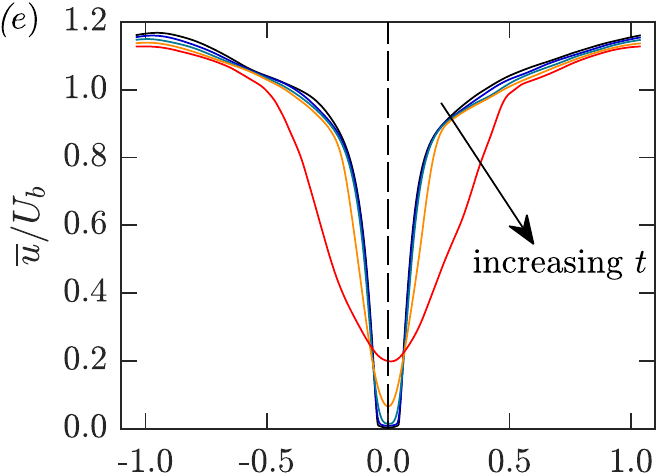}	&	
\includegraphics[scale=0.68,clip]{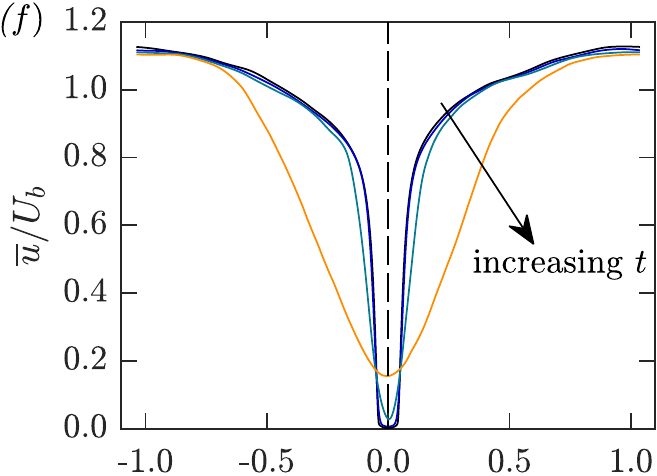} \\		
		
\includegraphics[scale=0.68,clip]{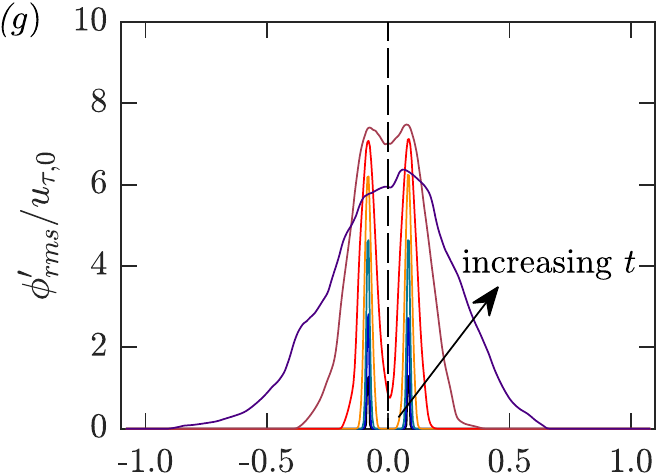} &		
\includegraphics[scale=0.68,clip]{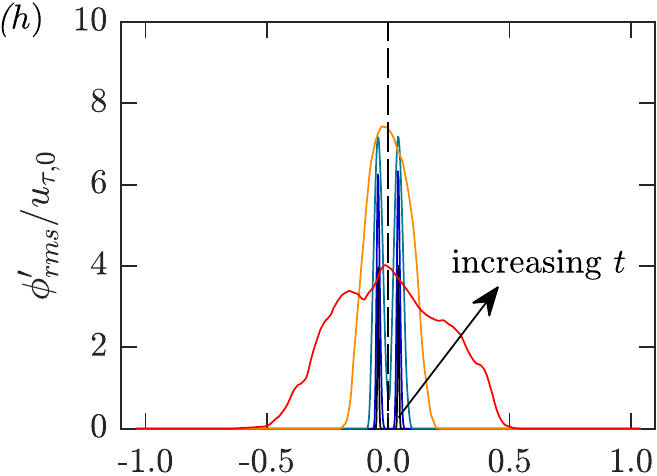}	&	
\includegraphics[scale=0.68,clip]{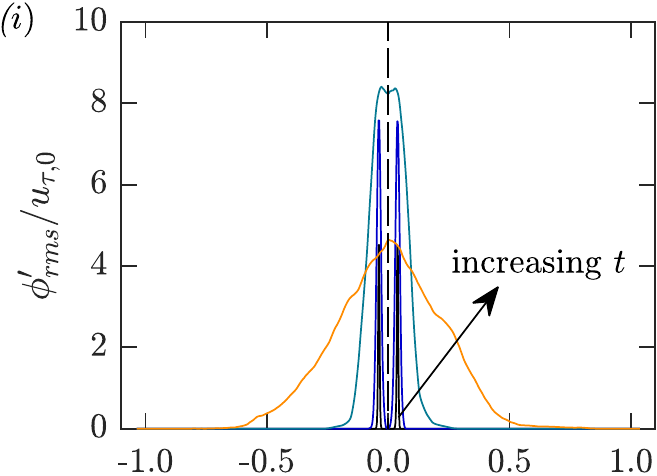} \\		
		
\includegraphics[scale=0.68,clip]{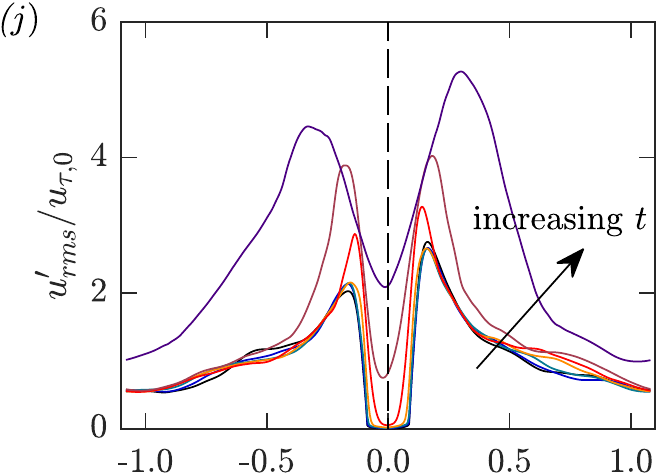}	&	
\includegraphics[scale=0.68,clip]{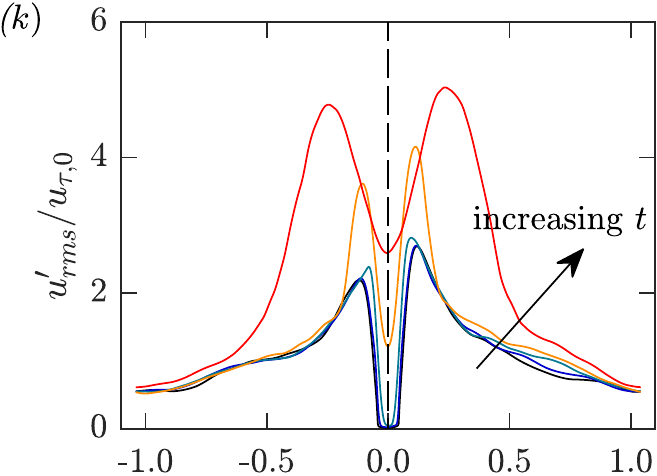} &		
\includegraphics[scale=0.68,clip]{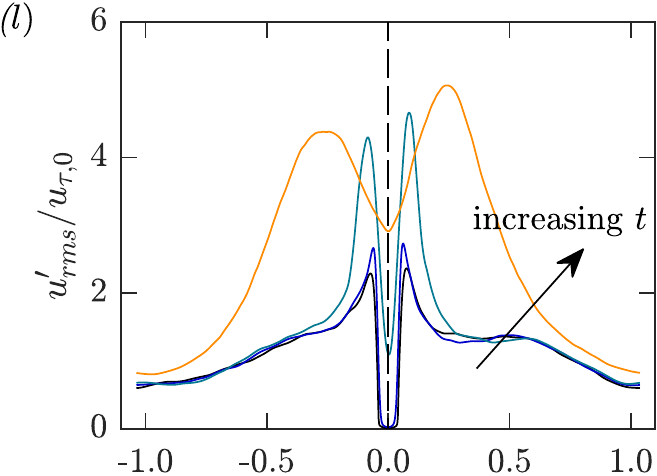}	\\	

\includegraphics[scale=0.68,clip]{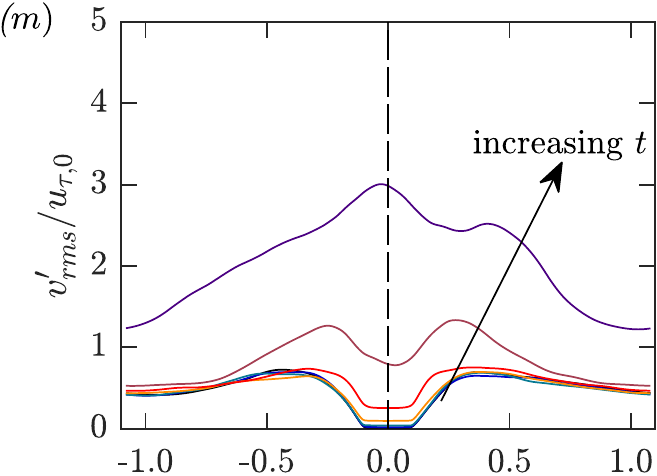}	&	
\includegraphics[scale=0.68,clip]{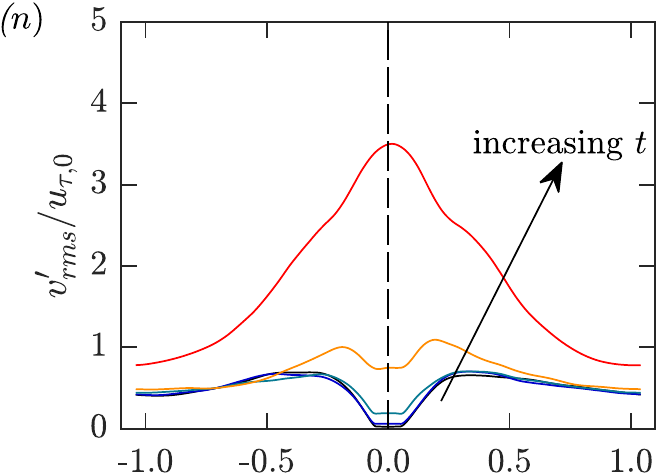} &		
\includegraphics[scale=0.68,clip]{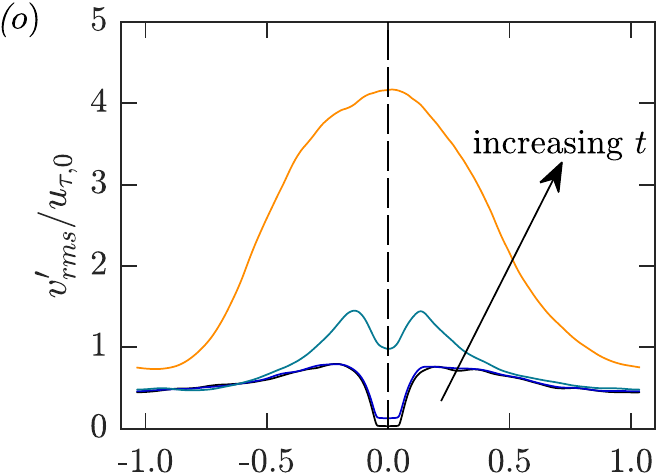}	\\	
		
\includegraphics[scale=0.68,clip]{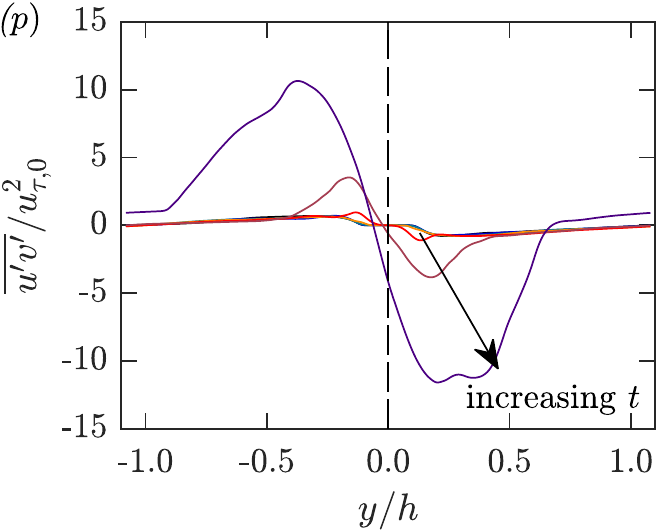} &		
\includegraphics[scale=0.68,clip]{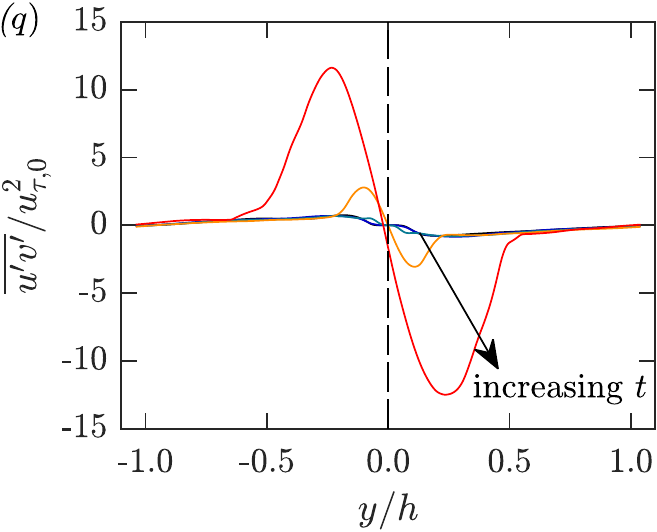} &		
\includegraphics[scale=0.68,clip]{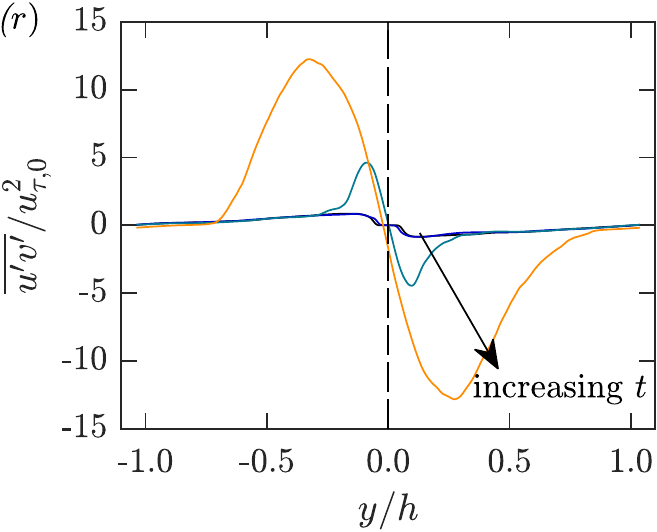}\\		
		
 \end{tabular}		
		\caption{Statistics deduced by averaging in the homogeneous $xz$-planes; left column, LoRe-med case, $\Rey_{\tau,0} = 180$ ($w_{\ell,0}/h = 1/6$); middle column, LoRe-thin case, $\Rey_{\tau,0} = 180$ ($w_{\ell,0}/h = 1/12$); right column, HiRe-thin case, $\Rey_{\tau,0} = 393$ ($w_{\ell,0}/h = 1/13$); curves colored by time: \textcolor{black}{\drawline{22}{1.5}}, $t = 1$; \textcolor{royal_blue}{\drawline{22}{1.5}}, $t = 2$;
		\textcolor{teal}{\drawline{22}{1.5}}, $t = 3$;
		\textcolor{burnt_orange}{\drawline{22}{1.5}}, $t = 4$;
		\textcolor{red}{\drawline{22}{1.5}}, $t = 5$;
	\textcolor{crimson}{\drawline{22}{1.5}}, $t = 6$; \textcolor{indigo}{\drawline{22}{1.5}}, $t = 7$; \dashed, domain centerline.}
		\label{fig:profiles_in_t}

	\end{center}
\end{figure}

Figure \ref{fig:profiles_in_t} plots various profiles averaged in the homogeneous $xz$-planes (the initial plane of the film) following that shown in Ref. \cite{desjardins2010detailed} for the liquid turbulent jet within a quiescent gas phase. The curves pertain to the same cases as those shown in figures \ref{fig:filmVisA} and \ref{fig:filmVisB}. This permits a more quantitative appraisal of the average evolution of the fields in time and the differing rates of evolution between the cases. The curves are colored by time, such that the same color in each subfigure corresponds to the same development time in bulk units after the VoF simulations are started for each case.

The plane-averaged volume fraction of the liquid $\phi$ is shown in figures \ref{fig:profiles_in_t}(a-c). The starting step-like profile of $\overline{\phi}$ is gradually smeared out as the film deforms moving fluid away from the centerline. This happens at a markedly different rate for the different cases. The value $\overline{\phi}$ at the centerline reduces from the starting $\overline{\phi} = 1$ to $\overline{\phi} \approx 0.5-0.6$ at $t = 6$ for the LoRe-med case, at $t = 4$ for the LoRe-thin case and at $t = 3$ for the HiRe-thin case. As seen above in the visualisations, both a thinner film and a higher gas-phase Reynolds number serve to increase the momentum transfer to the liquid phase increasing the rate of the liquid's spread away from the centerline. Despite different temporal evolutions, the qualitative trends appear to be similar amongst the cases, that is the liquid phase is spread in a similar manner throughout the domain in a mean sense. Figures \ref{fig:profiles_in_t}(d-f) show the development of the mean streamwise velocity. Changes to the boundary layer profiles are due to both momentum transfer to the liquid sheet (i.e. significant flattening of $\overline{\phi}$) and, likely to a smaller extent, dissipation in the gas phase, since there is no flow forcing in the VoF simulations. Figures \ref{fig:profiles_in_t}(g-i) show the root-mean-squared (rms) fluctuations in $\phi$. At the start of the VoF simulations, $\phi'_{rms}$ is zero everywhere since the film is initially perfectly quiescent. Regions of non-zero $\phi'_{rms}$ are only found near the film surface at early times, resulting in two sharp peaks about the centerline. After some time, fluctuations in $\phi$ are felt through the whole thickness of the film, and at later times spread away from the centerline as the film deformations grow. Ref. \cite{agbaglah2017numerical} conducted spatial large-eddy simulations of planar air/water air-blast atomization, and found similar development in the rms fluctuations within the liquid phase at positions downstream of the splitter plate, although their incoming gas phase was not turbulent. 

Figures \ref{fig:profiles_in_t}(j-o) show how the streamwise and spanwise velocity fluctuations evolve throughout the domain. Note they were also zero within the film at $t=0$ yet increase through the liquid phase as the film deforms. The initial peak values in both streamwise and film-normal velocity variance are amplified in time by the film's deformation. The streamwise velocity fluctuations $u'_{rms}$ remain double-peaked across the film even at late times, with the strongest fluctuations occurring at the film's surface on either side. The film-normal velocity fluctuations however tend to evolve such that there is a single peak roughly at the film's centerline. A large and sudden increase in the Reynolds stresses both through and about the film's edge is seen in figures  \ref{fig:profiles_in_t}(p-r) as significant momentum is transferred from the gas phase into the liquid film. Again the final profiles are very similar amongst the different cases, yet occur at different times. The film thickness and gas-phase Reynolds number therefore both have a role in determining the rate of film deformation and rupture.

\subsection{Quantifying the deforming film surface} \label{sec:filmScales}

\begin{figure}
\setlength{\tabcolsep}{-12pt} % change this to widen/narrow columns
\def\arraystretch{3}
\begin{center}	
	\begin{tabular}{ccc} 
		
\hspace{-5.5mm} LoRe-med & \hspace{-5.5mm} LoRe-thin & \hspace{-5.5mm}  HiRe-thin \\		
		
\includegraphics[scale=0.45,clip]{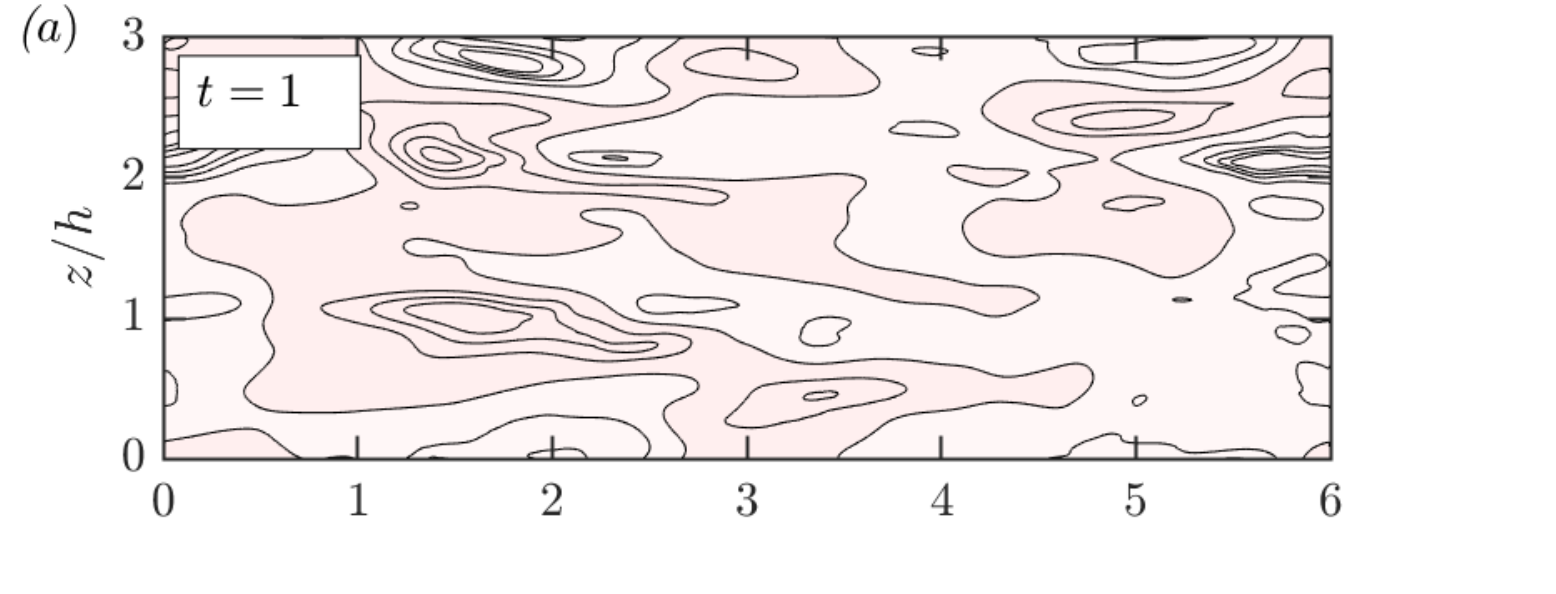}	& 
\includegraphics[scale=0.45,clip]{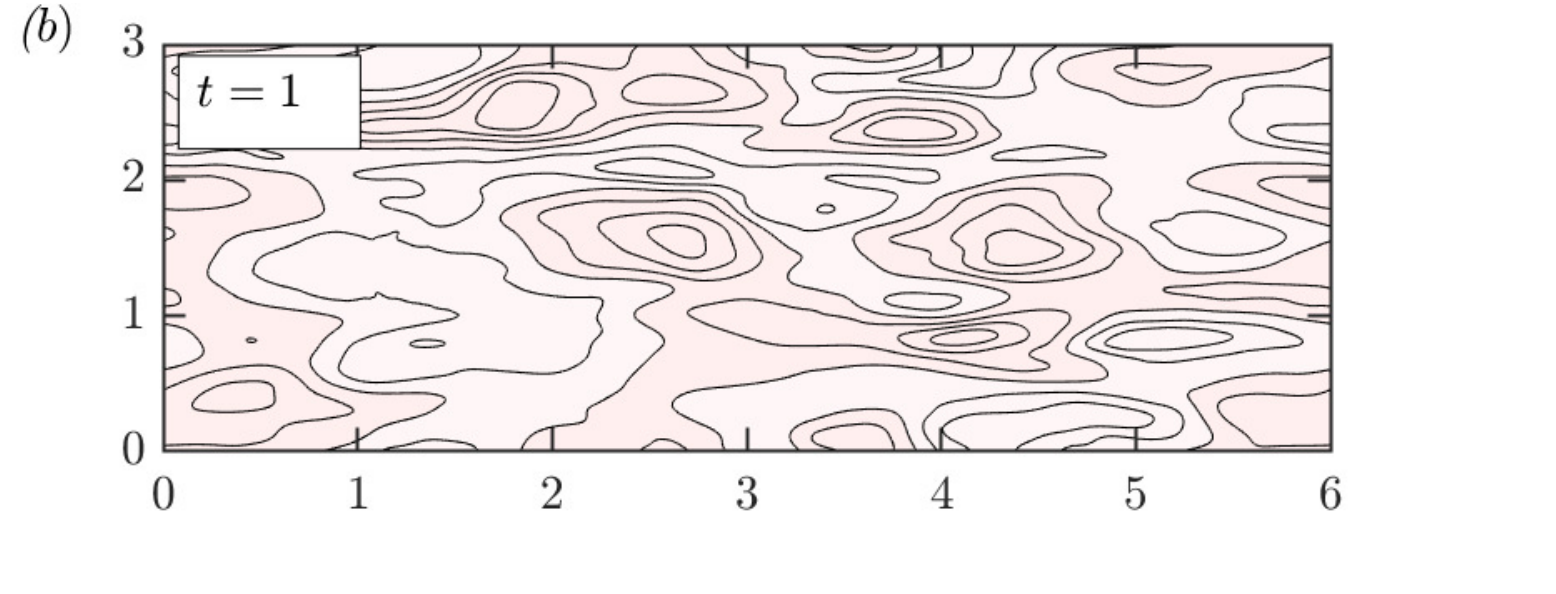} & 
\includegraphics[scale=0.45,clip]{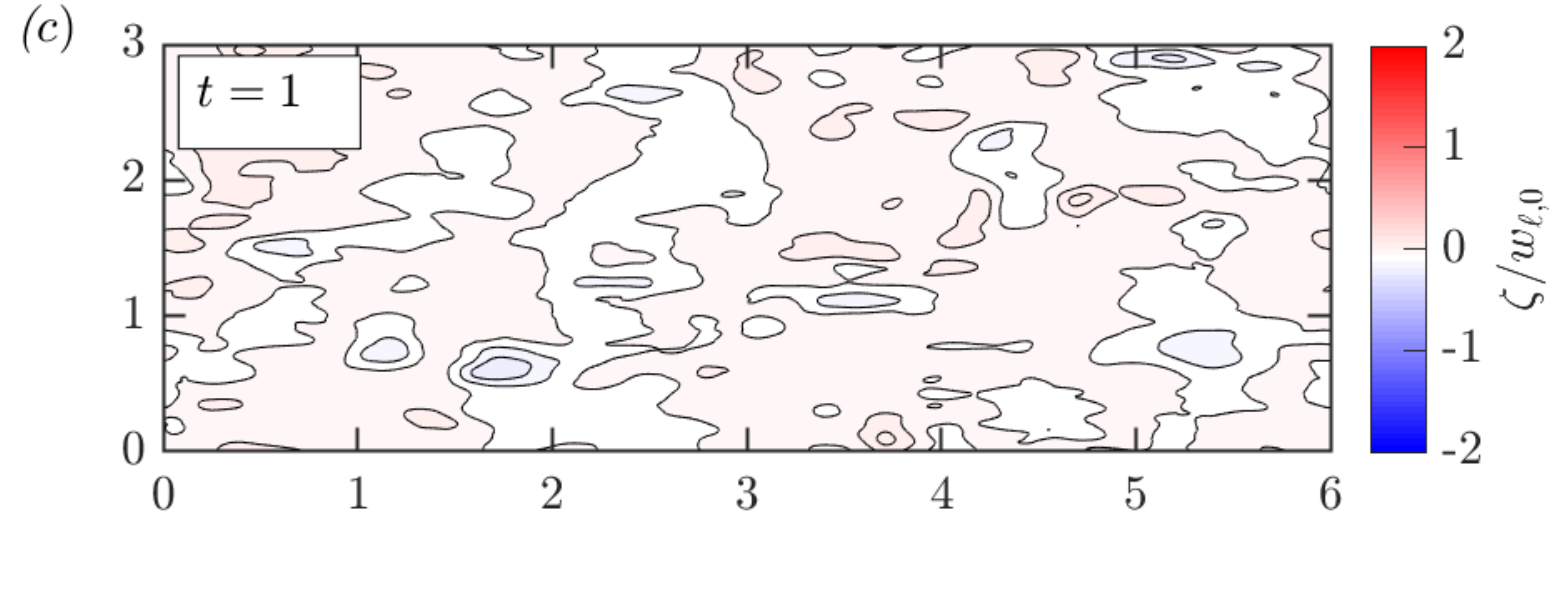} \\	

\includegraphics[scale=0.45,clip]{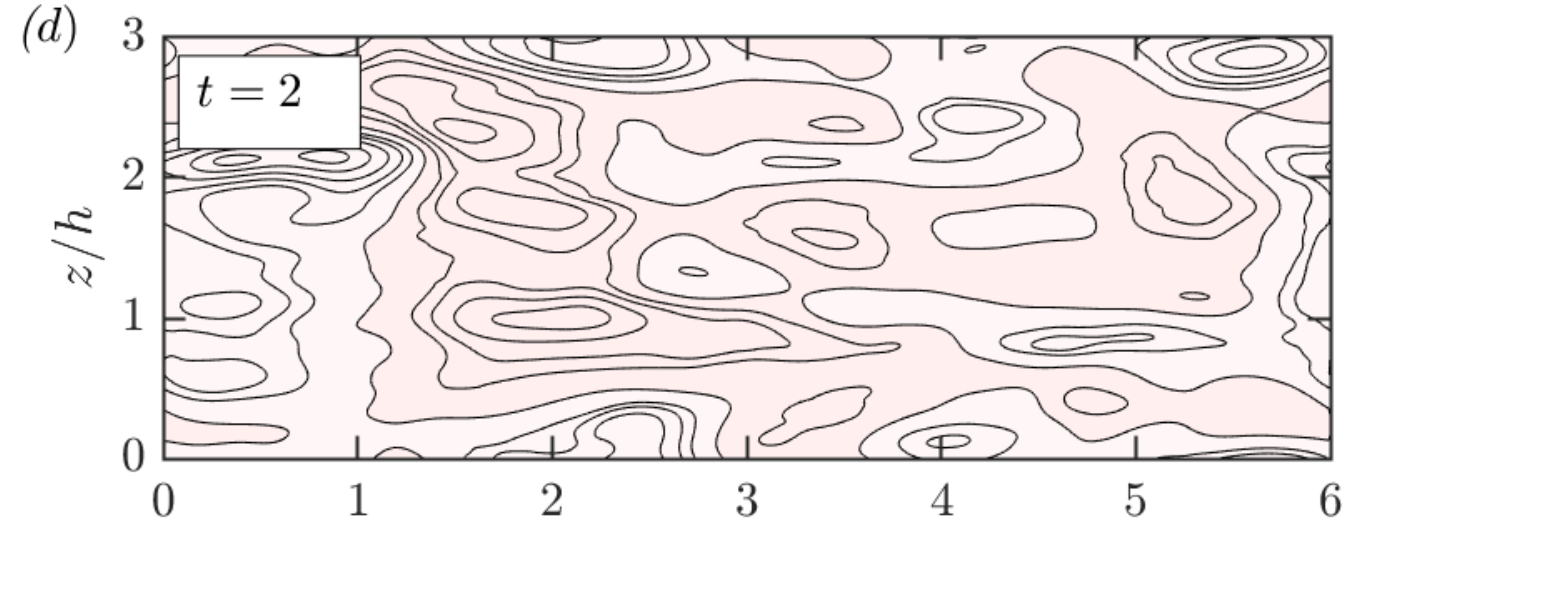}	& 
\includegraphics[scale=0.45,clip]{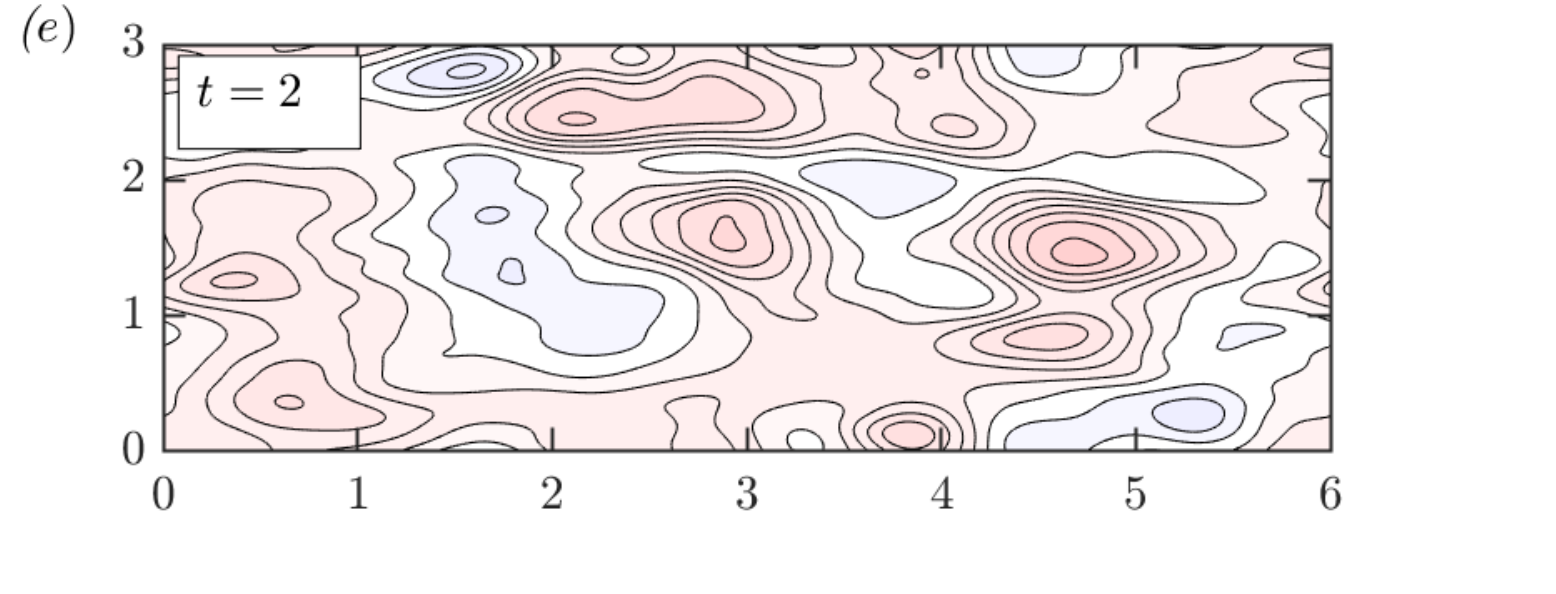} & 
\includegraphics[scale=0.45,clip]{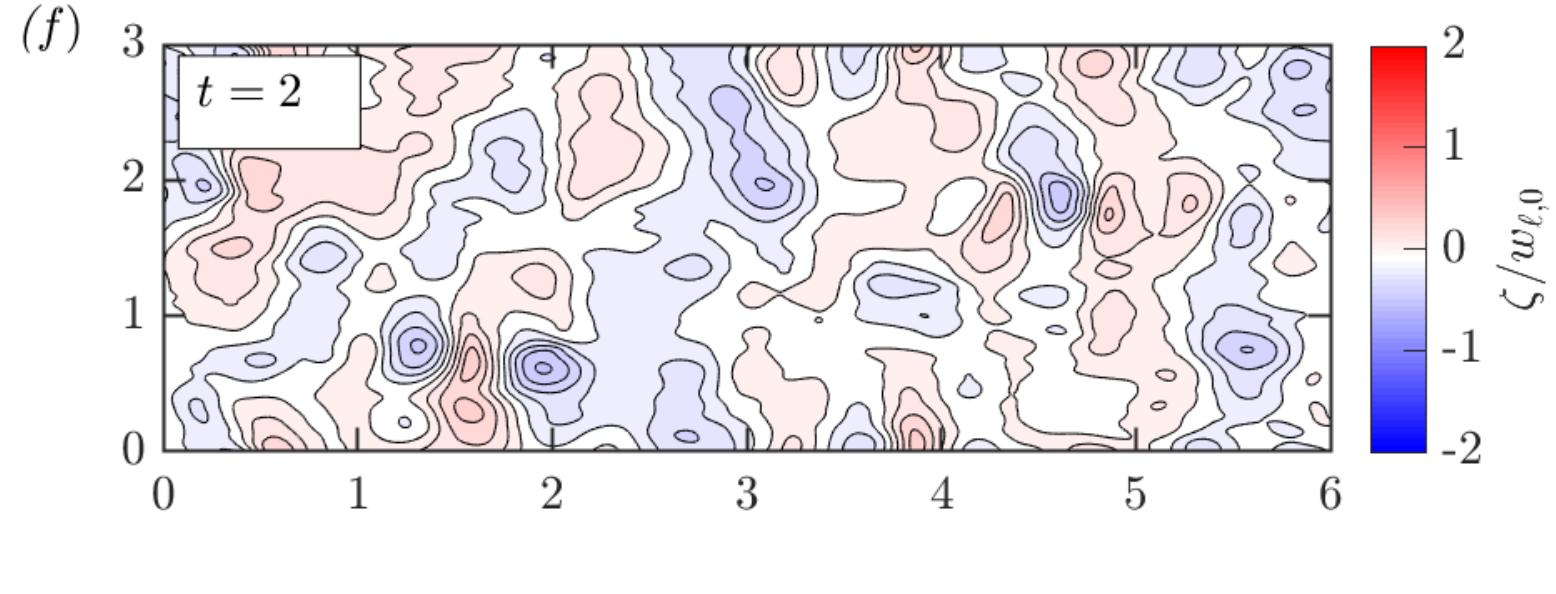} \\	

\includegraphics[scale=0.45,clip]{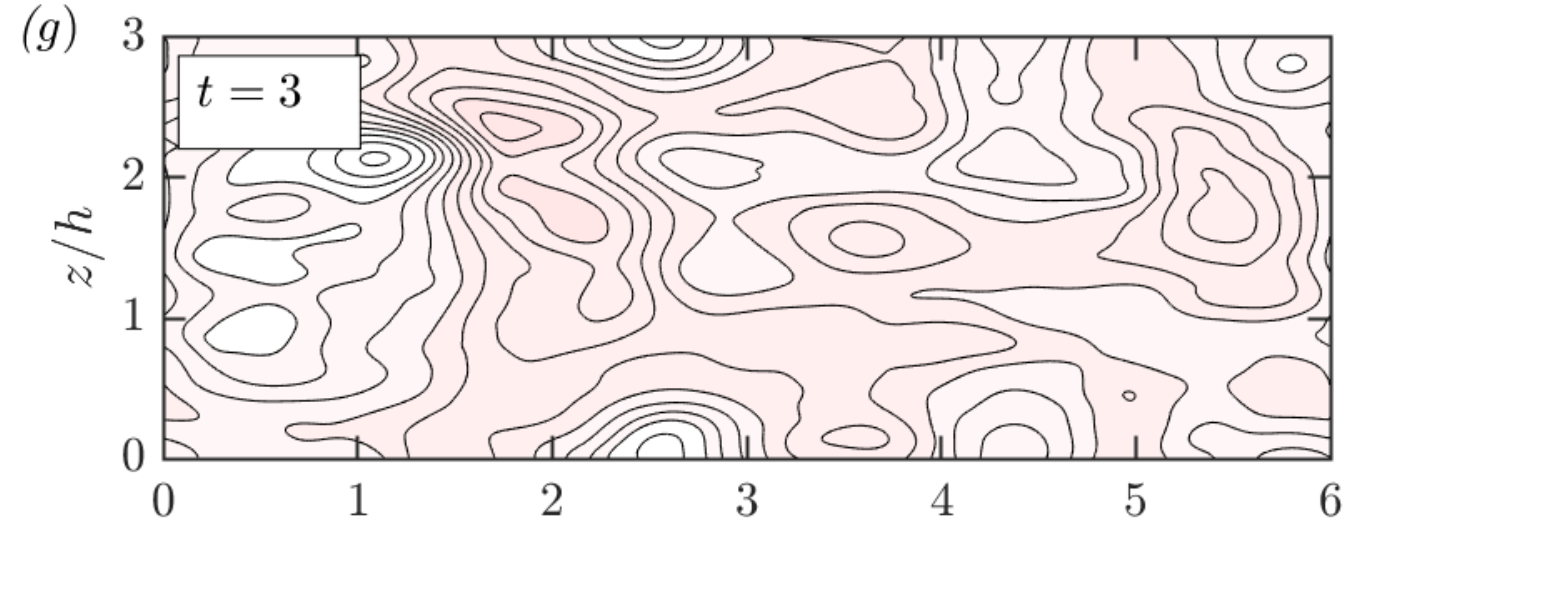}	& 
\includegraphics[scale=0.45,clip]{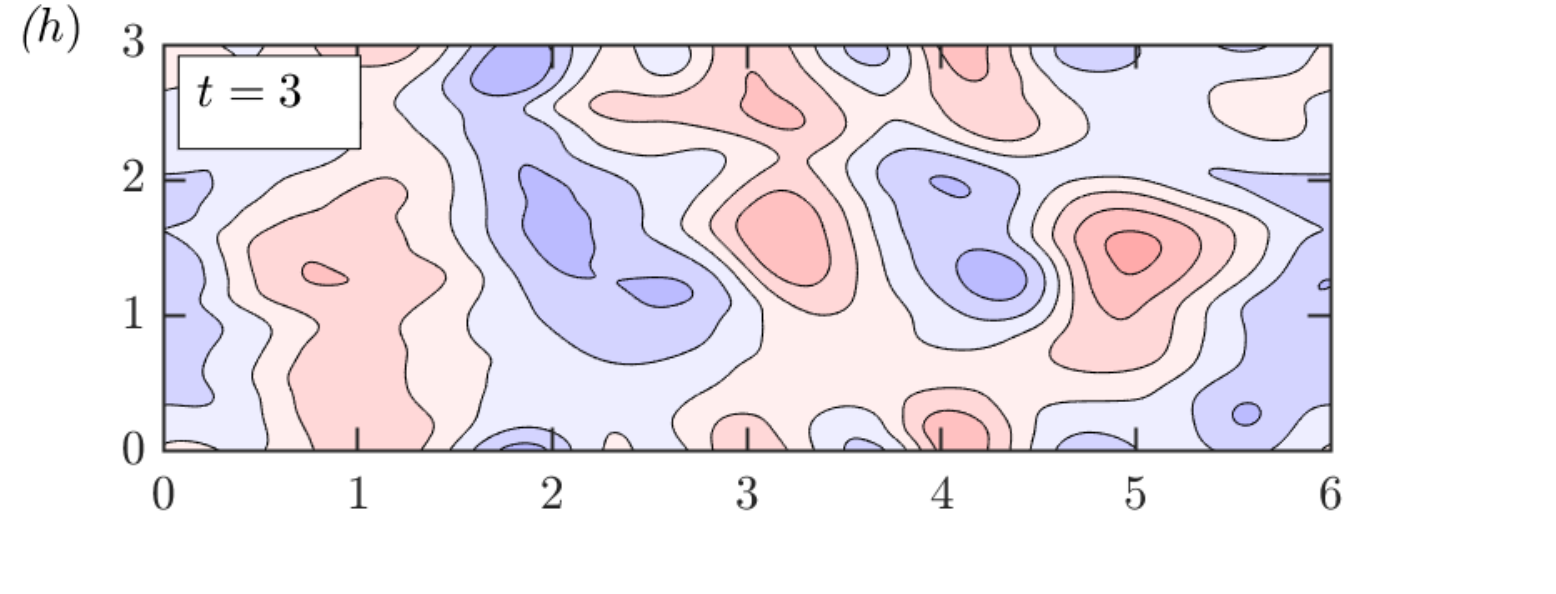} & 
\includegraphics[scale=0.45,clip]{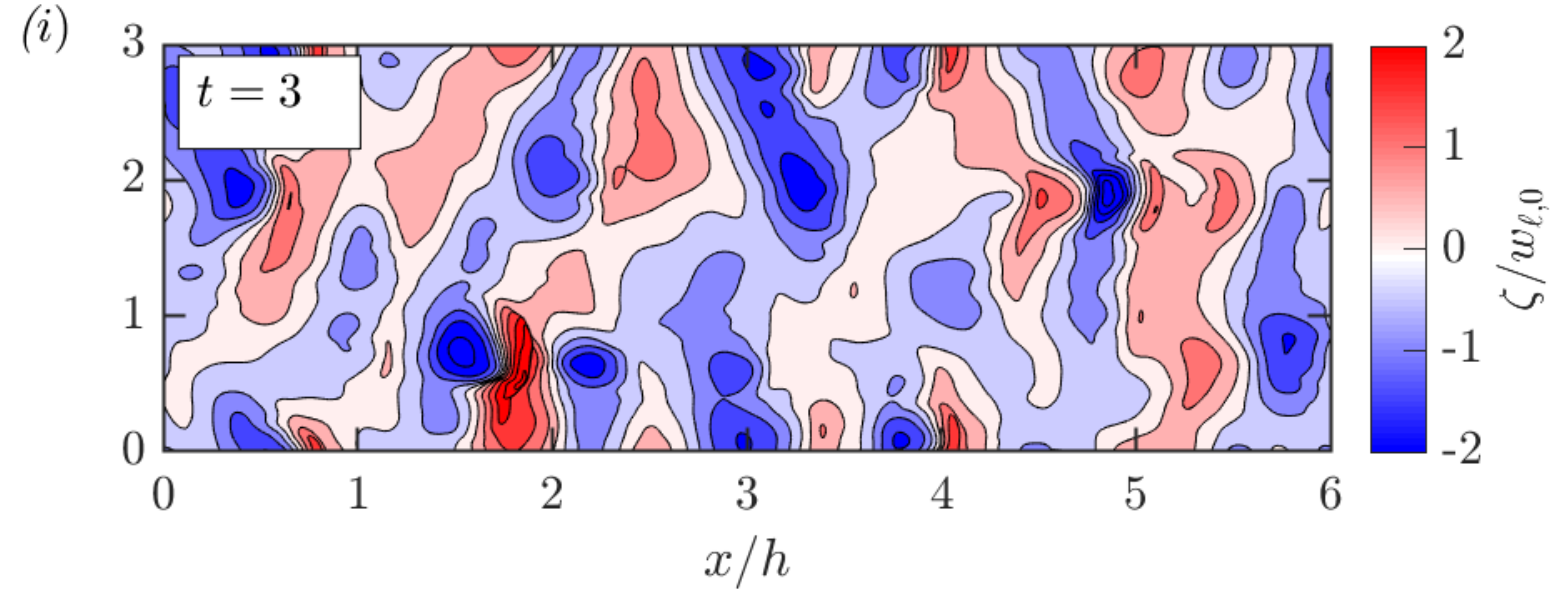} \\	

\includegraphics[scale=0.45,clip]{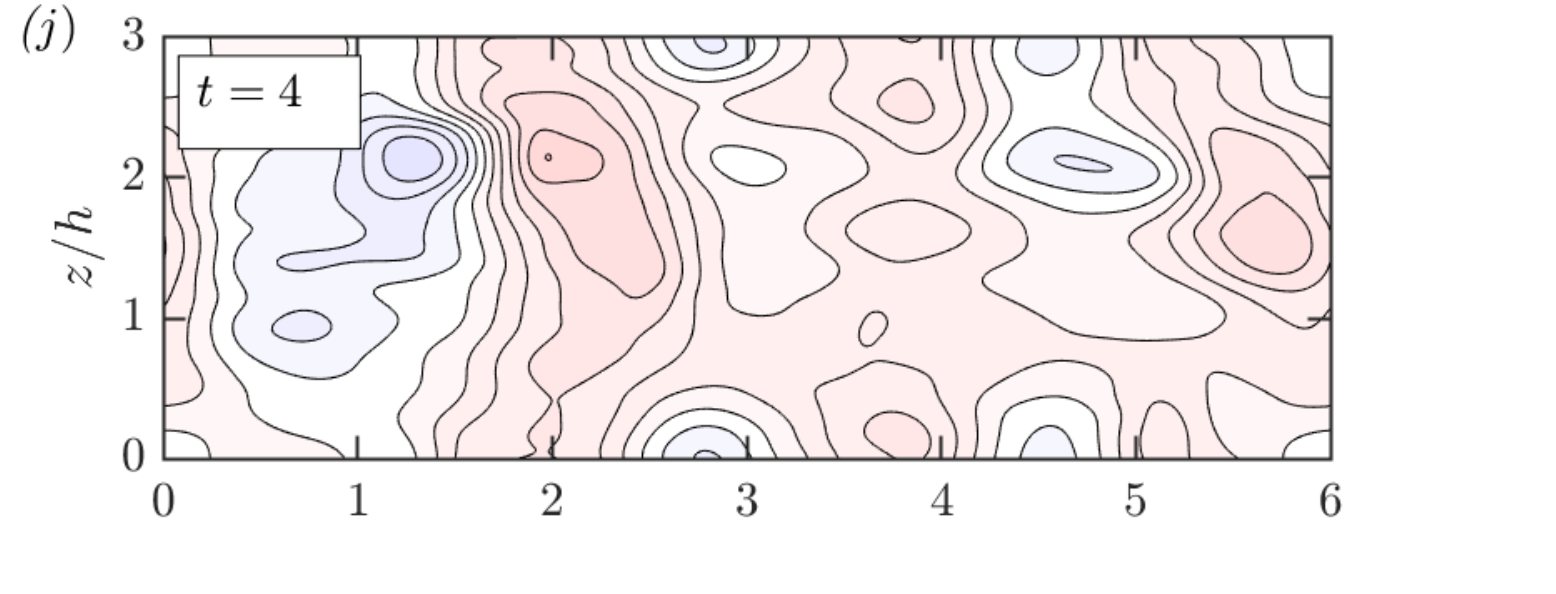}	& 
\includegraphics[scale=0.45,clip]{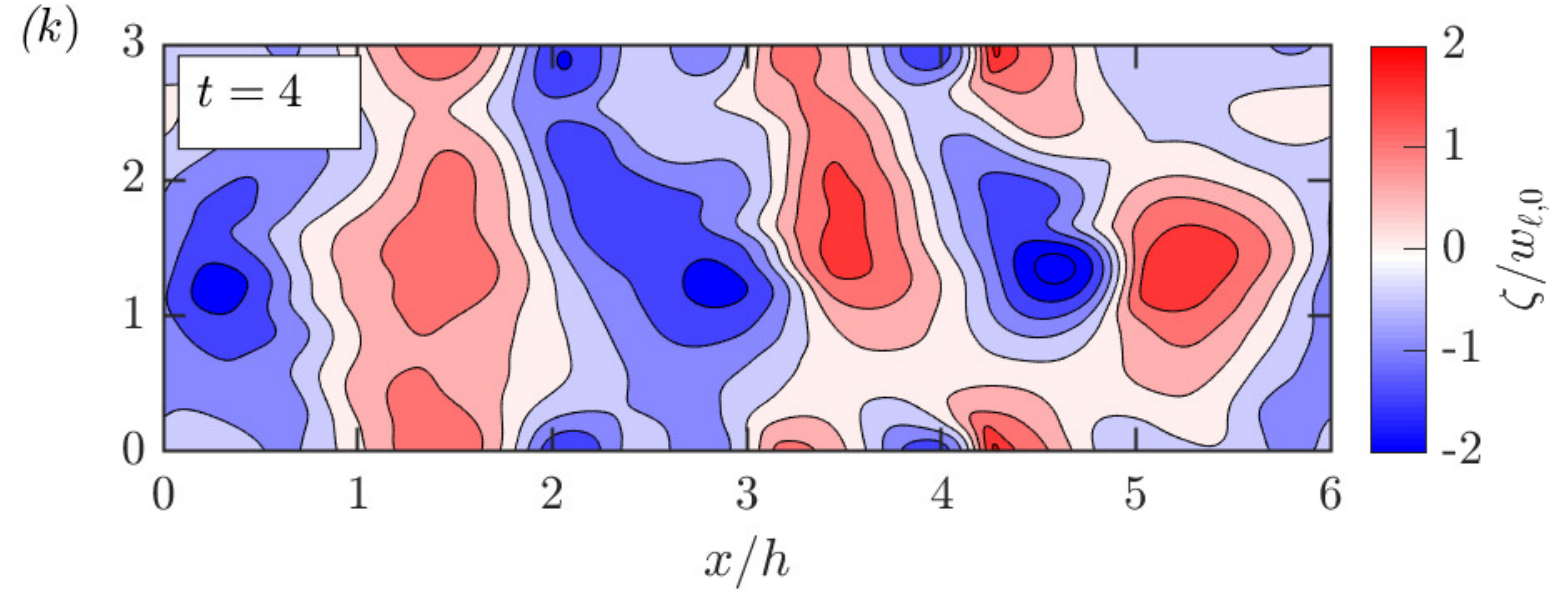} & 
\\	

\includegraphics[scale=0.45,clip]{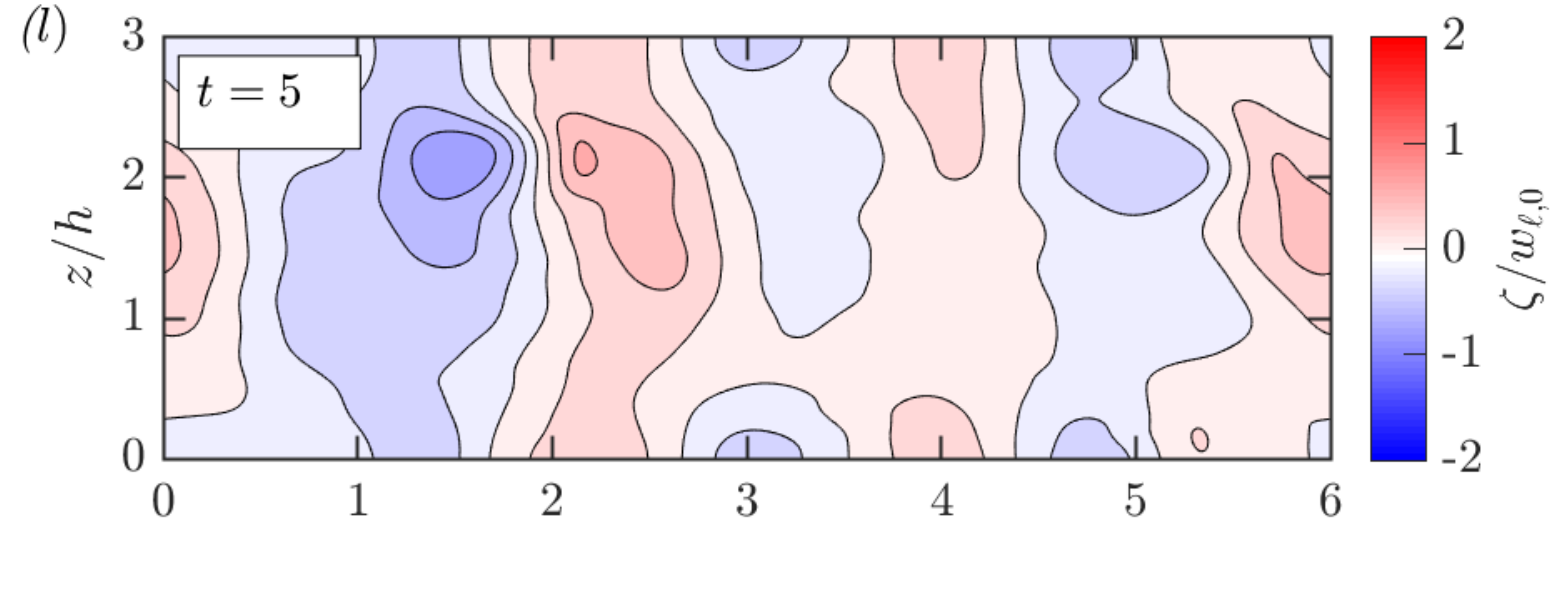}	& 
 & 
 \\	

\includegraphics[scale=0.45,clip]{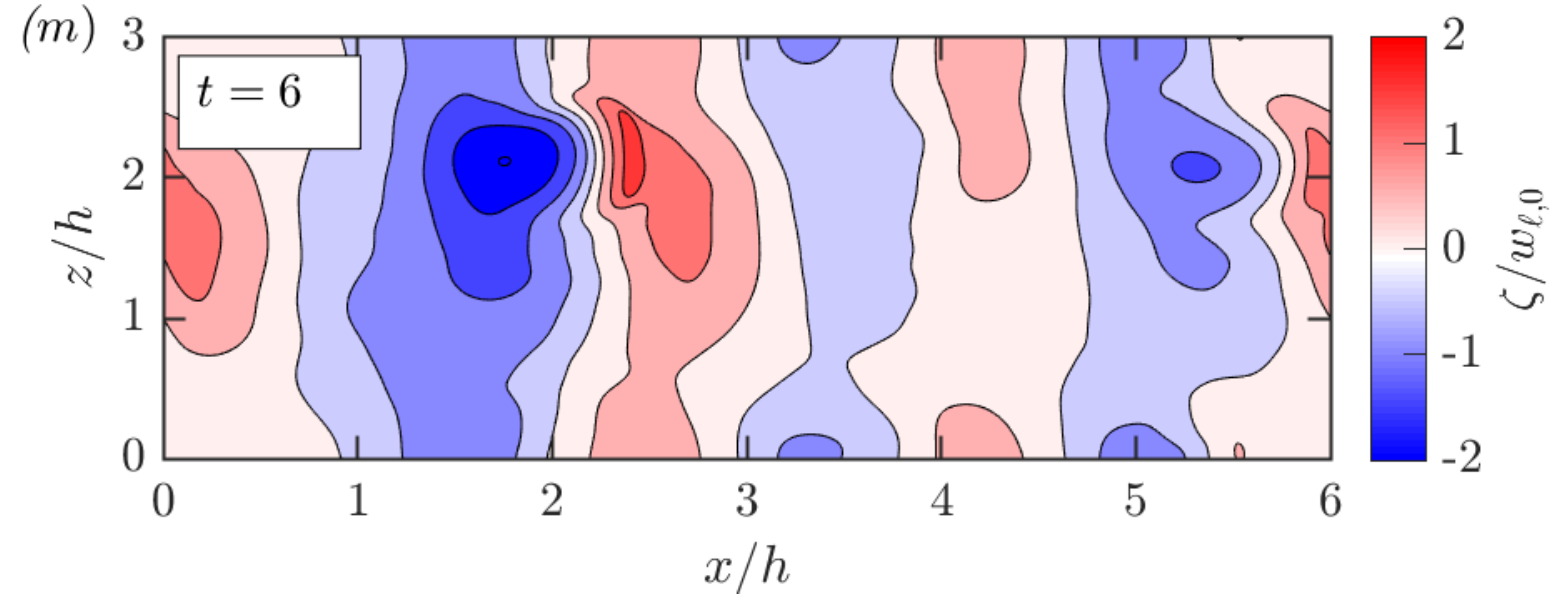}	& 
& 
\\	
	
 \end{tabular}
\caption{Contours of the deforming (top) film surface displacement $\zeta$  for the same three cases shown in figures \ref{fig:filmVisA}, \ref{fig:filmVisB} and \ref{fig:profiles_in_t}, scaled by initial film thickness $w_{\ell,0}$. Time advances downward in each column for each case. Contours are similar for the bottom film surface (not shown).}
\label{fig:filmSurfaceContours}
\end{center}	
\end{figure}

Returning to the visualisations in the $xy$-plane in figures \ref{fig:filmVisA} and \ref{fig:filmVisB}, figure \ref{fig:filmSurfaceContours} shows contours of the film surface as a deviation from the original film location, referred to as surface displacement $\zeta$ by analogy with that used in a wind-driven wave context (e.g. Ref. \cite{perrard2019turbulent}). Contours of $\zeta$ are shown in the $xz$-plane, computed with the aid of a reconstructed level-set field \cite{scapin2020volume}, and scaled with the initial film thickness $w_{\ell,0}$. At early times, deformations are small and incoherent, although their sizes show $Re$ dependence, and are due to the shear stresses and pressure fluctuations of the turbulent gas-phase boundary layer over the film's surface (which had developed over a solid wall in the channel simulations). Such deformations by a turbulent wind on a liquid surface, below the onset of wave generation, appear similar to the `wrinkles' in the wind-driven wave study of Ref. \cite{paquier2016viscosity}. At later times, a clear waveform in the streamwise direction emerges as the deviations from the original film location increase. These large deformations are those which eventually rupture the film. It is also interesting to compare the spanwise extent of this dominant streamwise waveform. In the LoRe-med and LoRe-thin cases, a single streamwise waveform extends over the whole span, whereas the film's surface in the HiRe-thin case is more three-dimensional although a clear waveform in the streamwise direction nonetheless emerges. A weaker spanwise waveform is also evident, as seen at later times in figure \ref{fig:render} for the LoRe-med case and giving rise to `cell' structures as previously reported \cite{park2004experimental}. A similar progression from disordered small surface displacements to large ordered streamwise waves was also observed for the surface of wind-wave generation in the numerical study of deep-water waves in Ref. \cite{lin2008direct}.

\begin{figure}
	\setlength{\tabcolsep}{-15pt} % change this to widen/narrow columns
	\def\arraystretch{0}
	\begin{center}	
		\begin{tabular}{ccc} 
			
	\hspace{-5.5mm} LoRe-med & 	\hspace{-5.5mm} LoRe-thin & 	\hspace{-5.5mm} HiRe-thin \\		
			
			\includegraphics[scale=0.45,clip]{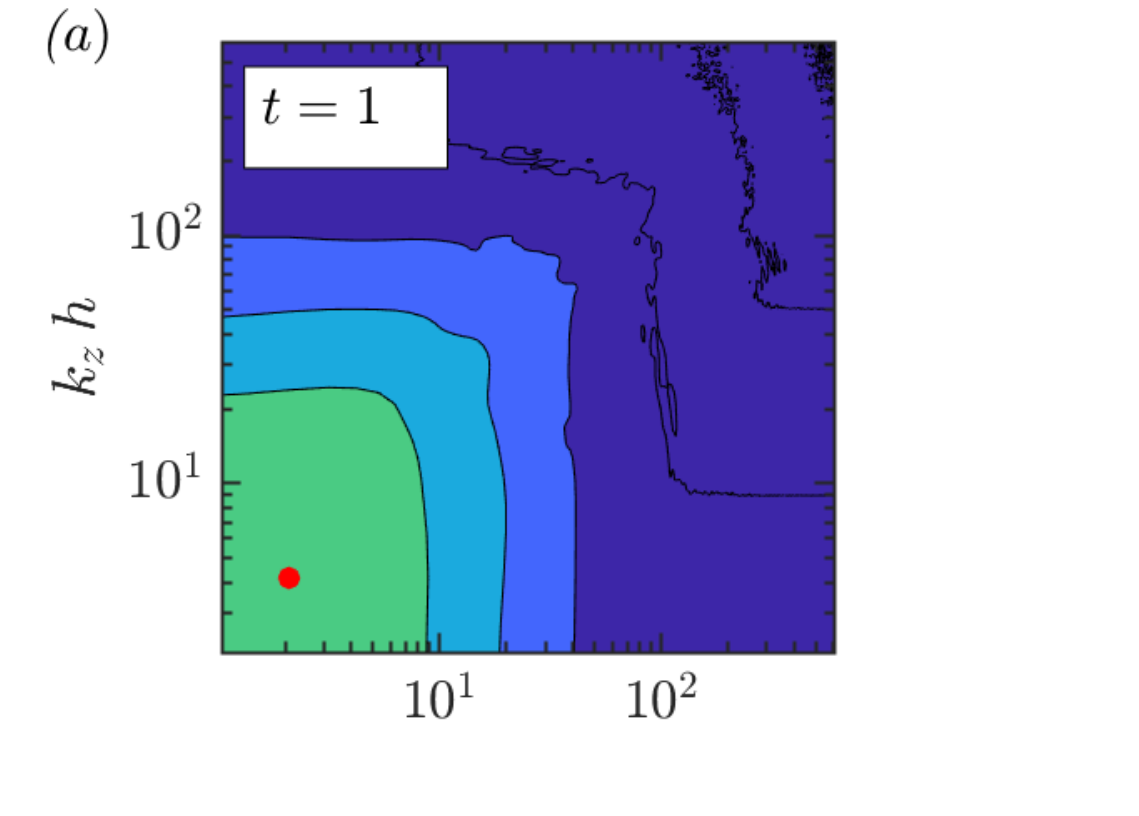}	& 
			\includegraphics[scale=0.45,clip]{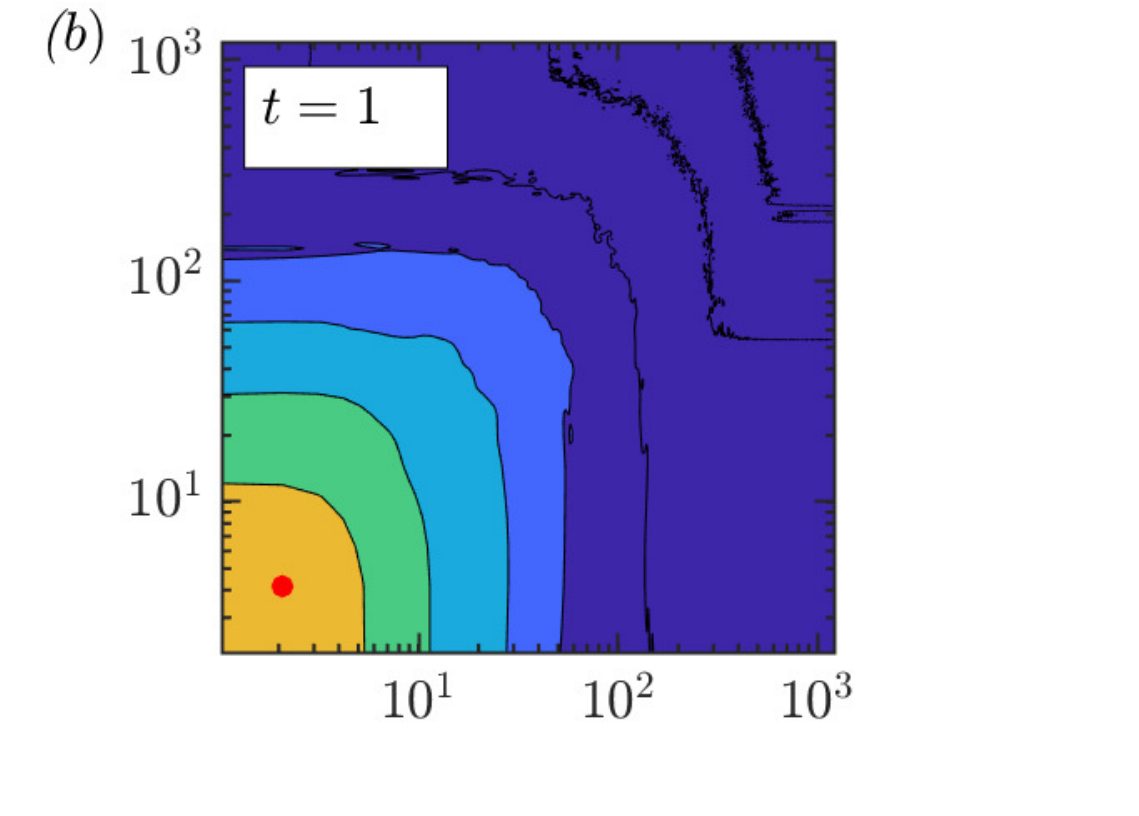} & 
			\includegraphics[scale=0.45,clip]{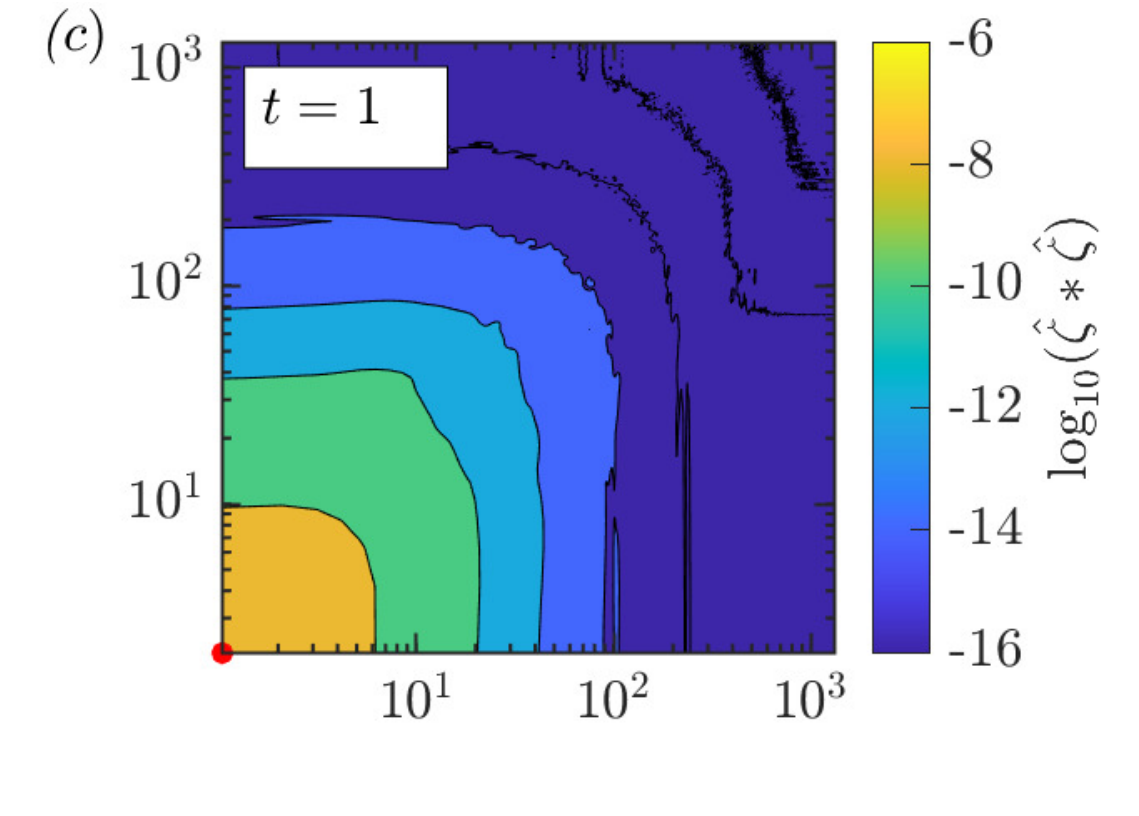} \\	
			
			\includegraphics[scale=0.45,clip]{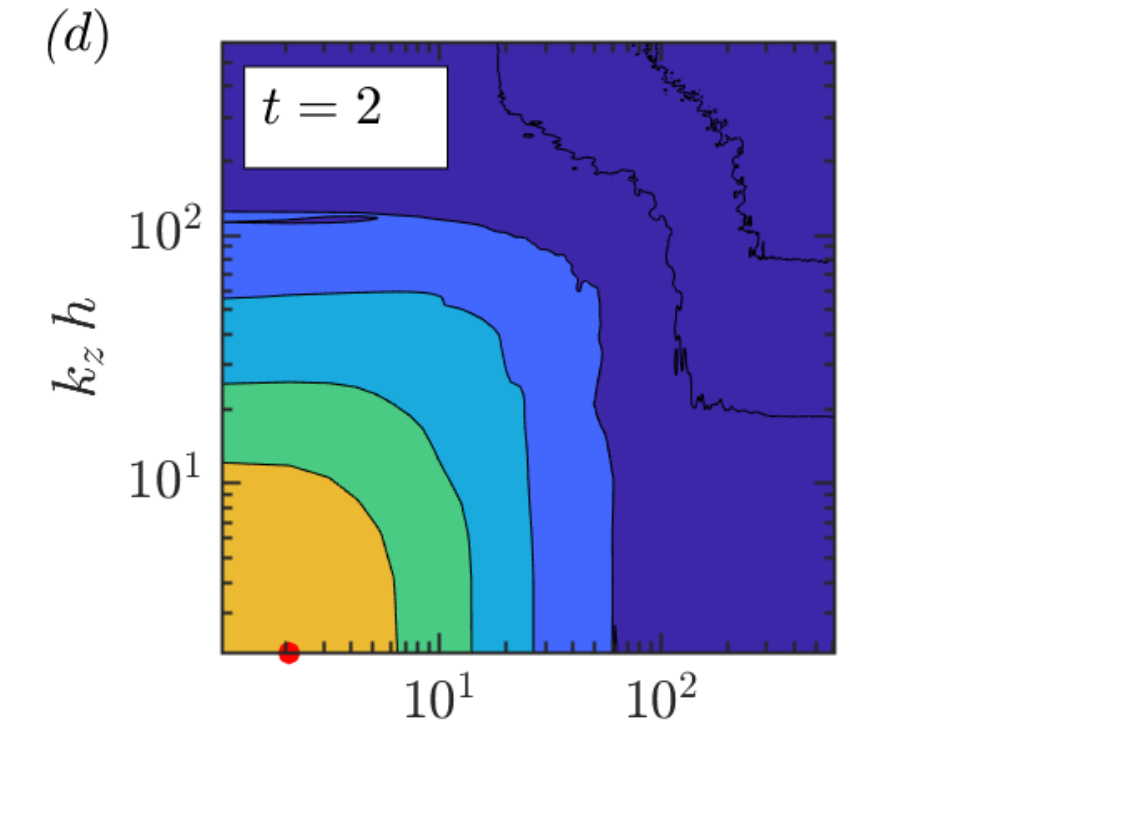}	& 
			\includegraphics[scale=0.45,clip]{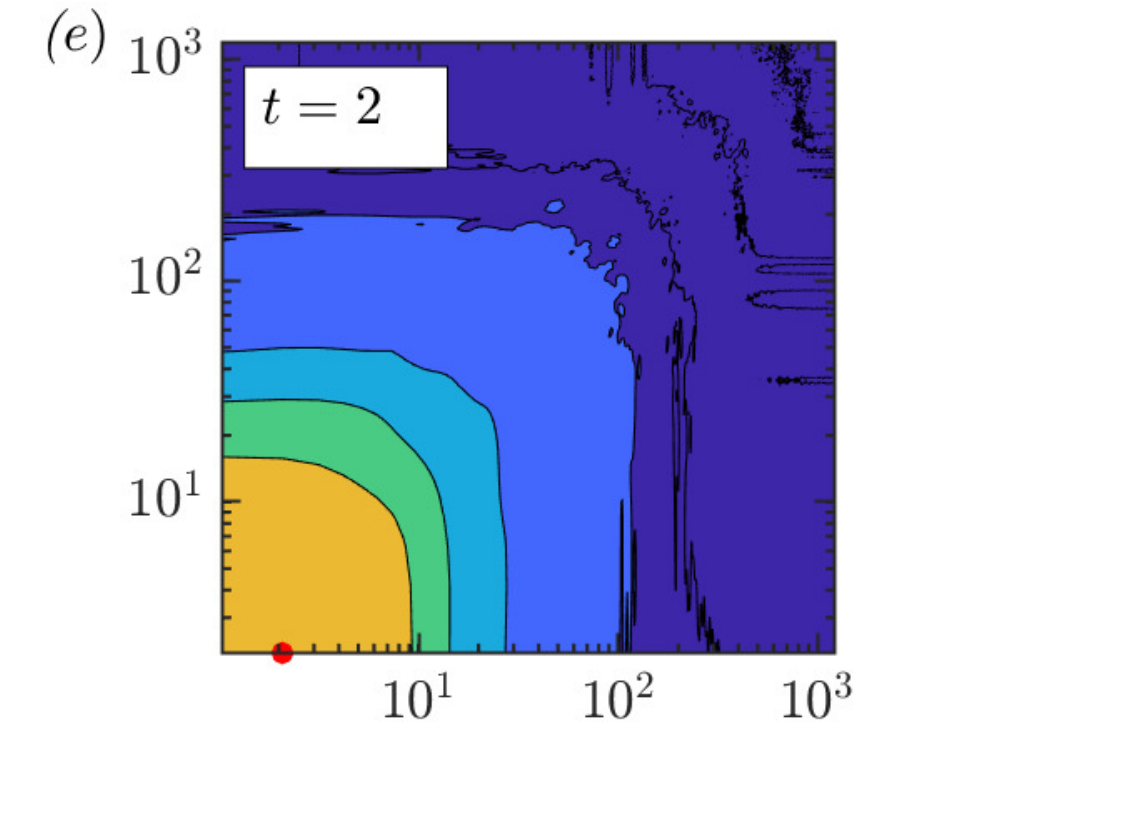} & 
			\includegraphics[scale=0.45,clip]{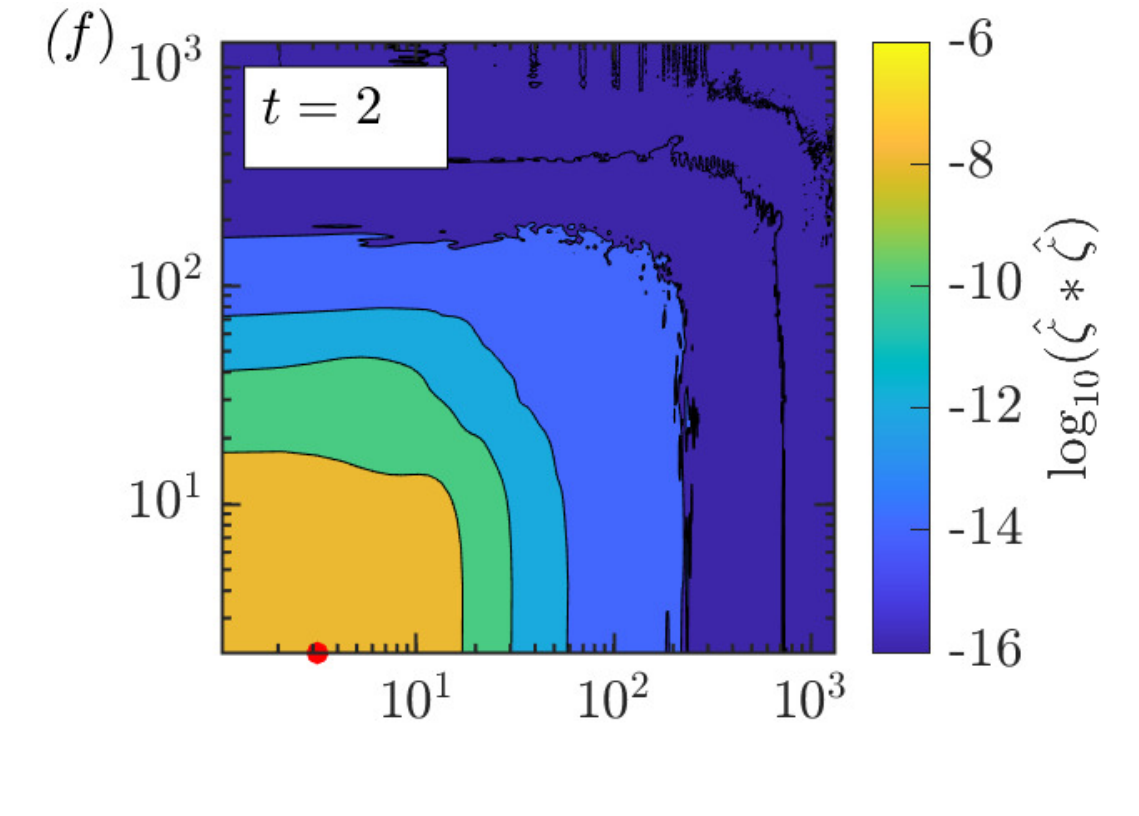} \\	
			
			\includegraphics[scale=0.45,clip]{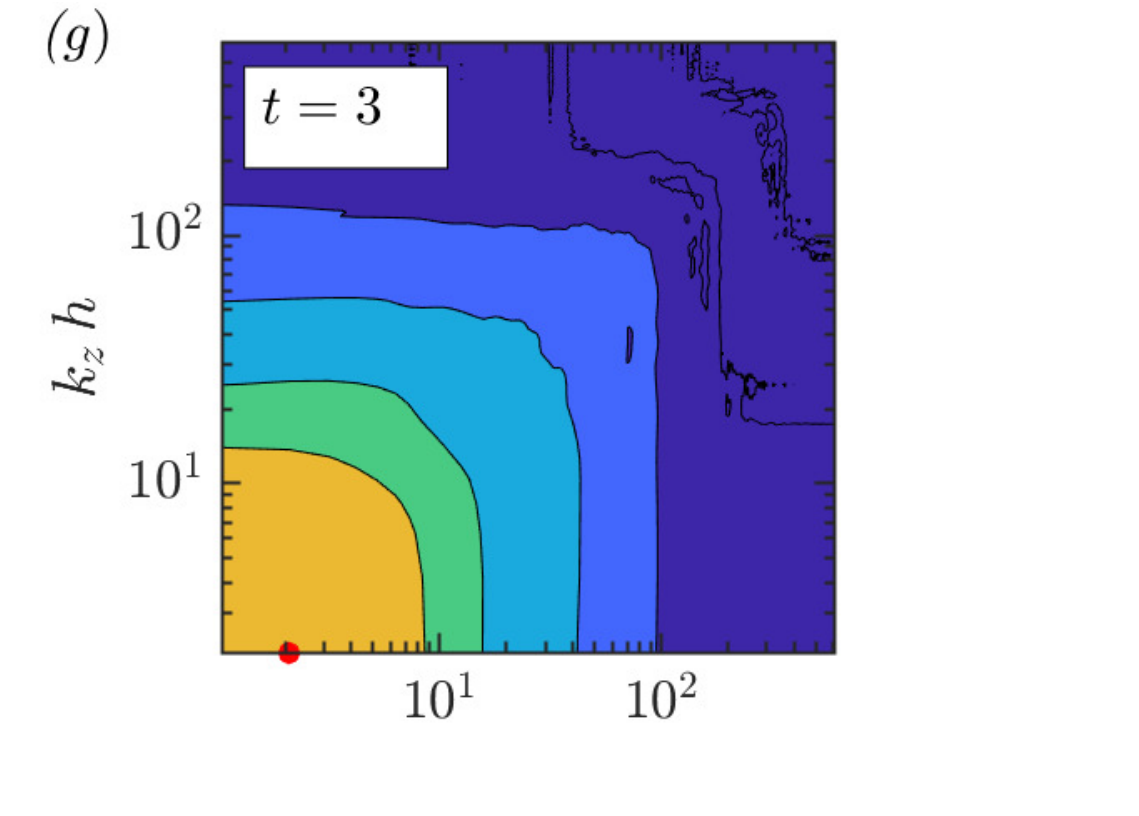}	& 
			\includegraphics[scale=0.45,clip]{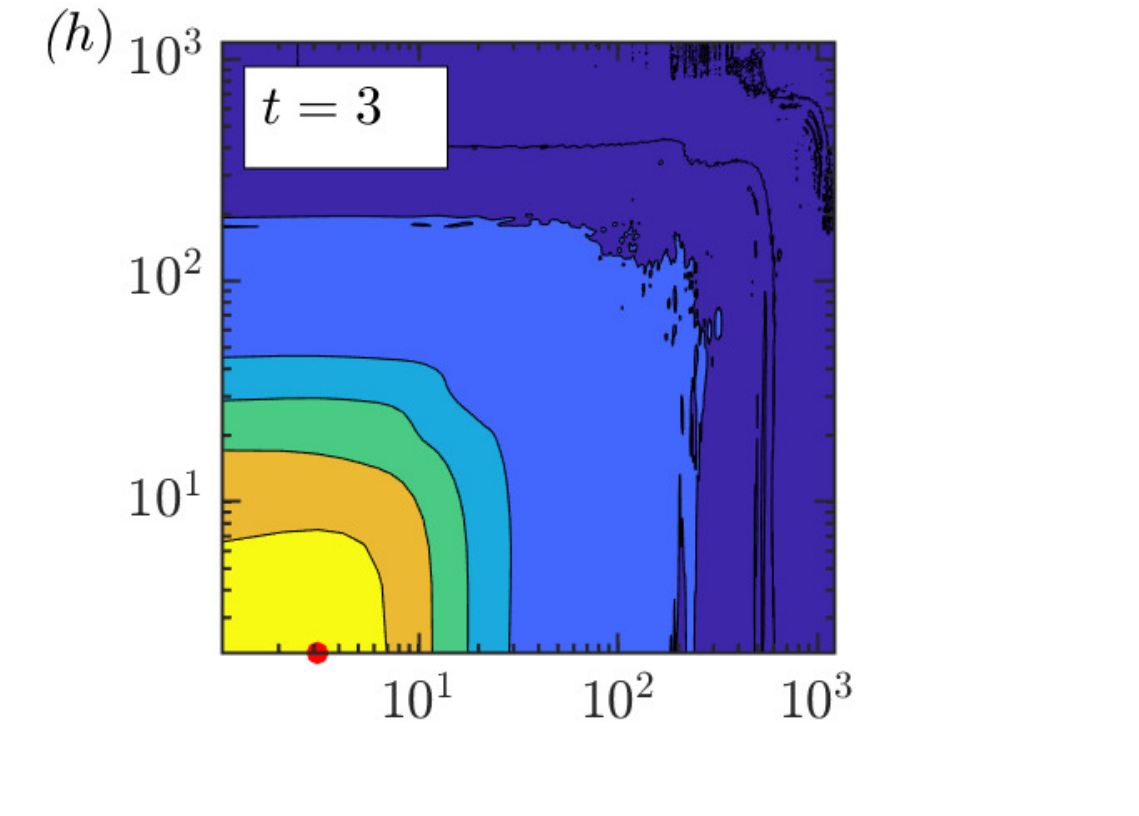} & 
			\includegraphics[scale=0.45,clip]{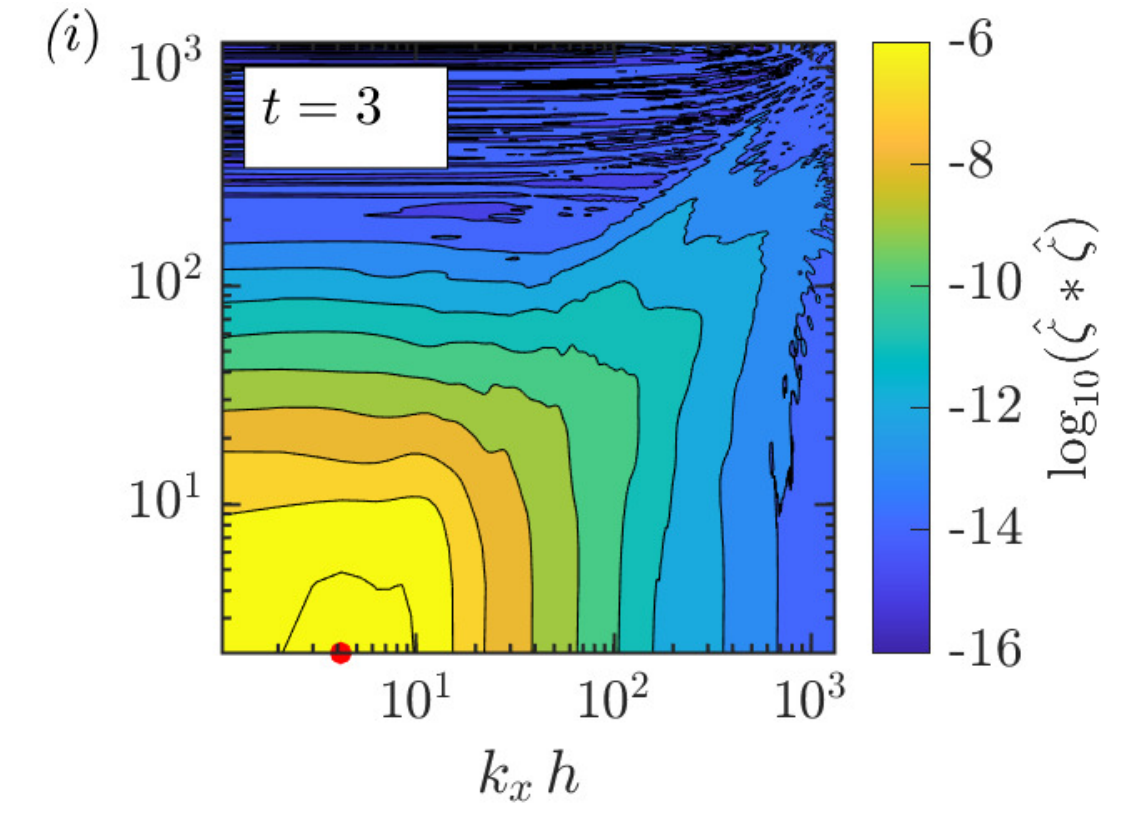} \\	
			
			\includegraphics[scale=0.45,clip]{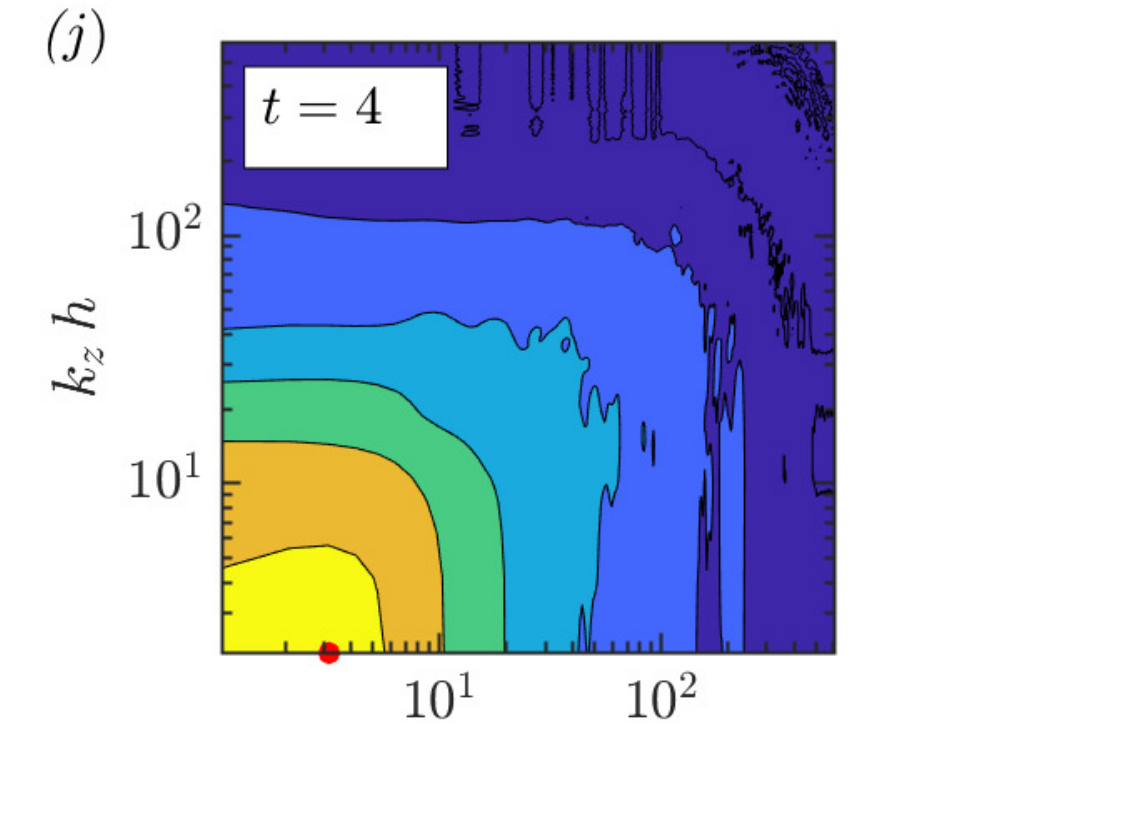}	& 
			\includegraphics[scale=0.45,clip]{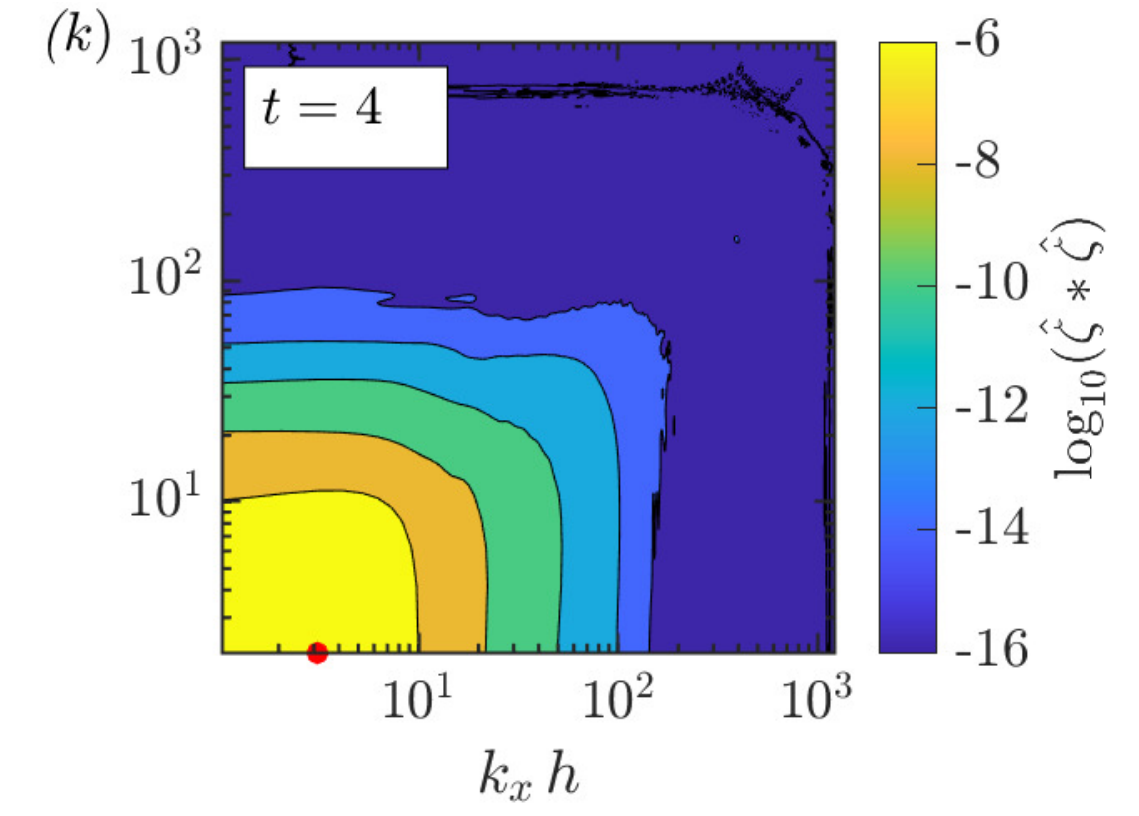} & 
			\\	
			
			\includegraphics[scale=0.45,clip]{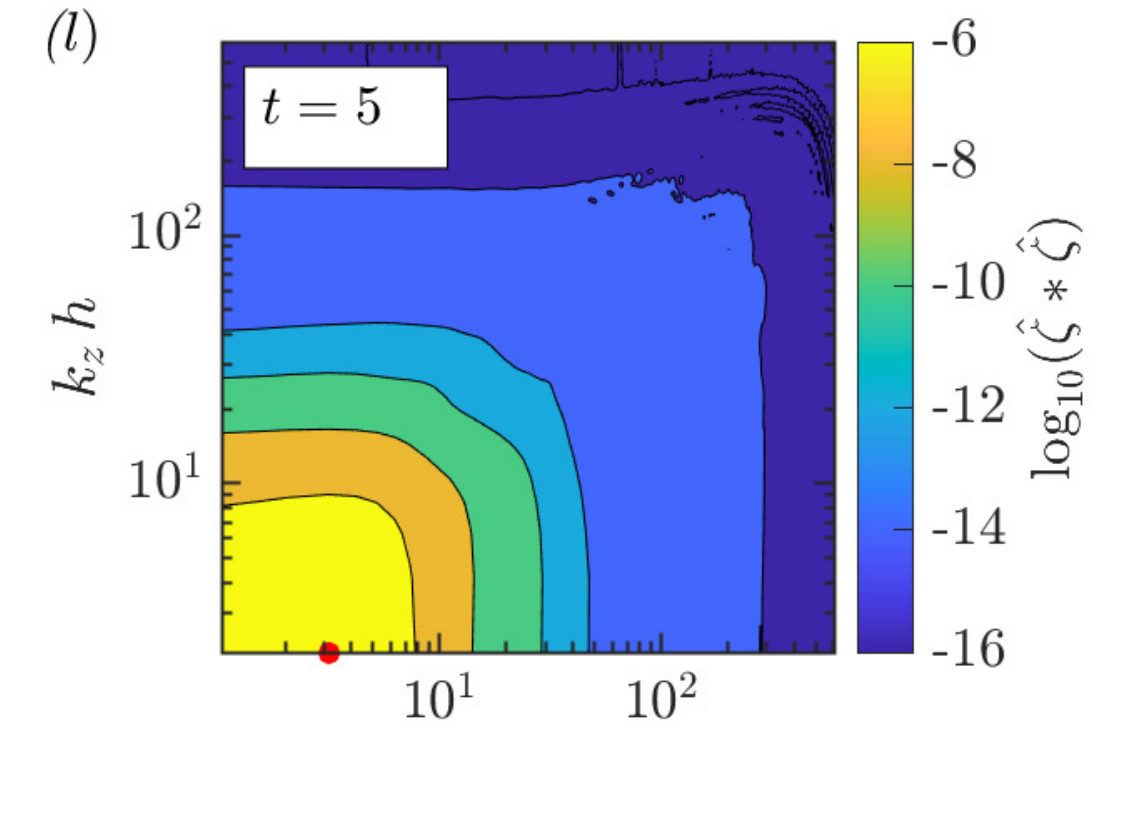}	& 
			& 
			\\	
			
			\includegraphics[scale=0.45,clip]{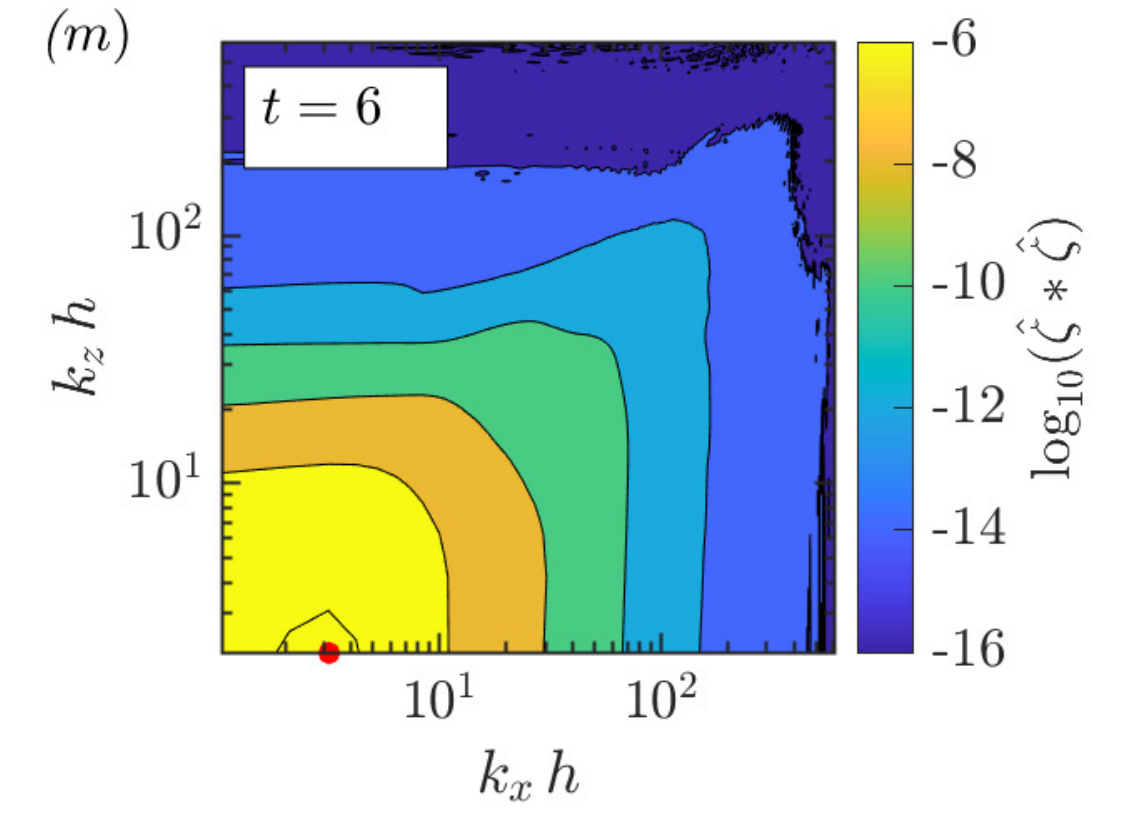}	& 
			& 
			\\	
			
		\end{tabular}
		\caption{Spectra of the surface displacement $\zeta$, computed for the same top film surfaces for the same three cases shown in figure \ref{fig:filmSurfaceContours}. The maximum in each subfigure is marked with a red bullet.}
		\label{fig:filmSurfaceSpectra}
	\end{center}	
\end{figure}

Figure \ref{fig:filmSurfaceSpectra} plots the temporal development of spectra of the surface displacement as a function of streamwise wavenumber $k_x$ and spanwise wavenumber $k_z$. The maximum is marked by a red bullet in each subfigure. The spectra are mostly rounded up to $t=2$, reflecting the early random surface displacements (figure \ref{fig:filmSurfaceContours}). At times $t \gtrsim 3$, the highest-energy contour of the spectra becomes elongated along the $k_x$ axis, although there is still significant energy along the spanwise $k_z$ axis, reflecting the secondary spanwise waveform seen in figure \ref{fig:filmSurfaceContours} at late times. A peak for small $k_x$ emerges for the LoRe-med (at $t = 6$) and HiRe-thin cases (at $t=3$), similar to that seen in the spectra of surface displacements for the deep-water wave simulations of Ref. \cite{lin2008direct}. It is centered at a larger $k_x$ for the HiRe-thin case indicating the smaller size of the emerging dominant streamwise waveform (and Reynolds number dependence on $k_x$). 

\begin{figure}
	\begin{center}
		
		\includegraphics[scale=0.5,clip]{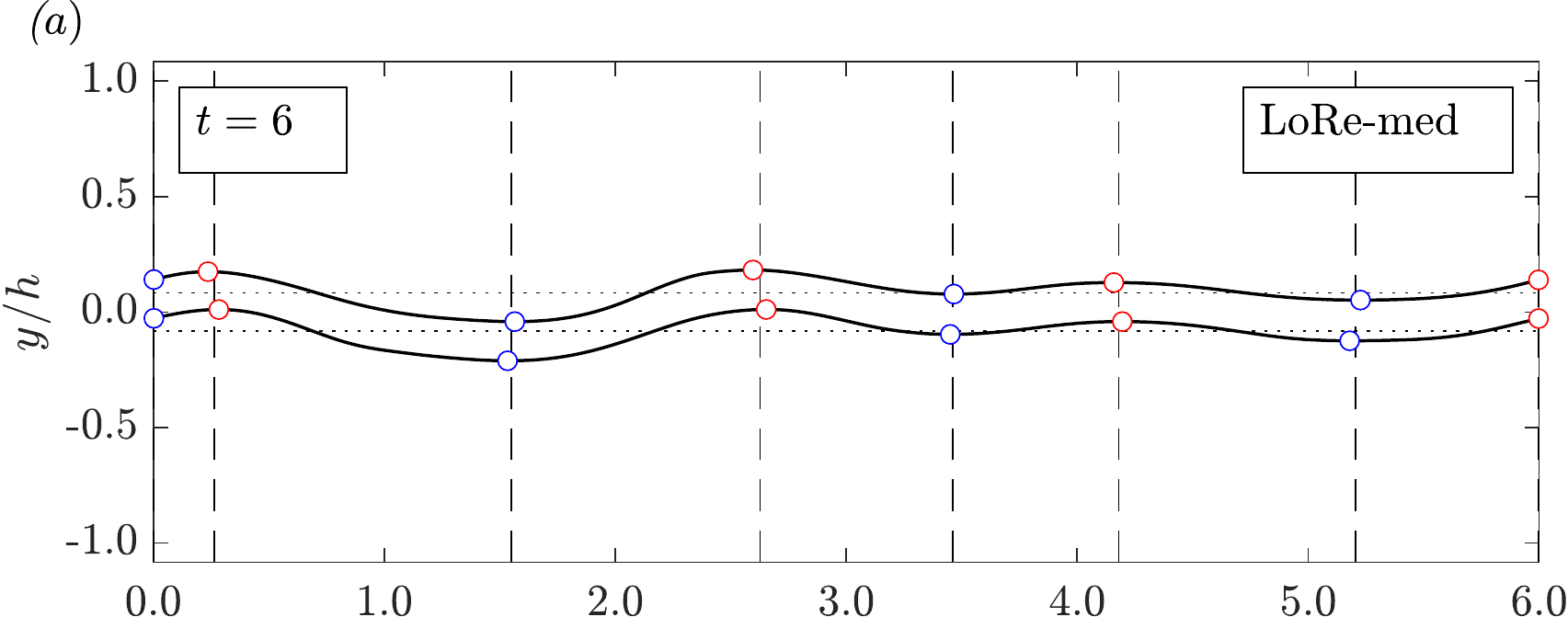}		
		\includegraphics[scale=0.5,clip]{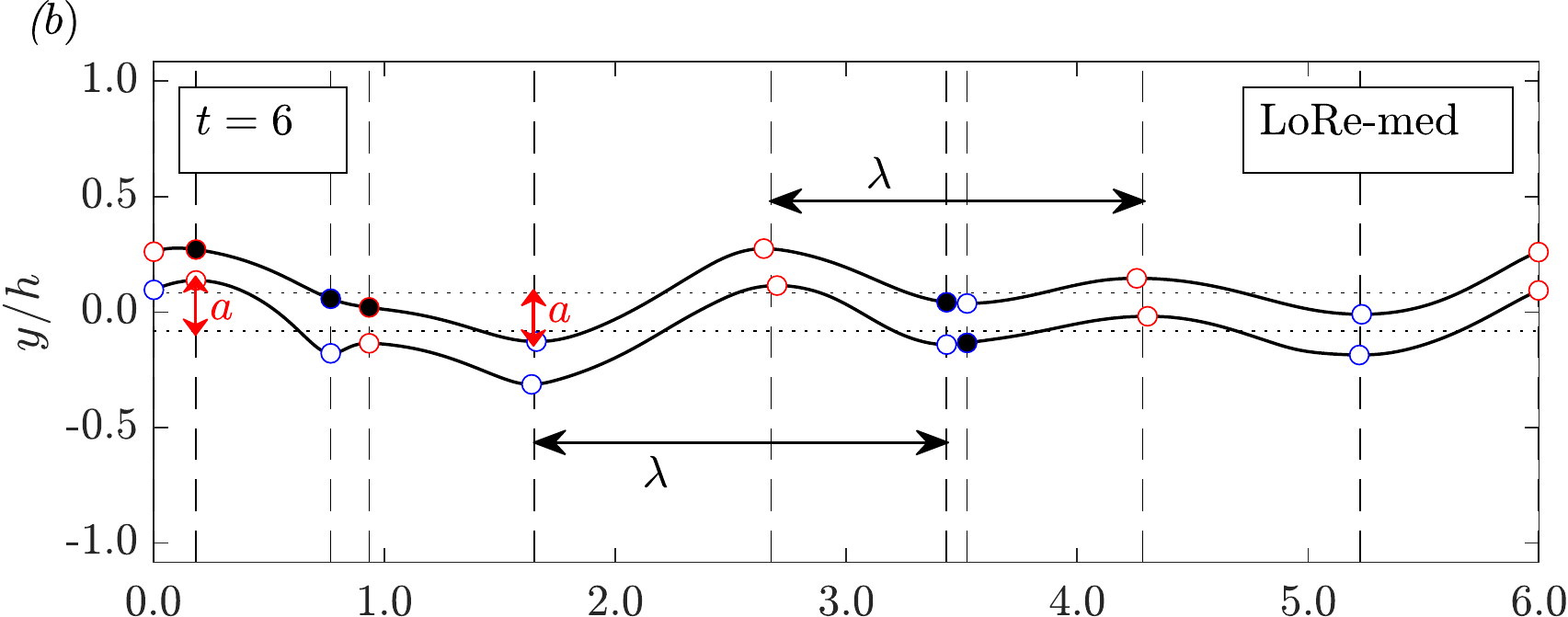}
		
		\vspace{3mm}
		
		\includegraphics[scale=0.5,clip]{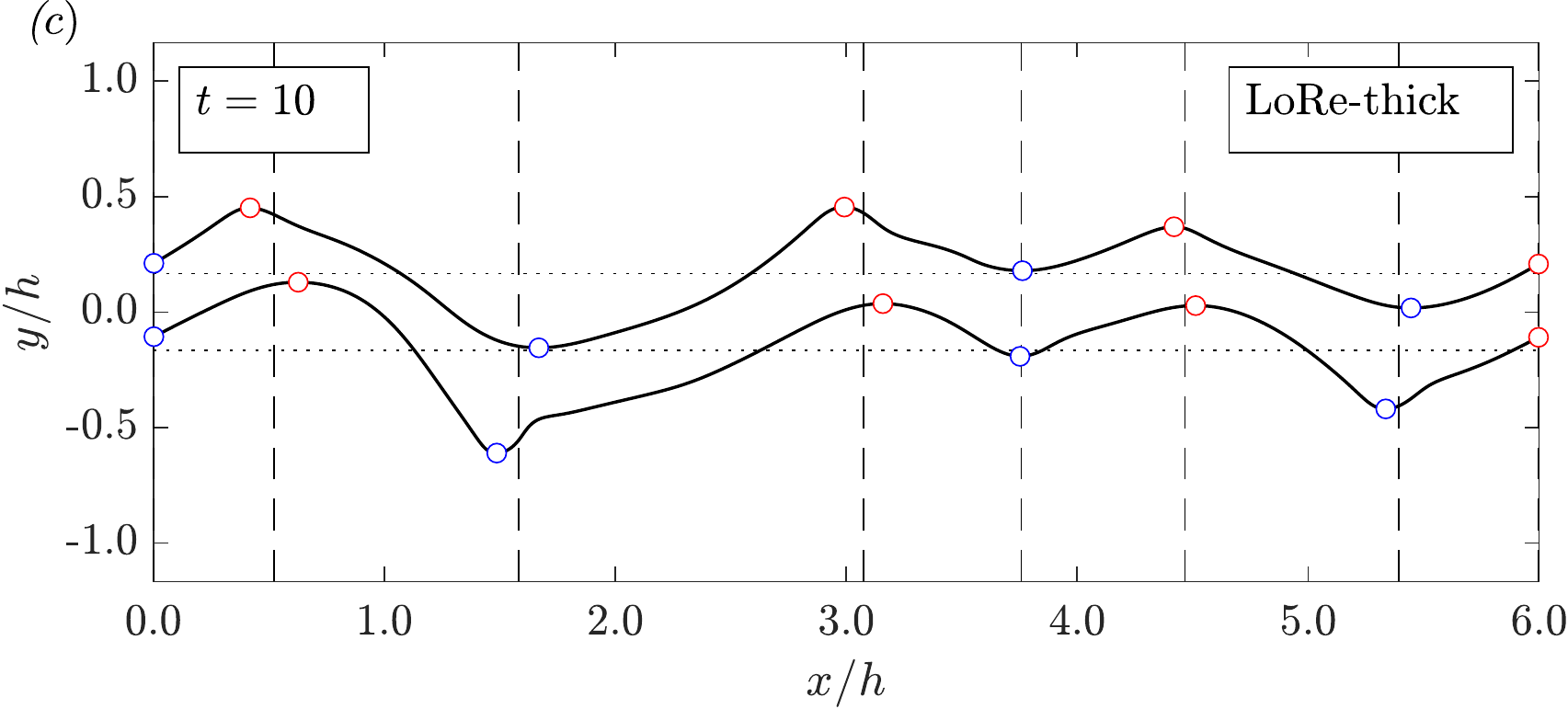}		
		\includegraphics[scale=0.5,clip]{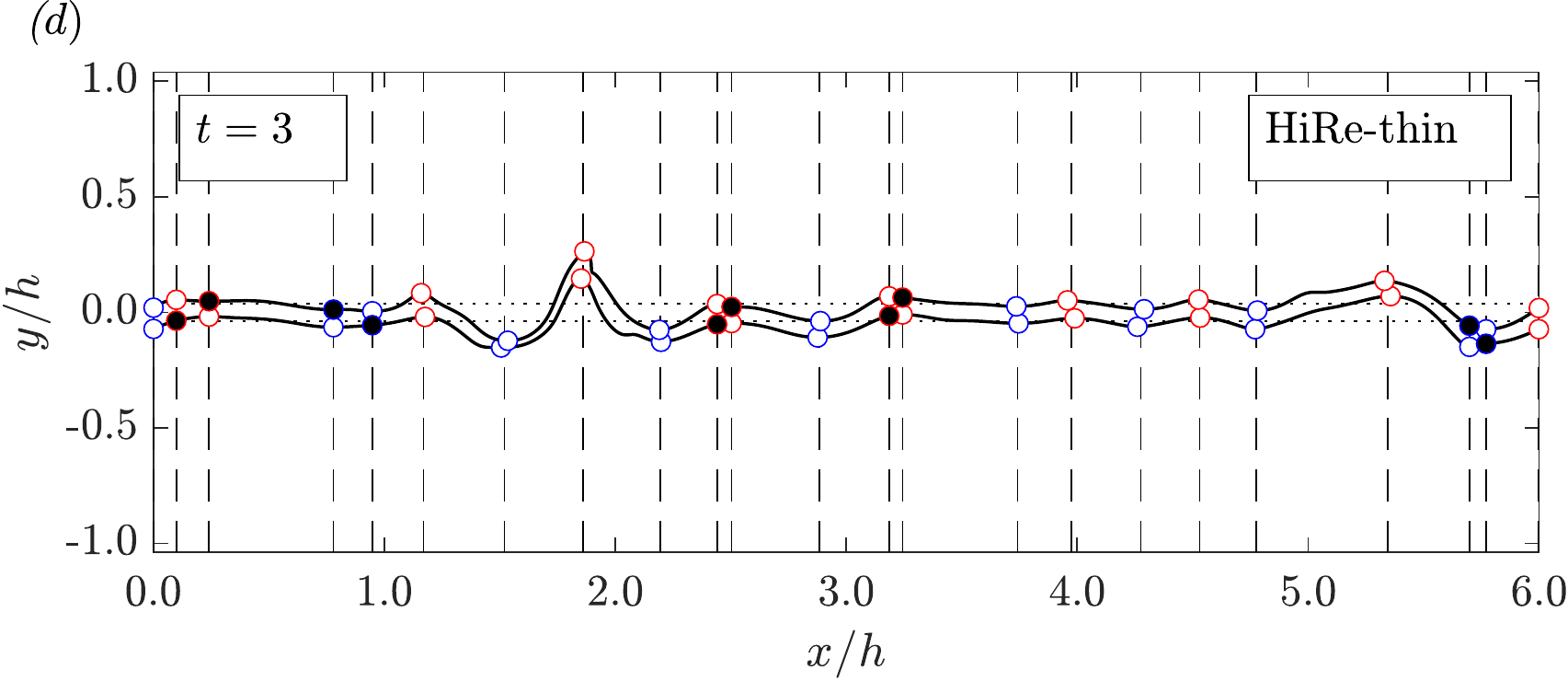}
		
		\caption{Indicative division of the deforming film into segments or panels, indicated by the vertical dashed lines, based on local minima and maxima (open bullets); (a,b) samples from the LoRe-med case at $t = 6$ at two different spanwise locations;  (c) LoRe-thick case at $t = 10$; (d) HiRe-thin case at $t = 3$. Filled bullets indicate points added from one side of the film to another if absent within a specified tolerance based on the starting film thickness. This procedure is carried out for each spanwise grid location. In some cases (e.g. subfigures (a) and (c)) no additional points need to be added, but in most cases (e.g. subfigures (b) and (d)) they are required in order to deduce the panel segments for the calculation of $C_{L,cumul.}$. For the purposes of calculating wavelengths ($\lambda$) and amplitudes ($a$) of the deforming liquid surface, only genuine local extrema (open bullets) are used; sample measurements marked in (b).}
		\label{fig:sampleSegments}
	\end{center}
\end{figure}

Several previous studies have attempted to measure the amplitude and wavelength of a deforming liquid surface. Ref. \cite{singh2020instability} extracted binary images of an atomizing liquid jet to deduce wavelengths ($\lambda$) and amplitudes ($a$) while the liquid core was still intact. An analysis based on local maxima and minima of the film surface extracting the streamwise $\lambda$ and $a$ is undertaken for the present cases. Figure \ref{fig:sampleSegments} shows contours of the film's surface in the $xy$-plane for several of the present cases making the emerging streamwise wavelength clear. The cases are approximately sinusoidal, although LoRe-thick case in figure \ref{fig:sampleSegments}(c) is somewhat dilational as slight film thickening is observed. Figure \ref{fig:sampleSegments} also indicates local maxima (red open bullets) and local minima (blue open bullets) on either side of the film. The distance between adjacent local minima or maxima is used to deduce a wavelength; since we can record this on either side of the film, we have up to 4 measurements of the wavelength for each $z$ (spanwise) location. The amplitude is defined as being the distance between the local extrema and the starting flat film location (horizontal dotted lines in figure \ref{fig:sampleSegments}). Additional closed bullets are added in order to deduce panels (vertical dashed lines in figure \ref{fig:sampleSegments}) for the aerodynamic coefficient calculations in Section \ref{sec:LiftDrag}. This was done by checking if minima or maxima found on one side of the film had a corresponding `partner' on the other side of the film within a certain tolerance based on $w_{\ell,0}$. If this was absent, the `orphan' minimum/maximum was copied to the other side of the film. 

\begin{figure}
	\begin{center}
		
	\includegraphics[scale=0.9,clip]{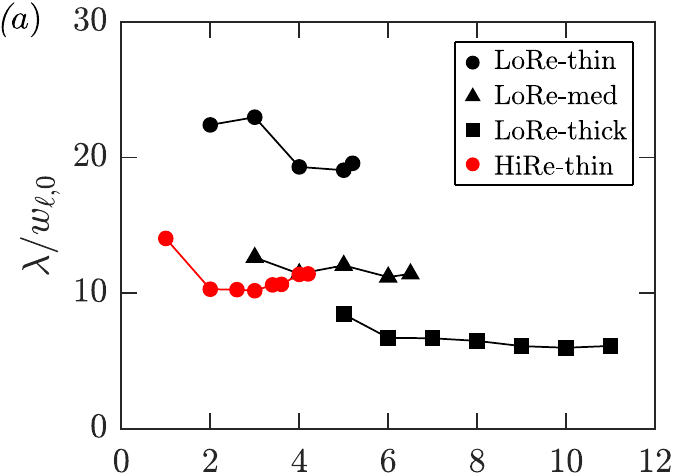}	
	\includegraphics[scale=0.9,clip]{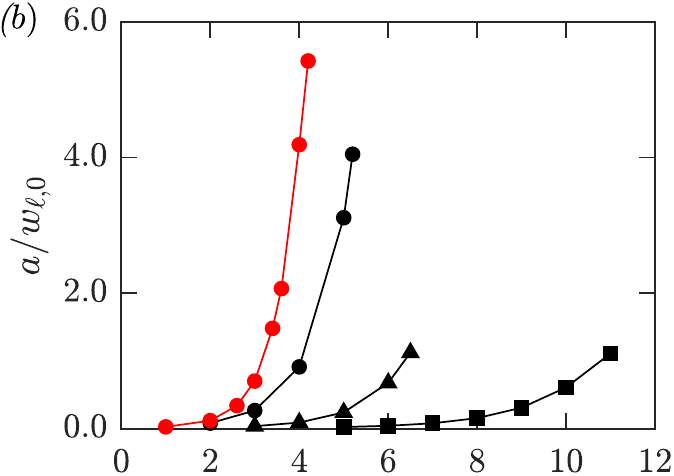}	
	
	\vspace{2mm}
	
	\includegraphics[scale=0.9,clip]{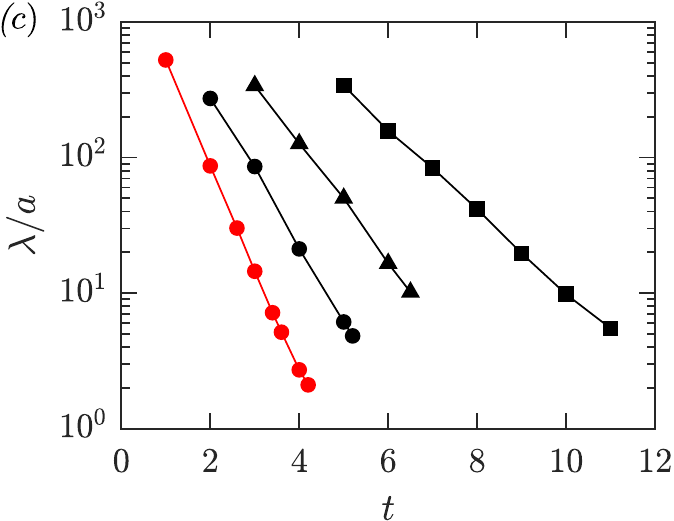}	
    \includegraphics[scale=0.9,clip]{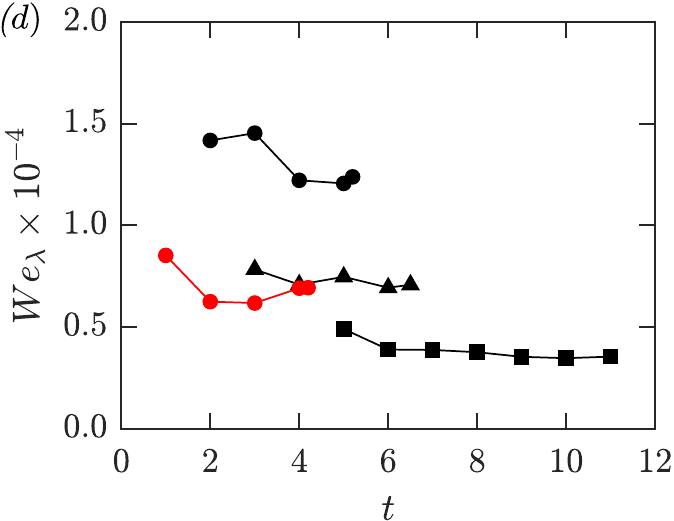}	
		
		\caption{Characteristics of the deforming film as a function of time; (a) streamwise wavelength of the film's surface $\lambda$; amplitude of surface deformation $a$ (measured as the difference from the original film's location on either side); (c) ratio $\lambda/a$; (d) $We_\lambda$, a Weber number defined using $\lambda$, the streamwise wavelength of the deforming film. Note that significant film deformation must occur to reliably measure of $\lambda$ or $a$, therefore the curves in figure \ref{fig:LambdaA} begin at different $t \geq 1$ depending on the film's rate of deformation. Symbols as per legend in (a).} 
		\label{fig:LambdaA}
	\end{center}
\end{figure}

The procedure demonstrated in figure \ref{fig:sampleSegments} allows us to calculate various parameters of the liquid sheet's deformation. Figure \ref{fig:LambdaA} plots the development of the streamwise liquid surface wavelength $\lambda$ and amplitude $a$ in time. Note that significant film deformation must occur to reliably measure of $\lambda$ or $a$, therefore the curves in figure \ref{fig:LambdaA} begin at different $t \geq 1$ depending on the film's rate of deformation. Interestingly, $\lambda$ does not vary significantly in time as shown in figure \ref{fig:LambdaA}(a). Despite their ubiquity, small liquid surface deformations have remained uncharacterised within experiments until recently due to their small size requiring very high surface-normal resolution. In a turbulent wind-driven water wave setup, Ref. \cite{paquier2016viscosity} found the amplitude of `wrinkles' emerging below the wave threshold to scale with $\sim \nu_\ell^{-1/2} u_\tau^{3/2}$, whereas their size (related to the present $\lambda$) remained virtually unchanged. The present simulations are clearly beyond the `wave threshold', yet the early small deformations on the film's surface may be similar to that observed by Ref. \cite{paquier2016viscosity}. Figure \ref{fig:LambdaA}(a) suggests the $\lambda$ that emerges is influenced by both the Reynolds number and film thickness for the present cases, yet once it emerges, it does not seem to change significantly as the film continues to deform away from the starting film location. Unlike the wavelength $\lambda$, $a$ increases exponentially with time as shown in figure \ref{fig:LambdaA}(b), which means the ratio $\lambda/a$ decreases exponentially in figure \ref{fig:LambdaA}(c). Ref. \cite{singh2020instability} found breakup of the liquid jet to occur a ratio of $\lambda/a \approx 2$ in their experimental study. In the present cases, where the curves extend until the film ruptures, we find that the HiRe-thin case (red bullets) achieves a minimum value of $\lambda/a \approx 2$, however in the other cases the film ruptures earlier. The initial Weber number $We_0$ was set to 500 for all the present cases in order to deduce the surface tension coefficient $\gamma$. However, figure \ref{fig:LambdaA}(d) considers an alternate Weber number defined using $\lambda$, the streamwise wavelength of the deforming film \cite{wu1993aerodynamic}:

\begin{equation}
We_\lambda = \frac{\rho_\ell U_\infty^2 \lambda}{\gamma},
\end{equation}
where $\gamma$ is that deduced from $We_0 = 500$. $We_\lambda$ is significantly larger than $We_0$, being in excess of 15,000 for the LoRe-thin case at early times. A higher Weber number means the external inertial forces acting on the liquid sheet are large compared to the restoring surface tension forces. As expected, the higher values of $We_\lambda$ for the LoRe-thin case are consistent with our observations that the thin film will be more susceptible to deforming inertial forces, and therefore rupture earlier, than the LoRe-med and LoRe-thick cases. 

\subsection{Aerodynamics of film rupture} 

As discussed in the Introduction, previous studies have suggested the central role of aerodynamic effects in liquid film breakup for a liquid fuel-to-gas density ratio of less than 500 \cite{wu1993aerodynamic}, the present ratio being 40. However, the role of the evolving pressure field has not been thoroughly investigated given the experimental difficulty, if not impossibility, of measuring the pressure field in the vicinity of a liquid surface. This section seeks to investigate these suggested aerodynamic processes using a novel application of classical aerodynamics to quantify the liquid film's deformation and rupture. Our analysis suggests the film's evolution can be broadly understood in terms of inviscid processes, where the liquid film can be effectively understood as a deformable surface until it ruptures. 

\subsubsection{Pressure fluctuations}

\begin{figure}
	\begin{center}
		
		\includegraphics[scale=0.7,clip]{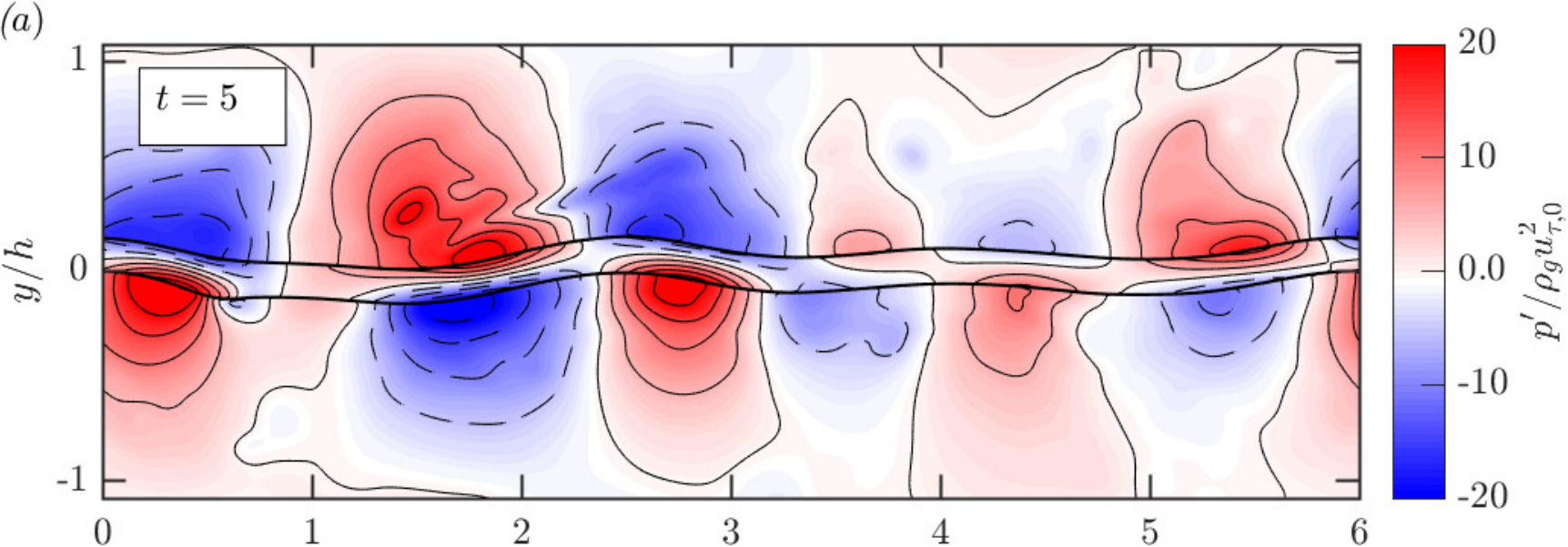}		
		
		\vspace{2mm}
		
		\includegraphics[scale=0.7,clip]{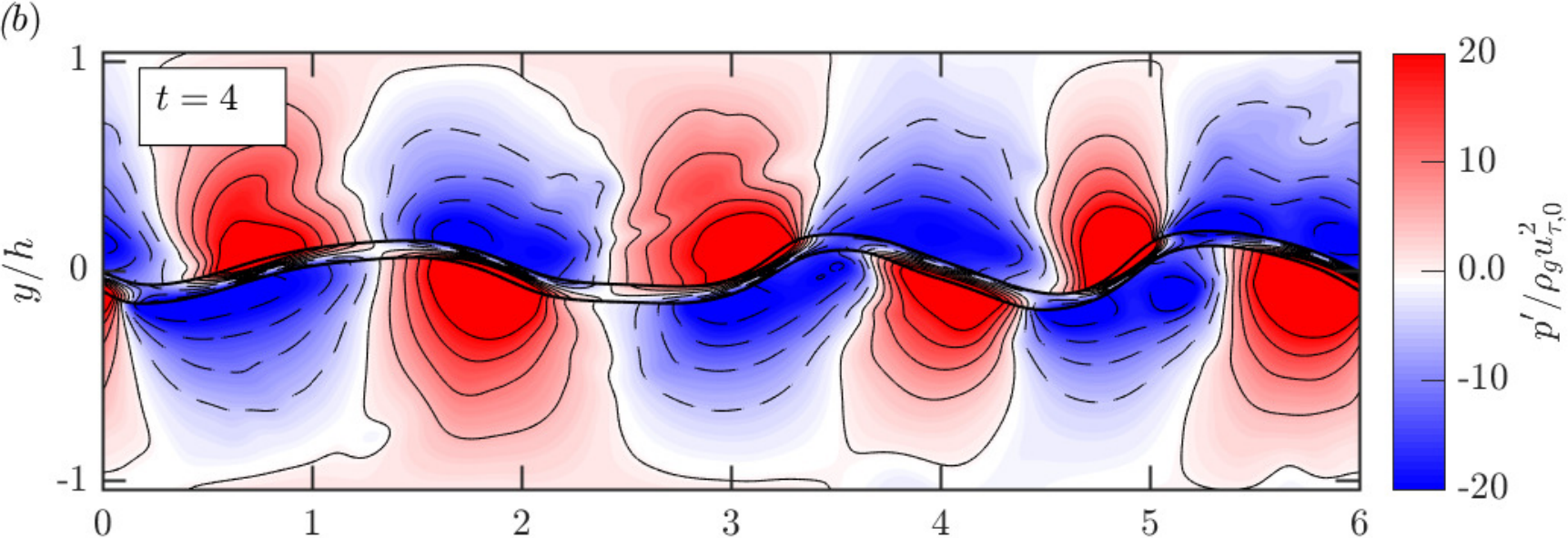}
		
		\vspace{2mm}
		
		\includegraphics[scale=0.7,clip]{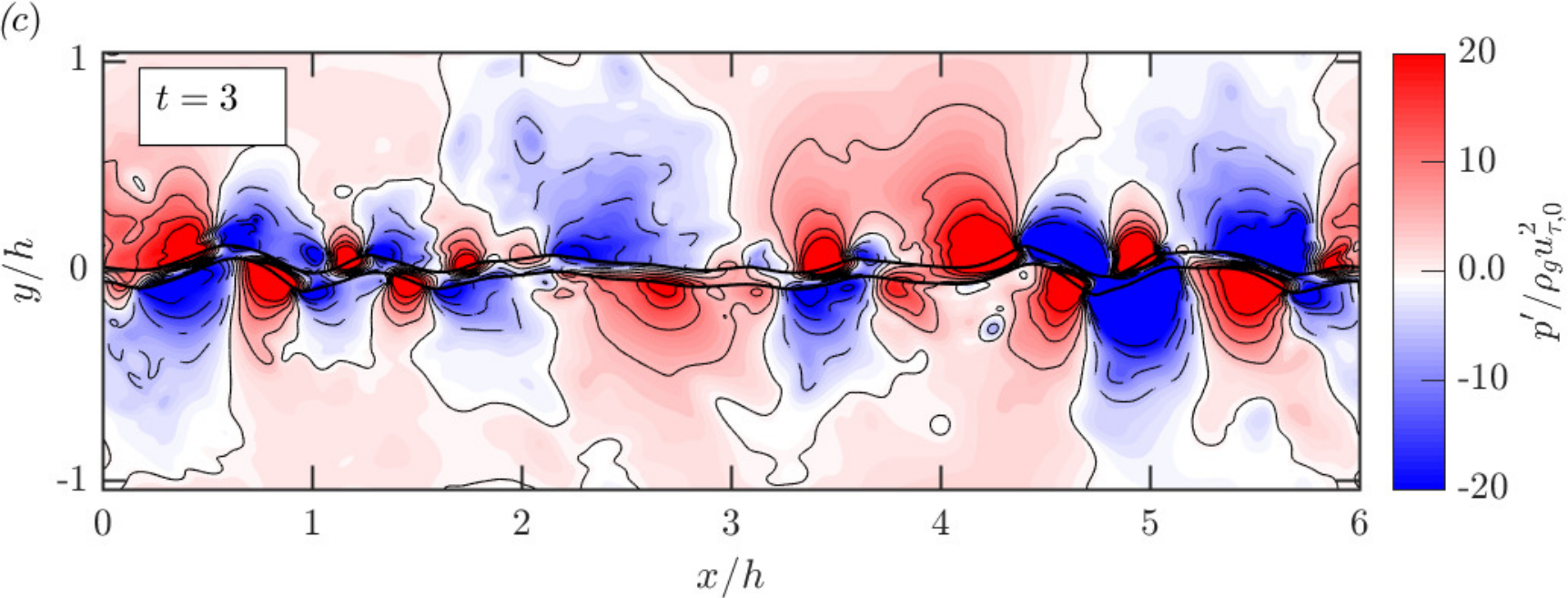}
		
		\caption{Indicative contours of the normalised pressure fluctuations over the deforming film for the different cases, contours are spaced at intervals of $5\, p'/\rho_g u^2_{\tau,0}$: (a) LoRe-med case, with $Re_{\tau,0} = 180$, $t = 5$; (b) LoRe-thin case with $Re_{\tau,0} = 180$, $t = 4$; (c) HiRe-thin case with $Re_{\tau,0} = 393$, $t = 3$. Solid lines denote positive contours and dashed lines negative contours in each case. Thicker black contours denote the surface of the liquid film.}
		\label{fig:pressureFields}
	\end{center}
\end{figure}

Figure \ref{fig:pressureFields} shows contours of fluctuations in the pressure field over the deforming film. They are shown for the same three cases as in figures \ref{fig:filmVisA} and \ref{fig:filmVisB}, but at different times, corresponding to the emergence of significant alternating pressure minima and maxima which directly precedes film rupture. Contours of the film are superimposed highlighting the pressure jump across the gas-liquid interface. Whilst the pressure fluctuations are somewhat weaker in the LoRe-med case (figure  \ref{fig:pressureFields}a), the pattern of minima and maxima nonetheless looks quite similar to that of the LoRe-thin case (figure \ref{fig:pressureFields}b). We note this common pattern emerges at an earlier time ($t=4$) for the LoRe-thin case, compared to the LoRe-med case ($t=5$) because the thicker film delays film deformation and also has a much smaller $We_\lambda$ (figure \ref{fig:LambdaA}d). For the higher Reynolds number case (figure \ref{fig:pressureFields}c) the alternating minima and maxima in the pressure field have a considerably smaller streamwise scale and a clear alternating pattern emerges earlier at $t=3$ as is reflected in the smaller and more contorted features seen in figure \ref{fig:filmVisB}. The alternating minima and maxima in the pressure field suggests that each segment of the deforming film can be considered to effectively function like an airfoil, which motivates the analysis below. Ref. \cite{yang2010direct} simulated the flow over a wavy solid surface to study waves of steepness sufficient to induce boundary layer separation and vortex generation in the wind flow, and considered form drag over the wavy boundaries. In the early stages of the film's deformation, it would seem to share similarities with the a wavy water surface. The simulations of Ref. \cite{yang2010direct} produced similar pressure fields to that shown in the present for the deforming liquid film.

\subsubsection{Lift and drag over the deforming film} \label{sec:LiftDrag}

As mentioned previously, a high Weber number means the deforming external inertial forces are large compared to the restoring surface tension forces. In the present configuration, the film's breakup is therefore likely to be dominated by aerodynamic effects, including the lift force \cite{desjardins2010detailed}. Following the insight gained from figure \ref{fig:pressureFields}, we investigate the development of aerodynamic quantities derived from the pressure field, namely the lift and drag coefficients, over the deforming film using a panel methodology applicable up to the point of film rupture. 

The total force exerted on the film surface may be calculated from the integral of the stress tensor over the film surface,
\begin{equation}
F_{film} = \int_{\Gamma}\left(\left(\boldsymbol{\tau}-p\mathbf{I}\right)\cdot\mathbf{n}\right)\,\mathrm{d}S,
\end{equation}
which can be approximated from the VoF field $\phi$ as:
\begin{equation}  \label{eqn:totalGasForce}
F_{film} = \int_{V}\left(\left(\boldsymbol{\tau}-p\mathbf{I}\right)\cdot\mathbf{n}\,\delta_{\Gamma}\right)\,\mathrm{d}V,
\end{equation}
where $\mathbf{n}\,\delta_{\Gamma}$ is a regularized Dirac delta function computed as $\nabla \phi$ \cite{tryggvason2011direct}. % p. 38 of the Tryggvason book

\begin{figure}
	\begin{center}
		
		\includegraphics[scale=0.8,clip]{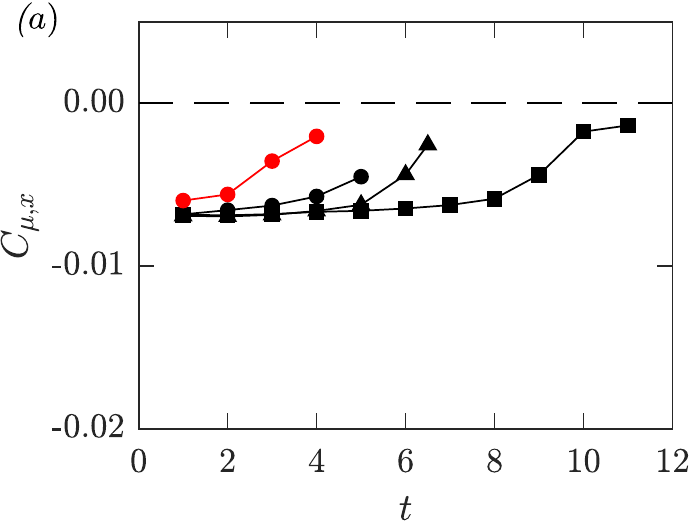}		
		\includegraphics[scale=0.8,clip]{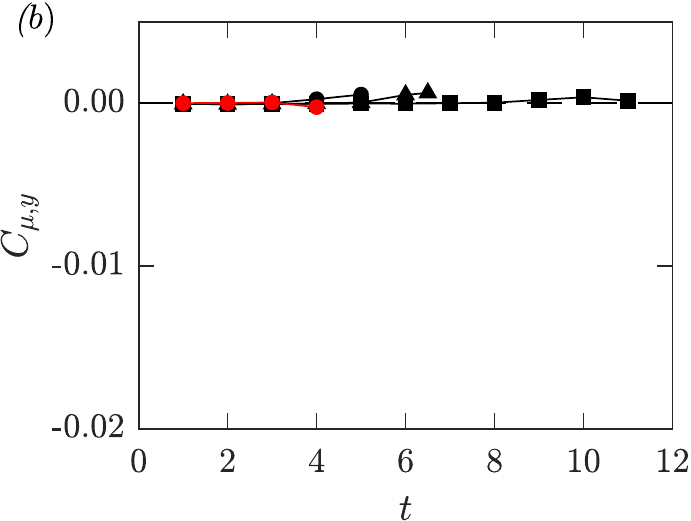}
		\includegraphics[scale=0.8,clip]{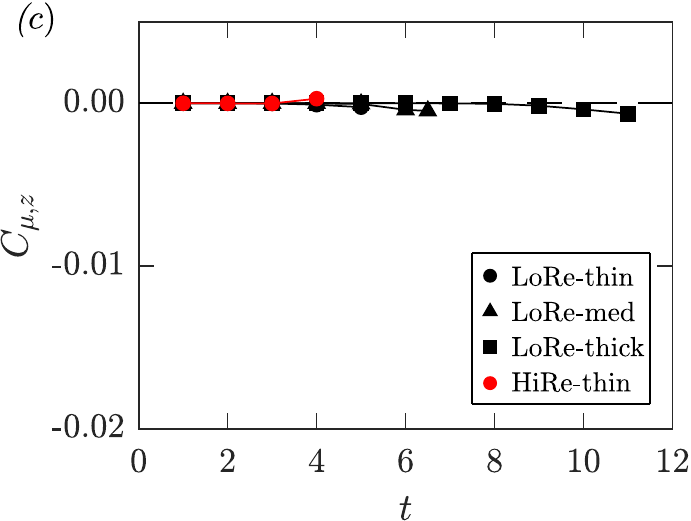}		
		
		\caption{Viscous stress over the deforming film plotted as skin friction drag coefficients, analogous to figure \ref{fig:AeroCoeffs} (note the extent of the vertical axis of these plots is only 1/10th that of the plots in figure \ref{fig:AeroCoeffs}); (a) streamwise $x$ component; (b) film-normal $y$ component; (c) spanwise $z$ component. Symbols as per legend in (c).}
		\label{fig:viscStress}
	\end{center}
\end{figure}

\begin{figure}
	\begin{center}
		
		\includegraphics[scale=0.9,clip]{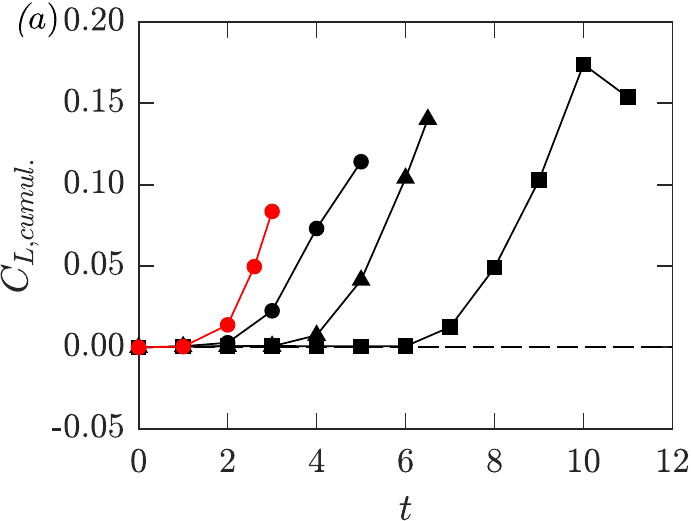}		
		\includegraphics[scale=0.9,clip]{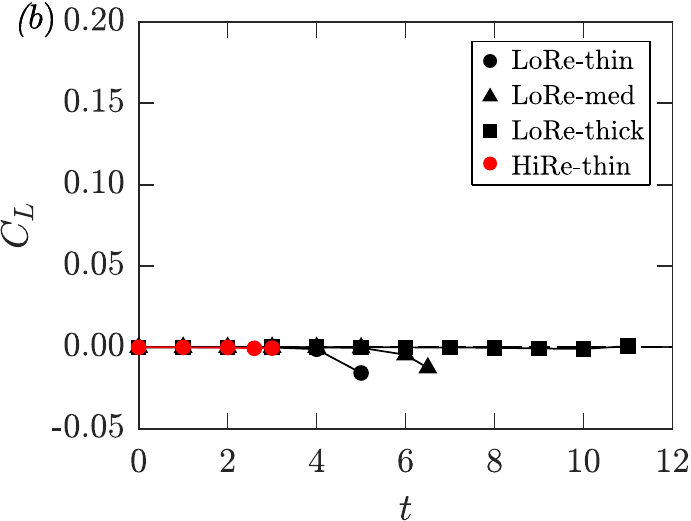}		
	
		\vspace{2mm}
		
		\includegraphics[scale=0.9,clip]{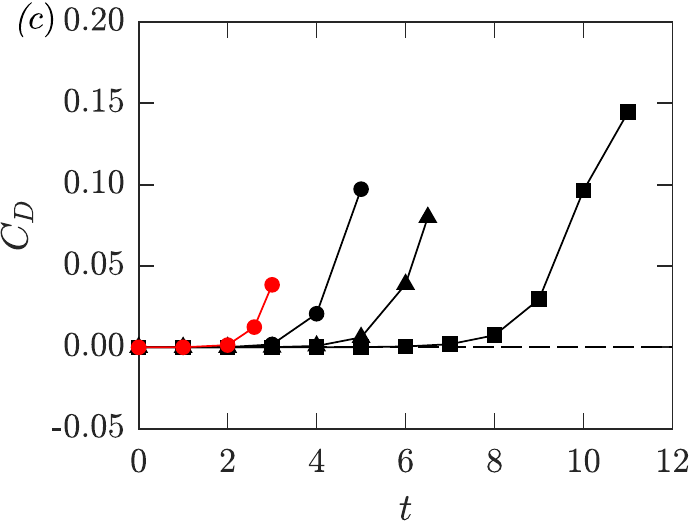}		
		\includegraphics[scale=0.9,clip]{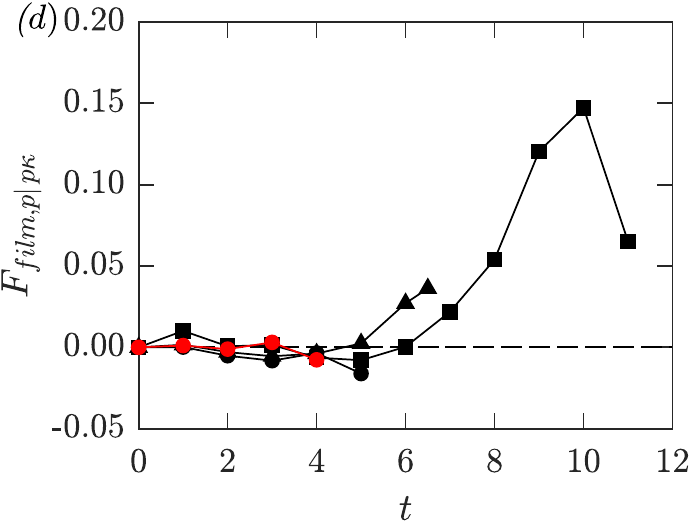}		
		
		\caption{Aerodynamic coefficients over the deforming film for the different cases. Symbols as per legend in (b).}
		\label{fig:AeroCoeffs}
	\end{center}
\end{figure}

The viscous stress components $F_{film, \mu, i}$ of this force (corresponding to the $\bm{\tau} \cdot \mathbf{n}\delta_{\Gamma} $ terms in (\ref{eqn:totalGasForce})) are shown in figure \ref{fig:viscStress}. They are plotted as skin friction drag coefficients (i.e. $C_{\mu,x} = 2\, F_{film, \mu,x}/ (\rho_g U_\infty^2 A_{film})$) to permit comparison to the drag and lift coefficients plotted in figure \ref{fig:AeroCoeffs}. All three viscous components are much smaller than the inviscid pressure components, with only the $x$-component non-negligible. The value of $C_{\mu,x}$ in figure \ref{fig:viscStress}(a) is set by the Reynolds number of the gas flow, therefore that of the HiRe-thin case is slightly higher than all other cases. This force tends toward zero with time as the mean shear profile decays over the duration of the simulations (cf. figures \ref{fig:profiles_in_t}d-f). Overall, these plots confirm the minor role of parasitic drag in the current simulations, and we therefore focus on the role of the inviscid forces, namely the lift and lift-induced drag, on the deformation and rupture of the liquid film.

For this section we make use of the panel segments shown in figure \ref{fig:sampleSegments}. Calculating the lift over the entire film is statistically zero since the problem is symmetric in the film-normal direction. However, it is clear from figure \ref{fig:pressureFields} that there is a local segment-wise lift serving to increasingly deform the film. The lift-induced drag ($F_{D,p}$) and lift ($F_L$) forces are defined as:

\begin{equation} \label{eqn:liftDrag}
F_{D,p} = \int p\, \mathbf{n} \cdot \mathbf{i}\, \mathrm{d} S, \ \ \ \ \  F_L= \int p\, \mathbf{n} \cdot \mathbf{j}\, \mathrm{d} S,
\end{equation}

where $\mathbf{n}$ is the inward point normal vector to the film's surface, $\mathbf{i}$ is the unit normal vector in the streamwise $x$ direction, $\mathbf{j}$ is the is the unit normal vector film-normal direction $y$, and $\mathrm{d}S$ is the elemental surface area of the film. Throughout, this analysis has been performed at each spanwise location. Figure \ref{fig:AeroCoeffs}(a) shows the segment-wise `cumulative' lift coefficient $C_{L,cumul.} = 2\, F_{L,cumul.}/ (\rho_g U_\infty^2 A_{film})$ which is seen to steeply increase with time as the film deforms in the different cases, pointing to the central role of pressure in amplifying film deformation.  The cumulative lift force $F_{L,cumul.}$ is defined as the sum of the absolute values of the lift force $F_L$ on each segment of the film, $U_\infty$ is in this case the (decaying) mean streamwise velocity and $A_{film}$ is the deforming film's surface area. As expected, figure \ref{fig:AeroCoeffs}(b) shows the typical lift coefficient $C_L = 2\, F_{L}/ (\rho_g U_\infty^2 A_{film})$ is zero during the simulations illustrating that the lift felt by each individual panel is cancelled out with that of its neighbor. The action of the lift force is therefore hidden if the lift analysis is not carried out using a panel method approach. Exponential growth in $C_{L,cumul.}$ is observed in the different cases, with all curves ending when the film's surface is no longer intact, since at that point the pressure integration over the film's surface is no longer well-defined. Interestingly, $C_{L,cumul.}$ is shown to decrease at late times just before the film ruptures for the LoRe-thick case. Integration of the $- p \mathbf{I} \cdot \mathbf{\nabla} \phi$ term in (\ref{eqn:totalGasForce}) yields plots virtually identical to figures \ref{fig:AeroCoeffs}(b) and \ref{fig:AeroCoeffs}(c), meaning that the contribution of the stress tensor on the liquid side to the net force on the film, a contribution omitted in the lift and drag calculated using the external pressure according to (\ref{eqn:liftDrag}), is much smaller than that on the gas side. The drag coefficient over the deforming film is defined as $C_D = 2\, F_{D,p}/ (\rho_g U_\infty^2 A_{film})$, where $F_{D,p}$ is the lift-induced drag force over both surfaces of the film. Figure \ref{fig:AeroCoeffs}(c) shows how $C_D$ develops in time for the present cases. $C_D$, calculated over the entire film, develops in a similar manner to the panel-wise  $C_{L,cumul.}$. The drastic increase in the lift-induced drag coefficient as the film deforms compared to the small, and relatively stable, values of the parasitic drag coefficient (figure \ref{fig:viscStress}) again points to the dominance of lift-induced drag and the minor role of parasitic drag. The increasing lift-induced drag and cumulative lift coefficients over the film as it deforms appear to scale consistently with film thickness and therefore may provide a useful means for ultimately predicting time to film rupture. 

We now consider the inviscid pressure component of (\ref{eqn:totalGasForce}), $F_{film, p}$, conditioned on the local curvature $\kappa$:
\begin{equation}
F_{film, p |p\kappa} = \int_{V}\left(\left[p\mathbf{I}\cdot\mathbf{n}\right]|(p\kappa)<0\right)\,\mathrm{d}V
\end{equation}
This quantity attempts to capture the amplification of deformation in the film's surface due to the pressure field, and we will here consider the film-normal $y$ component. If the curvature has the opposite sign of the $y$-component of $F_{film, p}$ (i.e. like pressing down on the inside of a parabola), then the lift force will tend to deform the film further. If it is of the same sign as the curvature, the lift force will tend to reduce the curvature of the film's surface. Figure \ref{fig:AeroCoeffs}(d) shows the vertical component of the conditioned inviscid vertical force. It follows the values of $C_{L,cumul.}$ for the LoRe-thick film case quite closely until $t=10$. It also increases at late times for the LoRe-med case, although it remains significantly smaller than $C_{L,cumul.}$. It is less effective as a measure of the action of the lift force for the thin film cases.
%, possibly because there is high curvature through the film itself which may be clouding the effect at the deforming surface, whereas in the LoRe-thick case high curvature is limited to the surface of the film. 

% for high Re case: film already seems to be breaking up by it = 40000, but there is a marked increas in the L_film_cumul between it = 20000 and 30000 (and a smaller increase betwee 10000 and 20000). Don't need evidence for the lift effect over the whole time - of course this will be small at the beginning, then large, and then no longer valid, but try to quantify when the large increase is seen, and how large the maximal L_film_cumul is. 

% for fine_n32 case, also see large jump in L_film_cumul until it = 50000 (large jump seems to happen between 40000 and 50000)

% could find if the above holds for the present: our "thin plates" also have limited aspect ratio since there is variation in the span - i.e. a 2D assessment might be inappropriate. Could deduce angle via knowledge of the location of markers on each segements

% for reference: old C_D figure from paraview: Cd_vs_t_thickness_study.pdf

% In laboratory measurements of the forces felt by thin flat plates of varying aspect ratios, Ref. \cite{ortiz2015forces} found the ratio of drag to lift closely followed the inverse tangent of the angle of incidence for virtually all measurements. This implies that the forces of interest are due largely to the instantaneous pressure distribution around the plate and are not significantly influenced by shear stresses. In the present film simulations, it would seem the higher-Reynolds number gas phase flow serves to create smaller-scale deformations within the film, yet it is ultimately still the inviscid pressure amplification process that results in the film's rupture. 

\begin{figure}
	\begin{center}
		
		\includegraphics[scale=0.9,clip]{fig/L_film_cumul_allCases.pdf}		
		\includegraphics[scale=0.9,clip]{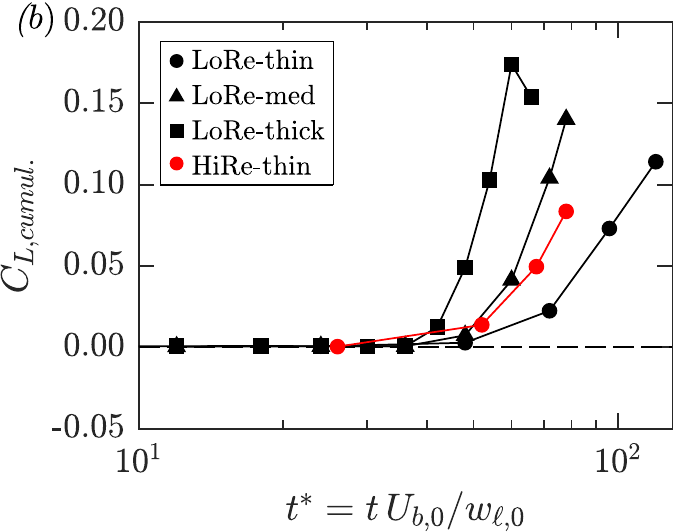}
		
		\includegraphics[scale=0.9,clip]{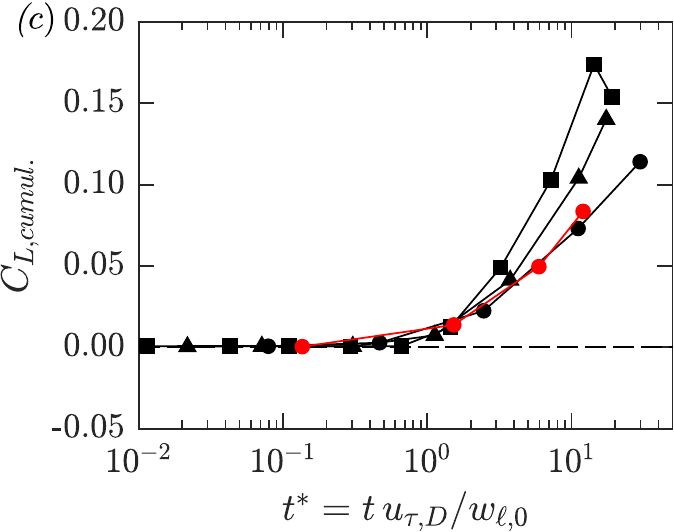}
		\includegraphics[scale=0.9,clip]{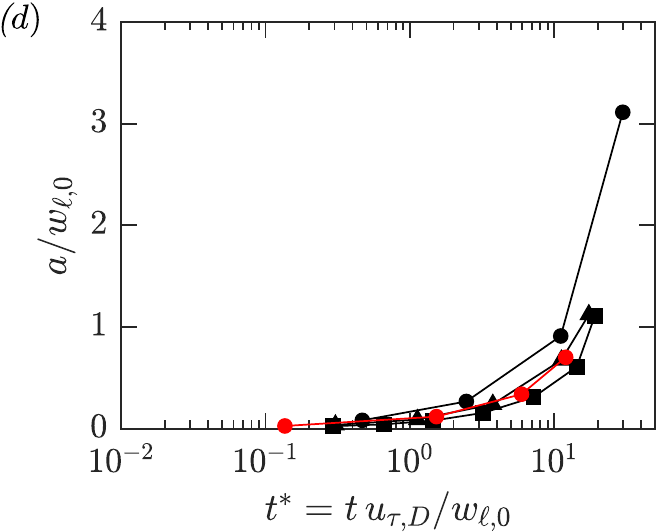}

		\caption{(a) Cumulative Lift coefficient $C_{L,cumul.}$ (same as figure \ref{fig:AeroCoeffs}a); (b) temporal development of $C_{L,cumul.}$ scaled with $t^*$ formed from the starting bulk velocity $U_{b,0}$ and the starting film thickness $w_{\ell,0}$; (c) temporal development of $C_{L,cumul.}$ scaled with $t^*$ formed from $u_{\tau,D}$, a friction velocity derived from the lift-induced drag force $F_{D,p}$ and $w_{\ell,0}$; (d) amplitude of film deformation scaled with $w_{\ell,0}$, measured as a deviation from the original film surface location (figure \ref{fig:LambdaA}b), plotted against the scaled time $t^*$ formed from $u_{\tau,D}$ and $w_{\ell,0}$ as in (c). Symbols as per legend in (b).}
		\label{fig:AeroCoeffs_localuTau}
	\end{center}
\end{figure}

The cumulative lift force $C_{L,cumul.}$ captures the emerging pressure minima and maxima forming on the film surface which results in film rupture. We now seek to collapse the temporal development of $C_{L,cumul.}$ (figure \ref{fig:AeroCoeffs}a) for the different cases. Scaling time with the bulk velocity $U_{b,0}$ and starting film thickness $w_{\ell,0}$ does not appear to collapse the late-time development of $C_{L,cumul.}$ (figure \ref{fig:AeroCoeffs_localuTau}b), although the start of the steep increase in $C_{L,cumul.}$ occurs at $t^* = t U_{b,0} /w_{\ell,0} \approx 40$ for all the cases.
% Figure \ref{fig:AeroCoeffs_localuTau}(b) scales time $t$ with the local friction velocity $u_{\tau,local}$, calculated normal to the deforming film, and the starting film thickness $w_{\ell,0}$. For the purposes of calculating $u_{\tau,local}$, a local mean velocity is defined over each panel (figure \ref{fig:sampleSegments}) of the deforming film. At $t=0$, $u_{\tau,local}$ is equal to the friction velocity of the precursor gas-phase channel simulations, $u_{\tau,0}$. 
As the boundary layer over each panel varies, an alternate velocity scale $u_{\tau,D}$ is calculated from the lift-induced drag force $F_{D,p}$, where $F_{D,p}$ is plotted in the form of a drag coefficient $C_D$ in figure \ref{fig:AeroCoeffs}(c). We define this velocity scale as:

\begin{equation}
u_{\tau,D} = \sqrt{\frac{F_{D,p}}{\rho_g \,A_{film}}}.
\end{equation}

This velocity scale can be interpreted as a kind of mean streamwise inertial force over the film. Against logarithmic horizontal scales, $u_{\tau,D}$ collapses both $C_{L,cumul.}$ and the film's amplitude with some success, even at late times (figures \ref{fig:AeroCoeffs_localuTau}c and \ref{fig:AeroCoeffs_localuTau}d), noting $u_{\tau,D}$ is only defined so long as the film's surface remains intact. 

% from same paper: Our investigation suggests that above the slow wave, there exist two types of coherent vortical structures, namely quasi-streamwise vortices and horseshoe vortices, as shown in figure 29(a).
% idea: deforming film is akin to a steeper wave (higher ak)

\subsubsection{Flow reversal} \label{sec:FlowRev}
% FIGURES
% flow reversed regions correlate to regions of low pressure - or regions of high drag? 

\begin{figure}[h]
	\begin{center}
		
		\includegraphics[scale=0.9,clip]{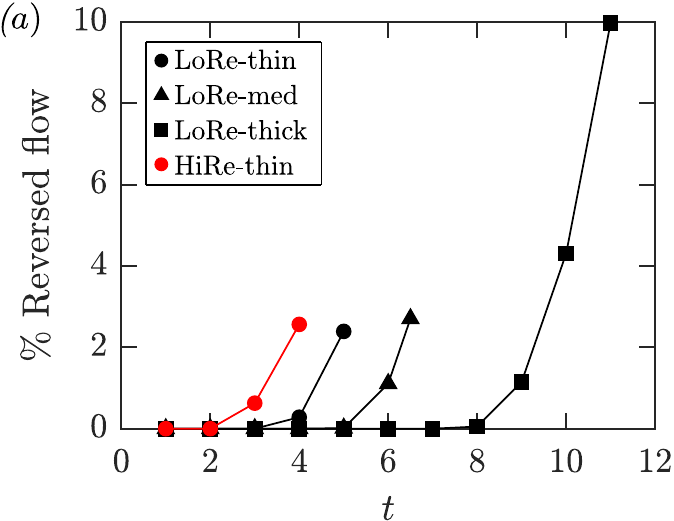}
		\includegraphics[scale=0.9,clip]{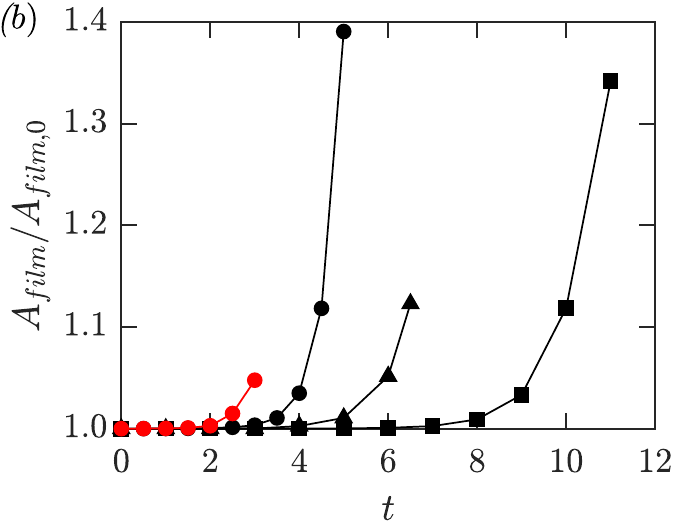}		
		\caption{(a) Percentage of reversed flow in time; (b) increasing total film surface area with time. Symbols as per legend in (a).}
		\label{fig:FlowRev}
	\end{center}
\end{figure}

Deformations of large enough amplitude are accompanied by flow reversal as the gas-phase turbulent boundary layer separates from the liquid film surface, as reported in \cite{ling2019two}. Figure \ref{fig:FlowRev} shows the percentage of reversed flow for the different cases as a function of time. As the film deforms, low pressure regions form (figure \ref{fig:pressureFields}) correlating with recirculation of the flow (white regions in figure \ref{fig:filmVisB}). Significant flow reversal only appears to occur directly before the film ruptures, suggesting its role in this process. The LoRe-thick case can withstand a much larger percentage of reversed flow than the other cases. The total area of the film's surface is shown in figure \ref{fig:FlowRev}(b). The area's increase in time appears to closely follow that of the percentage of reversed flow of figure \ref{fig:FlowRev}(a) with the exception of a much higher relative value of the film's area at $t=5$ for the LoRe-thin case. 

\subsubsection{Impulse on the liquid sheet: linking pressure to film displacement}
% FIGURES
% correlation between vertical displacement of VOF and local lift force
% correlation between change in momentum within the VoF (impulse) and the lift force

% the impulse is calculated at two times - are the current fields spaced closely enough? Although film is initially stationary so the cumulative change in momentum is easily calculated - but the pressure mean is arbitrary (?) so need to look at \Delta P (since there seems in some cases to be a jump early on in the background pressure)

\begin{figure}
	\begin{center}
		
		\includegraphics[scale=0.9,clip]{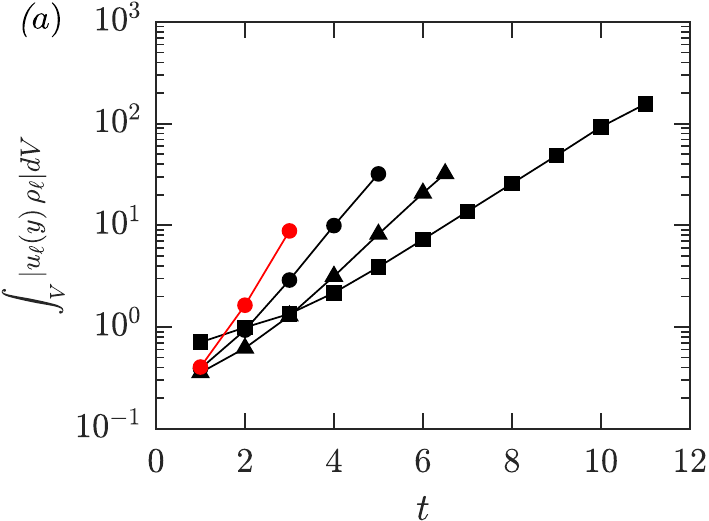}		
		\includegraphics[scale=0.9,clip]{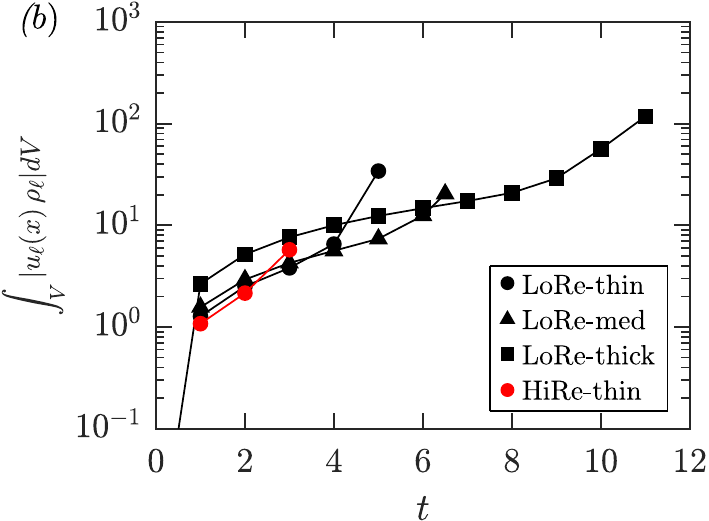}	
		\caption{Impulse on the liquid film over time (change in momentum); (a) film-normal impulse; (b) streamwise impulse. Symbols as per legend in (b).}
		\label{fig:Impulse}
	\end{center}
\end{figure}

The pressure impulse theory for incompressible liquid impact problems was described in Ref. \cite{cooker1995pressure} as an inviscid process where the pressure impulse is $P(x,y,z,t) = \int_{t_b}^{t_a} p(x,y,z,t)\, \mathrm{d}t$, for times $t_a$ and $t_b$ immediately before and after `impact'. In the present work the `impact' refers to the largely inviscid action of the gas phase on the quiescent liquid film. $P$ is related to the change in momentum of the liquid phase as:

\begin{equation}
(\mathbf{u}_{\ell,a}  - \mathbf{u}_{\ell,b})\rho_\ell = -\nabla P.
\end{equation}

where $\mathbf{u}_{\ell,a} = u_\ell(x,y,z,t_a)$ and $\mathbf{u}_{\ell,b}  = u_\ell(x,y,z,t_b)$ are the velocities of the liquid film phase at times $t_a$ and $t_b$. Since the liquid film is initially at rest, in the present context the integrated pressure impulse at any given time is equal to the momentum of the liquid phase. Figure \ref{fig:Impulse} plots the momentum of the liquid film in time (i.e. between the contours marking the top and bottom surface of the film as shown in figure \ref{fig:sampleSegments}). Both subfigures in figure \ref{fig:Impulse} plot the sum of the absolute value in the change of momentum within the film calculated for each panel. Similar to the lift coefficient, the change in the $y$-momentum (film-normal direction) would be statistically zero if the panel methodology is not used. Figure \ref{fig:Impulse}(a) shows the change in $y$-momentum grows exponentially in time in a similar manner to the cumulative lift coefficient (figure \ref{fig:AeroCoeffs}a) and lift-induced drag coefficient (figure \ref{fig:AeroCoeffs}c). In contrast, there is a more steady growth in the change of $x$-momentum (streamwise direction), with more similar behaviour for the different cases at early times. Note the uptick in the $x$-momentum seen for all the cases at late times. This results from the large emerging film deformations causing large changes in the streamwise flow, including flow reversal. For example this occurs by $t \approx 3$ for the HiRe-thin case, but doesn't occur until $t \approx 9$ for the LoRe-thick case. This points to different origins in the change in momentum in the different directions: a change in $y$-momentum can only occur if the film deforms since the problem is symmetric. In contrast, even a film that is not deforming will have $x$-momentum imparted to it by the `dragging' motion of the co-flowing gas-phase boundary layers on either side of it. 

\section{Conclusions}  \label{sec:conclusions}

The aerodynamically-driven rupture of a liquid film has been simulated using a VoF method. The present results demonstrate the utility of this novel numerical setup, utilizing realistic fully developed turbulent gas-phase shear layers from a channel simulations, to parametrically investigate the role of important physical parameters such as film thickness and gas-phase Reynolds number in liquid film deformation and rupture. Aligning with results of the results of Ref. \cite{warncke2019new}, the presence of more turbulent scales at the higher Reynolds number considered here hastens film rupture, which come about due to the presence of smaller scales in the high-Reynolds number gas phase. Rupture of the liquid sheet appears to occur due to the amplification of aerodynamic lift over segments of the deforming film, at a scale much larger than instabilities predicted by linear stability analysis. A  cumulative lift force is introduced to capture the alternating pressure minima and maxima forming over the film which amplify to eventually rupture the film. A velocity scale derived from the lift-induced drag force collapses the development of this cumulative lift force as well as the amplitude of film deformation with some success for the different film thicknesses and Reynolds numbers considered in the present work. This setup will be used in future simulation campaigns building on the present `cold' simulations to study the transfer of heat and species via deforming and moving gas-liquid interfaces to understand ensuing interfacial thermodynamics within a turbulent environment.

\section*{Acknowledgements}

The simulations were performed on resources provided by UNINETT Sigma2 – the National Infrastructure for High Performance Computing and Data Storage in Norway. Post-processing was performed on resources provided by the NTNU IDUN/EPIC computing cluster \cite{sjalander+:2019epic}. PC acknowledges funding from the University of Iceland Recruitment Fund grant no. 1515-151341, TURBBLY.

\section*{Appendix: resolution of thin film case}

The thin film case is simulated at two different resolutions the effect of which is considered here. The LoRe-thin case is that presented in the main body of the paper, and here we consider another lower resolution of the same physical setup named LoRe-thin-loRes, which has the same starting grid resolution as the LoRe-med, LoRe-thick and HiRe-thin cases of table \ref{tab:FilmSimulations}. The two thin film cases are shown here to broadly evolve in a similar manner. Some of the differences between the two cases may be due to the fact that the present simulations are only modestly statistically converged, since the current dataset relies on single realisations of each case.

\begin{table}[h]
	\setlength{\tabcolsep}{7pt} % change this to widen/narrow columns
	\def\arraystretch{1.1}
	\begin{center}	
		\begin{tabular}{ccccS[table-format=1.1]ccccS[table-format=1.2]S[table-format=1.2]S[table-format=1.2]}  
			\hline \hline 
			case name & symbol & $\Rey_{\tau,0}$   & $w_{\ell,0}/h$  & $w_{\ell,0}^{+}$ & $N_x$ & $(N_{\ell, y})_0$ & $N_{y}$ & $N_z$ &  $(\Delta x^{+})_{0}$ & $(\Delta y^{+})_{0}$  & $(\Delta z^{+})_{0}$ \\  
			\hline
			LoRe-thin &\textcolor{black}{$\bullet$} & 180               &   $1/12$          & 15.0     & 2304    & 32     	    & 800    		  &  1152    & 0.47         & 0.47   & 0.47\\ 	
		    LoRe-thin-loRes & \textcolor{blue}{$\bullet$} & 180               &   $1/12$         & 15.0      & 1152    & 16             & 400    	      &  576     & 0.94         & 0.94    & 0.94 \\ 	
		    \hline \hline 
		\end{tabular}
		\caption{Parameters for the companion thin liquid film simulations. Subscript `$0$' refers to parameters at the instant the film simulations are started. Fully developed turbulent velocity fields from the channel simulations (table \ref{tab:ChannelSimulations}) form the gas phase; $\Rey_{\tau,0}$ refers to the Reynolds number of this gas phase. The viscous scaled initial film thickness is $w_{\ell,0}^+ = \Rey_{\tau,0} (w_{\ell,0}/h)$. Both simulations have domain size $L_x/h = 6$, $L_y/h = 2 + w_{\ell,0}$, $L_z/h = 3$.}
		\label{tab:FilmSimulationsThin} 
	\end{center}	
\end{table}

\begin{figure}		
	\setlength{\tabcolsep}{10pt} % change this to widen/narrow columns
	\def\arraystretch{3}
	\begin{center}	
		\begin{tabular}{ccc} 
		
\includegraphics[scale=0.68,clip]{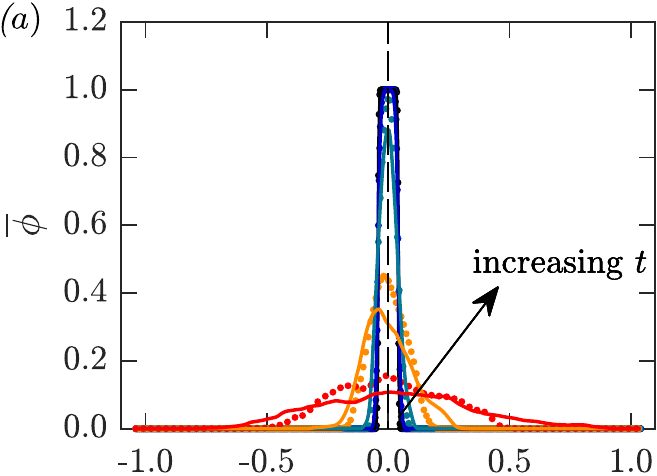} &
\includegraphics[scale=0.68,clip]{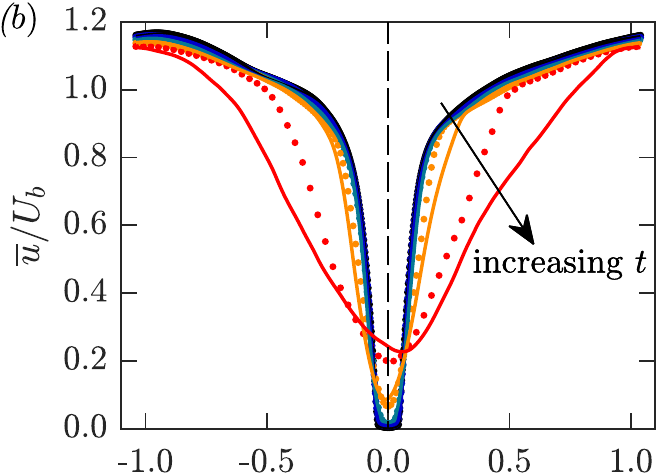} &
\includegraphics[scale=0.68,clip]{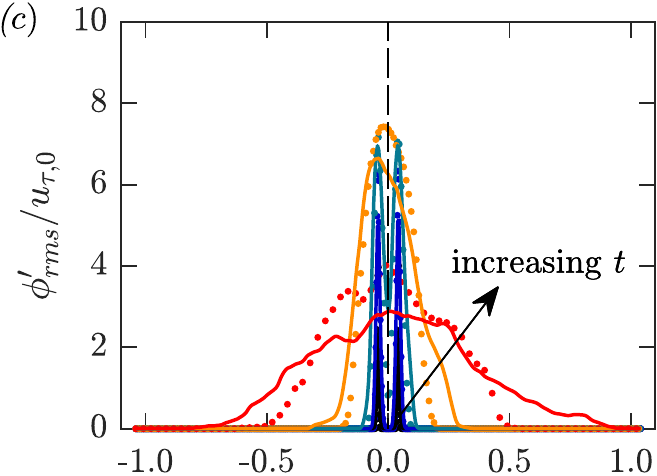} \\

\includegraphics[scale=0.68,clip]{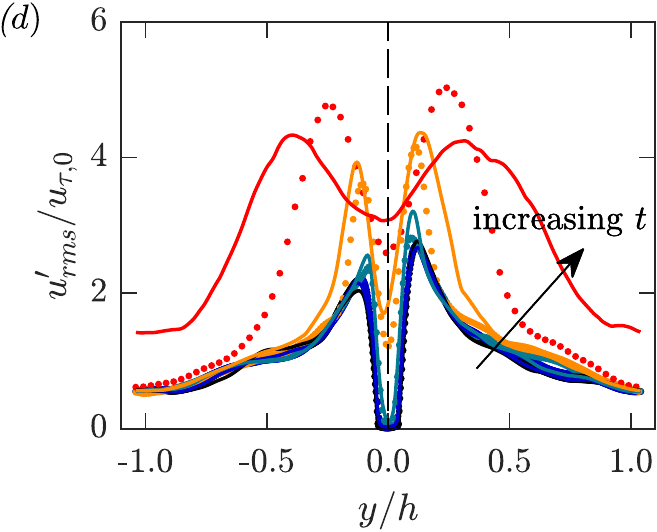} &
\includegraphics[scale=0.68,clip]{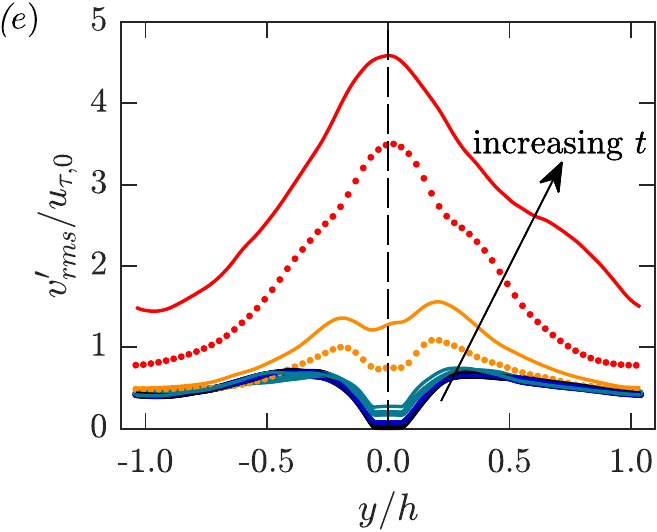} &
\includegraphics[scale=0.68,clip]{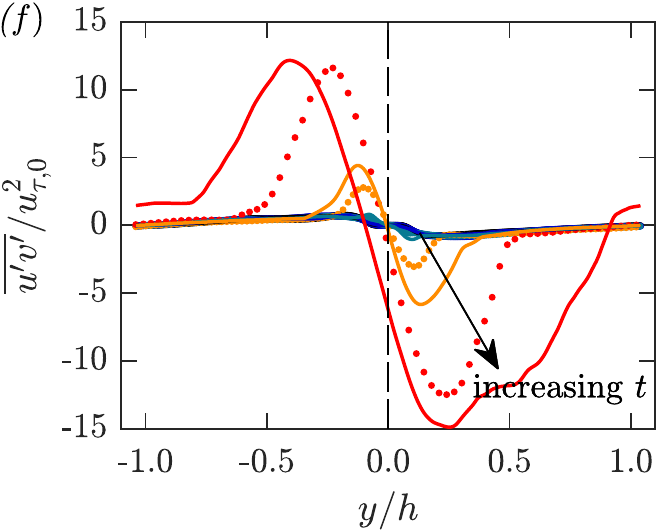} \\

 \end{tabular}	
		\caption{Comparing the temporal development of statistics averaged in the homogeneous $xz$-planes for the thin film at two different simulation grid resolutions; bullets, LoRe-thin case (same as in figure \ref{fig:profiles_in_t}); lines, LoRe-thin-loRes case. Curves colored by time: \textcolor{black}{\drawline{22}{1.5}}, $t = 1$; \textcolor{royal_blue}{\drawline{22}{1.5}}, $t = 2$;
			\textcolor{teal}{\drawline{22}{1.5}}, $t = 3$;
			\textcolor{burnt_orange}{\drawline{22}{1.5}}, $t = 4$;
			\textcolor{red}{\drawline{22}{1.5}}, $t = 5$; \dashed, domain centerline.}
		\label{fig:CompareThinFilmTemporal}
	\end{center}
\end{figure}

\begin{figure}		
	\begin{center}		
	\includegraphics[scale=0.8,clip]{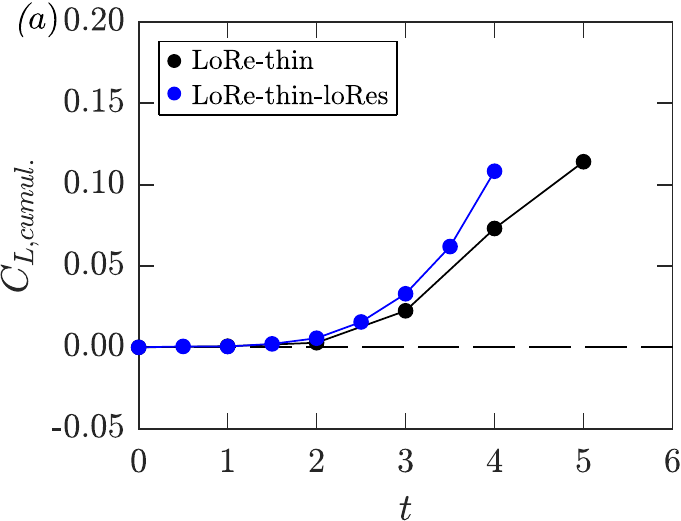}	
	\includegraphics[scale=0.8,clip]{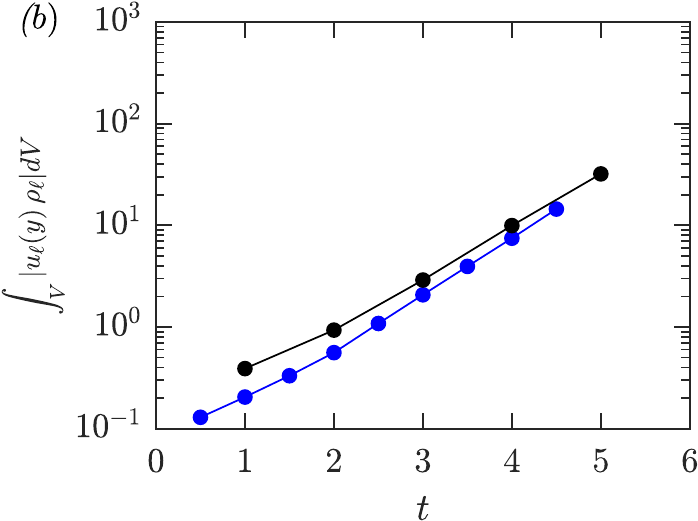}	
	
    \includegraphics[scale=0.8,clip]{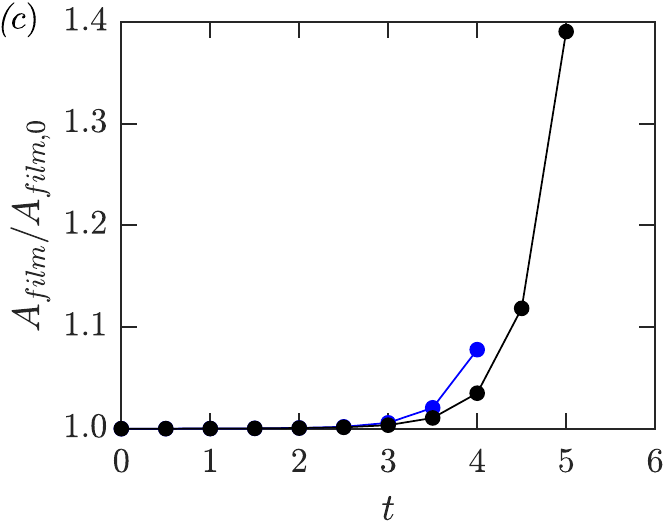}	 
	\includegraphics[scale=0.8,clip]{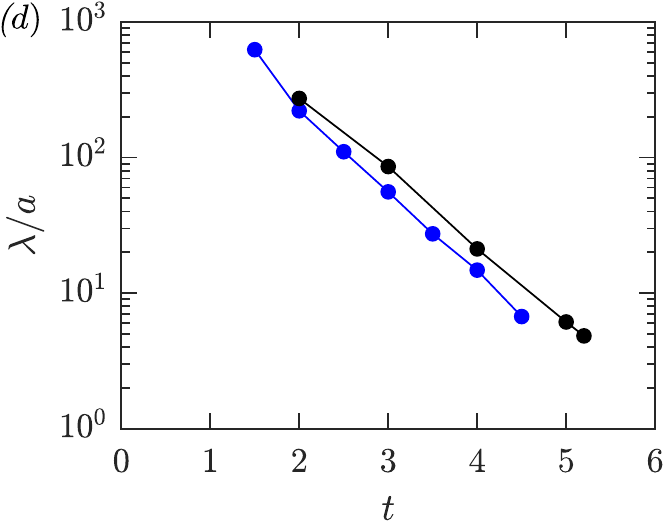}	 				

		\caption{Comparing the temporal development of bulk quantities for the thin film at two different simulation grid resolutions. Symbols as per legend in (a).}
		\label{fig:CompareThinFilmBulk}
	\end{center}
\end{figure}

Figure \ref{fig:CompareThinFilmTemporal} compares the temporal development of the LoRe-thin and LoRe-thin-loRes cases, reproducing the plots of figure \ref{fig:profiles_in_t}. There is some disagreement at late times, however the curves at specific times are overall similar, especially in the context of the large differences found between the cases differing in film thickness and Reynolds number in figure \ref{fig:profiles_in_t}. Larger differences are found for second order statistics (figures \ref{fig:CompareThinFilmBulk}d, \ref{fig:CompareThinFilmBulk}e and \ref{fig:CompareThinFilmBulk}f) at later times ($t=4$ and $t=5$), where $t=5$ is approximately the time the film ruptures. More uncertainty is expected upon film rupture, since for the present VoF method topology changes take place (primarily occurring due to local film rupture or thread breaking) when the resolution of a film or thread is sufficiently low \cite{lu2019multifluid} meaning the size of the liquid structure approaches that of the grid size. 

Figure \ref{fig:CompareThinFilmBulk} compares the temporal development of bulk quantities for the LoRe-thin and LoRe-thin-loRes cases. Small discrepancies are found for late times ($t=4$ onwards) when the LoRe-thin-loRes has already begun to rupture due to fine structures reaching the size of the grid spacing triggering a change in topology. Nevertheless, good agreement is found for the two cases.

\vspace{20mm}

\bibliography{LiquidFilmBib}

\end{document}